\documentclass[%
 reprint,
 amsmath,amssymb,
 aps,
]{revtex4-2}

\usepackage{graphicx}
\usepackage{dcolumn}
\usepackage{bm}

\begin{document}

\preprint{APS/123-QED}

\title{Cluster dynamics studied with the phase-space Minimum Spanning Tree approach}

\author{Viktar Kireyeu}
\affiliation{%
Joint Institute for Nuclear Research, Joliot-Curie 6, 141980 Dubna, Moscow region, Russia\\
Helmholtz Forschungsakademie Hessen für FAIR (HFHF), GSI Helmholtzzentrum für Schwerionenforschung, Campus Frankfurt, Max-von-Laue-Str. 12, 60438 Frankfurt am Main, Germany
}%

\date{\today}

\begin{abstract}
The origin of weakly bound objects like clusters and hypernuclei, observed in heavy-ion collisions, is of 
theoretical and experimental interest. It is in the focus of the experiments at RHIC and LHC since it is not evident how such weakly bound objects can survive in an environment whose hadronic decay products point to a temperature of the order of 150 MeV. It is as well one of the key research topics in the future facilities of FAIR and NICA which are under construction in Darmstadt (Germany) and Dubna (Russia), respectively. We present here first results on the cluster dynamics within the model-independent cluster recognition library "phase-space Minimum Spanning Tree" (psMST) applied to different transport approaches: PHQMD, PHSD, SMASH and UrQMD. The psMST is based on the "Minimum Spanning Tree" (MST) algorithm for the cluster recognition which exploits correlations in coordinate space, and it is extended to correlations of baryons in the clusters in momentum space.  We show the sensitivity of the cluster formation on the microscopic realization of the n-body dynamics and on the potential interactions in heavy-ion collisions. 
\end{abstract}

\maketitle


\section{\label{sec:1}Introduction}
Novel accelerator complexes provide us with the interesting possibility to produce and to study a new type of matter which can not exist under the "normal" conditions -- the quark-gluon plasma (QGP). A "soup" of quarks and gluons, which existed at the first microseconds after the Big Bang, can be created at facilities like the Relativistic Heavy Ion Collider (RHIC) and the Large Hadron Collider (LHC). In future experiments at lower energy, at the Nuclotron based Ion Collider fAcility (NICA) in Dubna and at the Facility for Antiproton and Ion Research (FAIR) in Darmstadt, extend these experiments towards a strongly interacting matter of high net baryon density ~\cite{Kekelidze:2016hhw}.

Cluster and hypernuclei production, experimentally observed in a wide energy range  \cite{Schuttauf:1996ci,Sfienti:2006zb,Nebauer:1998fy,Reisdorf:2010aa,Rappold:2015una,Anticic:2016ckv,Abelev:2010rv,Agakishiev:2011ib,Adam:2015yta,Adam:2015vda,Acharya:2017bso}, is interesting for several reasons: first of all, the mechanism of cluster formation in nucleus-nucleus collisions is not well understood and requires further investigations. For deuterons and tritons having a small binding energy (of several MeV) compared to freeze-out temperature (around 100-150 MeV), it is very unlikely that they will survive a collisions with another hadron. 
Consequently, it is most probable that the observed deuterons and tritons, as well as the significant fraction of few-nucleon bound states registered near midrapidity, are produced in the late stage of the reaction close to the freeze-out point. Thus, the light nuclei, observed in the experiment and formed near the freeze-out, may provide information on the space-time structure of this late stage of the collision. 
The capture of the produced hyperons by clusters of nucleons leads to hypernucleus formation which is a very rare process at strangeness threshold energies. Hypernuclei are an unique probe for improving our knowledge on the strange particle-nucleon interaction in a many-body environment and under the controlled conditions. The theoretical description of the cluster and hypernuclei formation is quite sensitive to the strange baryon-nucleon interaction which can therefore be tested in such heavy ion collisions.

Most of the transport approaches, developed for the dynamical description of heavy-ion collisions, use statistical \cite{Botvina:2011jt,Botvina:2016wko} or coalescence models \cite{Gutbrod:1988gt,Gosset:1976cy,Lemaire:1980qw,Sato:1981ez} to describe the cluster formation, or omit it. Recently these approaches were applied to the Ultrarelativistic Quantum Molecular Dynamics approach \cite{Gaebel:2020wid,Sombun:2018yqh}.
Another possibility is to use the "Minimum Spanning Tree" (MST) \cite{Aichelin:1991xy} procedure. While the coalescence model requires a multitude of parameters for each isotope, the MST uses only the coordinate space information of the baryons to identify clusters at the end of the reaction when free and bound nucleons are well separated in space. 

To overcome the MST limitation that clusters can be formed only at the end of reaction, the Simulated Annealing Clusterization Algorithm (SACA) \cite{Puri:1996qv,Puri:1998te} was developed based on the idea of Dorso and Randrup \cite{Dorso:1992ch}, that the most bound configurations of clusters and free nucleons, identified during the collision evolution, have a large overlap with the final distribution of clusters and free nucleons. Thus SACA can be applied to identify clusters at earlier times, shortly after the passing time, when interactions between nucleons are still on-going and the nuclear density is high. For each possible configuration of clusters and unbound nucleons SACA calculates the total binding energy (neglecting interactions among clusters). The most bound configuration is found by the simulated annealing technique using a "Metropolis" algorithm. An extended version of SACA, the "Fragment Recognition In General Application" (FRIGA) \cite{Fevre:2019lll} is under development now. The FRIGA includes the symmetry and the pairing energy as well as hyperon-nucleon interactions.

The extended study of the cluster formation mechanisms within the MST and SACA procedures have been performed recently employing the Parton-Hadron-Quantum-Molecular Dynamics approach \cite{Aichelin:2019tnk}. In this study, the importance of the dynamical n-body description of the nucleon-nucleon interactions has been demonstrated. It allows to conserve spacial and momentum correlations of nucleons and hyperons, which yields to the dynamical formation of (hyper)nuclei which later on can be recognized by MST and SACA.

The application of the MST ideas to other transport models would extend our understanding on the cluster formation and on the consequences of different realizations of nucleon-nucleon interactions. In particular, it allows to study whether the n-body quantum-molecular dynamics (QMD) approach provides for clusters at midrapidity different results than approaches in which the one body phase space density is propagated or which do not at all include potential interactions. Furthermore, this also allows studying of the impact of different matter equations of state (EoS) on the cluster formation process. 

Therefore we advance the "phase-space Minimum Spanning Tree" (psMST) approach for the simulation of the cluster production, which can be applied to any transport model and allow addressing multiple physics phenomena in the (hyper)nuclei formation process.  In addition, it can be also used for numerous feasibility study tasks for future experimental setups at NICA and FAIR.

This paper is organized as follows: in Section \ref{sec:2} a description of the psMST library is provided. A brief description of transport approaches to which psMST is applied is presented in Section \ref{sec:3}. First results for the cluster formation using psMST for four transport approaches are shown in Section \ref{sec:4}. Finally, conclusions are presented in Section \ref{sec:5}.

\section{\label{sec:2}psMST}
The phase-space Minimum Spanning Tree (psMST) is based on the idea of the MST algorithm for searching bound nucleon systems in dense hadronic matter \cite{Aichelin:2019tnk,Aichelin:1991xy}.
Since it is an independent library, it can be applied  to all transport models which propagate hadrons. This includes models based on n-body dynamics like the different flavours of the quantum-molecular dynamics (QMD) approach as well as models based on mean-field (MF) dynamics or cascade approaches in which no potential interactions between nucleons takes place. Thus, a model+psMST combination allows us to study the impact of particular realization of nuclear dynamics on the clusters formation process. Moreover, in addition to the spacial correlations, used in the default MST version for cluster recognition, psMST can be used to study the influence of the momentum correlations of nucleons and hyperons for the formation of (hyper)nuclei. The psMST library is a open source code \cite{psMSTlinkGitLab}, which can be used either in the stand-alone mode or can be integrated in any software framework for detector simulation and analysis.

In this work we study three different scenarios of using the psMST library:
\begin{itemize}
\item Scenario 1.
 Similar to the MST procedure, the psMST algorithm is only dealing with the coordinate information to find clusters: a pair of particles $(i,j)$ forms a cluster if the distance between candidates is less than the "clusterization" radius which is chosen to be $r_{clust} = 4$ fm, i.e. $\Delta r \leq r_{clust}$. 
The distance $\Delta r$ is calculated in the pair center of mass frame. 
 Moreover, a particle is assigned to the existing cluster if it meets the same condition as above with at least one  particle of the cluster. In this mode psMST results are identical to the original MST procedure.
\item Scenario 2.
 The first step is identical to the "scenario 1", but then, once all possible combinations for clusters are found, an additional momentum cut is applied for candidates. First, the cluster velocity is calculated as 
 \begin{eqnarray}
  {\bf V} = \frac{\sum_{i=1}^{n} {\bf p_{i}}}{\sum_{i=1}^{n} E_{i}}
 \end{eqnarray}
  where ${\bf p_{i}}$, $E_{i}$ are the $i$-th particle momentum and energy in the calculational frame of heavy-ion collisions
  (which is often taken as the $N+N$ center of mass frame),
  $n$ is the number of particles assigned to the cluster.
 Then the momentum of each particle assigned to the cluster is boosted to the cluster center of mass frame by the corresponding Lorentz transformation with the velocity ${\bf V}$: 
 \begin{eqnarray}
 {\bf p'}=L({\bf V}) {\bf p}.
 \end{eqnarray}
 In this scenario we apply the momentum cut for all found clusters in order to investigate the deviation of their momentum distribution from the Fermi momentum distribution expected for stable clusters. We exclude a cluster from the analysis if at least 
 one its particle has a momentum ${\bf p'} > 300$~MeV in the cluster rest system.
\item Scenario 3.
In this scenario the clusterization criterion of the spacial MST (scenario 1) is extended by a cut in momentum space
depending on the relative momentum of cluster particles:
\begin{itemize}
 \item As in the "scenario 1", the algorithm is looking first for the coordinate space information. Particles are selected  as a cluster candidates if the distance between two particles $(i,j)$ (or with at least one particle of the cluster in case of already existing cluster) $\Delta r \leq r_{clust }$. The distance $\Delta r$ is calculated in the particles pair center of mass frame $(i+j)$. 
 \item Then an additional momentum cut is applied to each particle: a particle can be added to the cluster only if each particle of this cluster has a relative momentum ${\bf p'} < 300$ MeV in the cluster center of mass frame. This procedure is repeated after a new particle is assigned to the cluster after the proximity $\Delta r$ check.
\end{itemize}
 Contrary to the scenario 2, the momentum cut is applied during the clusterization procedure. This leads to a different cluster distribution since the particles rejected from one cluster by the momentum cut can become a member of another cluster.
\end{itemize}

\section{\label{sec:3}Application of the psMST to transport models}
In order to study the similarity and possible differences of the cluster formation within different transport models (based on mean-field, cascade  or QMD dynamics) the psMST algorithm was applied to four transport approaches: PHSD-4.0, PHQMD-2.0, SMASH-2.0 and UrQMD-3.4 at two energies $\sqrt{s_{NN}} = 2.52$ GeV and $\sqrt{s_{NN}} = 8.8$ GeV. The basic ideas of these approaches are briefly mentioned here.

$\bullet$ The Parton-Hadron-String Dynamics (PHSD) \cite{Cassing:2008sv,Cassing:2009vt} is a microscopic off-shell transport approach that describes the evolution of a relativistic heavy-ion collision from the initial hard scatterings and string formation through the dynamical deconfinement phase transition to the quark-gluon plasma as well as hadronization and the subsequent interactions in the hadronic phase. It is based on the solution of Kadanoff–Baym equations in first-order gradient expansion employing "resummed" propagators from the dynamical quasiparticle model (DQPM) \cite{Cassing:2007nb, Cassing:2007yg} for the partonic phase. At lower energies it reduces to a hadronic transport model. PHSD incorporates a density dependent Skyrme potential at low (SIS) energies and the covariant momentum dependent potential at high energies.

$\bullet$ The Parton-Hadron-Quantum-Molecular Dynamics (PHQMD) \cite{Aichelin:2019tnk} is a n-body dynamical transport approach which is designed to provide a microscopic dynamical description for the formation of light and heavy clusters and hypernuclei as well as for hadrons in relativistic heavy-ion collisions. 
The propagation of baryons is based on the n-body QMD dynamics while the description of mesons and of the QGP dynamics as well as the collision integral was taken from the PHSD model. The PHQMD includes mutual 2-body density dependent Skyrme type potentials for the interaction among baryons. The attractive interaction binds clusters with a binding energy of about 8 MeV/N.

$\bullet$ The Simulating Many Accelerated Strongly-interacting  Hadrons (SMASH) \cite{Weil:2016zrk,dmytro_oliinychenko_2020_4336358}  model is a hadronic transport approach which provides a dynamical description of heavy-ion reactions in the low and intermediate beam energy range. The relativistic Boltzmann equation with hadronic degrees of freedom is solved including an (optional) density dependent Skyrme type mean field potential. In this study the SMASH model is used in its default version without potential.

$\bullet$ The Ultrarelativistic Quantum Molecular Dynamics (UrQMD) \cite{Bass:1998ca,Bleicher:1999xi} model is a microscopic transport approach which describes hadronic reactions at low and intermediate energies in terms of collisions among hadrons and their resonances. At higher energies the multi-particle production within UrQMD model is dominated by the excitation of color strings and their subsequent fragmentation into hadrons. We used here the default version of UrQMD without potentials.

Thus, we use in our study one model based on quantum-molecular dynamics (PHQMD with a Skyrme potential for baryons), one mean field model (PHSD, based on the Kadanoff–Baym equation including potentials for baryons) and two cascade approaches (UrQMD and SMASH).
We apply the psMST algorithm to "snapshots" from these models at different times and compare the results for the yields of clusters. A "snapshot" means that at a given time all coordinates and momenta of all baryons are stored for a further psMST analysis. We point out that all resonance states are excluded  from the cluster recognition process.
However, we note  that the standard output from SMASH and UrQMD contains also particles under formation time and those are also included in the clusters finding procedure. 
The PHSD and PHQMD provide a separate output for the formed baryons only, used for the clusters recognition algorithm. 
A study with PHQMD shows that including unformed nucleons in the MST, additionally to the formed nucleons, gives only small differences at early time steps.

\section{\label{sec:4} Results for the cluster formation}
\begin{figure*}[h!]
    \resizebox{\textwidth}{!}{
        \includegraphics{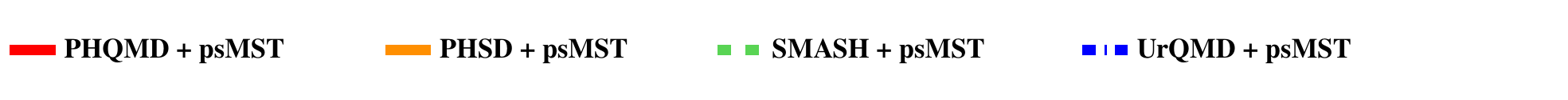} 
    } \\
    \resizebox{\textwidth}{!}{
        \begin{tabular}{ccc}
          \includegraphics{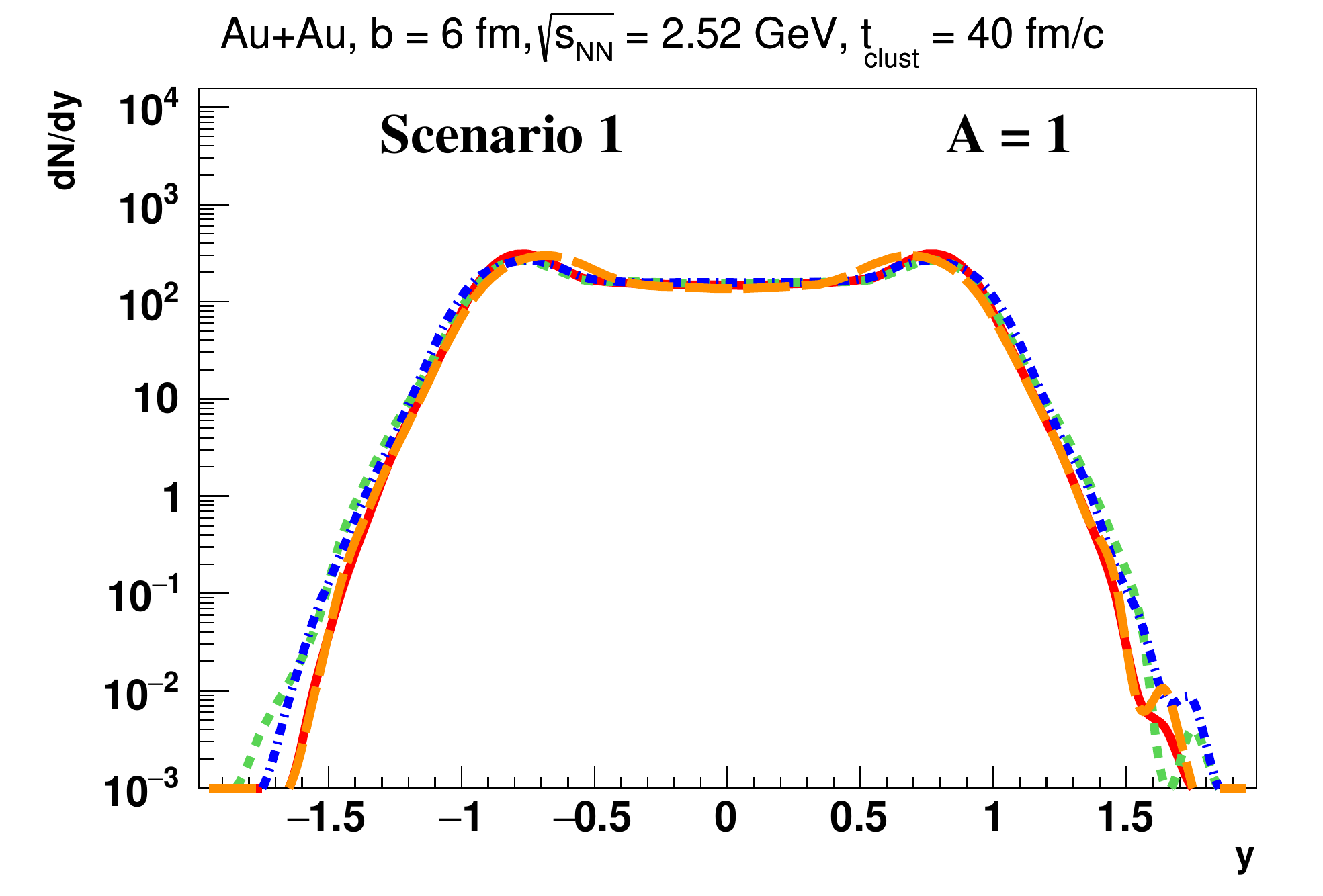} &
          \includegraphics{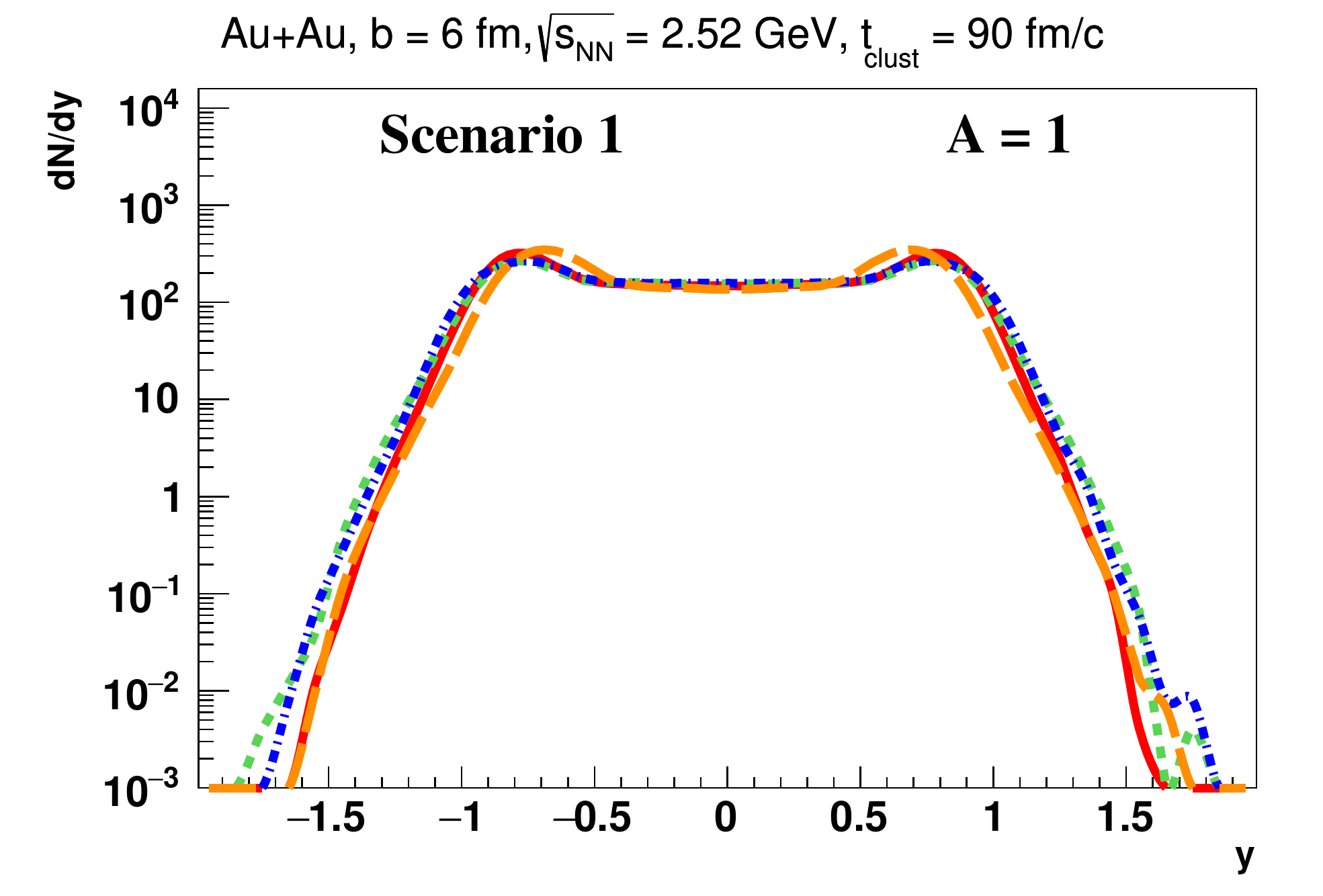} &
          \includegraphics{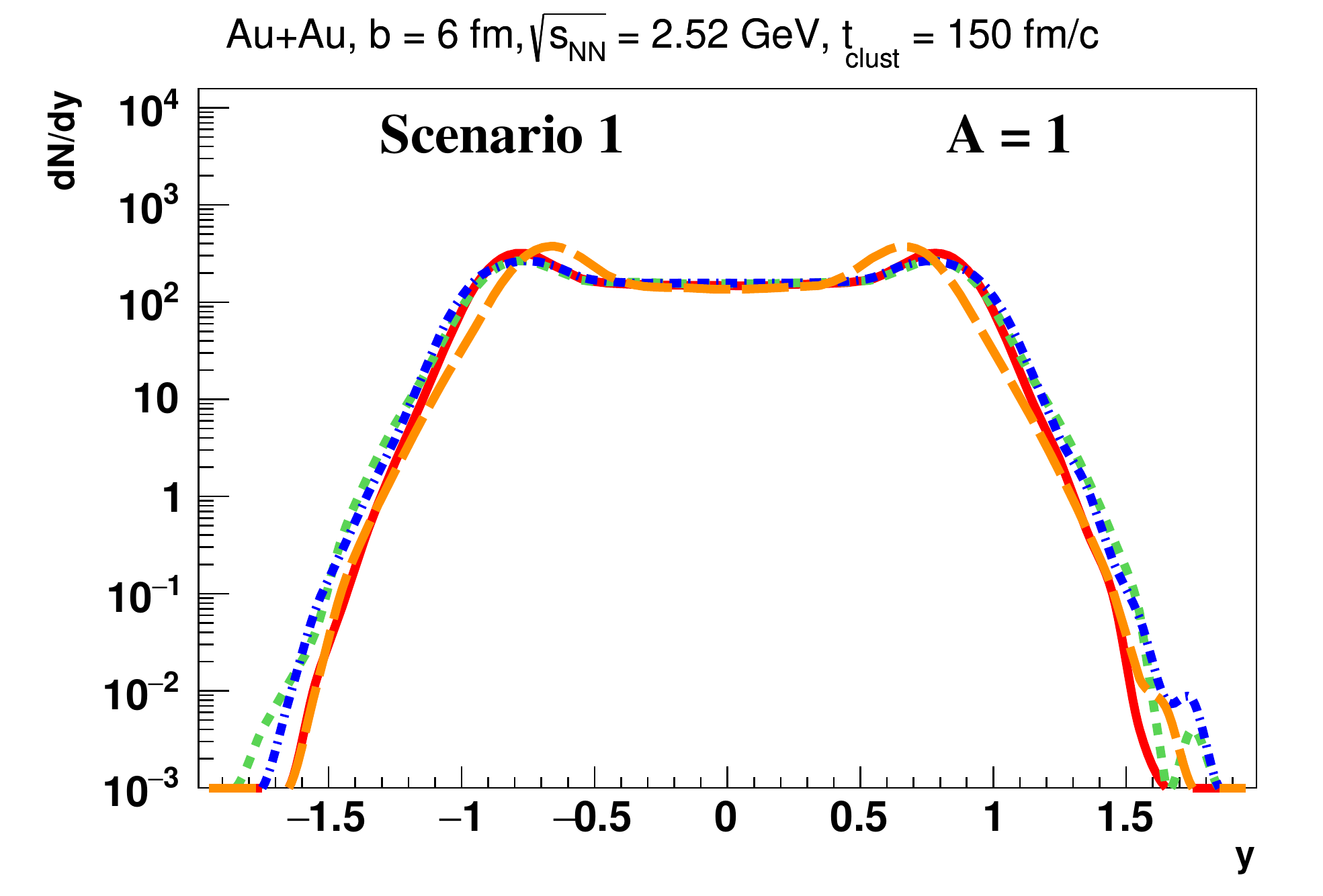} \\
        \end{tabular}
    }
\caption{\label{fig:2.52dndyA1} The rapidity distributions of unbound baryons ($p$, $n$, $\Lambda$ and $\Sigma^{0}$) in semi-peripheral ($b=6$ fm) $Au+Au$ collisions at $\sqrt{s}=2.52$ GeV. The left panel: the $y$-distributions at 40 fm/c, the central panel -- at 90 fm/c, the right plot -- at 150 fm/c.
The red solid lines show the PHQMD  results, the orange long dashed lines indicate the PHSD results, the green dashed lines correspond to the SMASH results and the blue dot dashed lines -- to the UrQMD results.
}
\end{figure*}

Figure.~\ref{fig:2.52dndyA1} presents the results of calculations for the rapidity spectra of $A = 1$ baryons ($p$, $n$, $\Lambda$ and $\Sigma^{0}$) in semi-peripheral ($b=6$ fm) $Au+Au$ collisions at $\sqrt{s}=2.52$ GeV (Scenario 1). The left plot represents the distributions at 40 fm/c, the central column -- at 90 fm/c, the right column -- at 150 fm/c. The red solid lines show the PHQMD results, the orange long dashed lines indicate the PHSD results, the green dashed lines correspond to the SMASH results and the blue dot dashed lines show the UrQMD predictions. 

\begin{figure*}[h!]
    \resizebox{\textwidth}{!}{
        \includegraphics{plots/scenario2/header.pdf} 
    } \\
    \resizebox{\textwidth}{!}{
        \begin{tabular}{ccc}
          \includegraphics{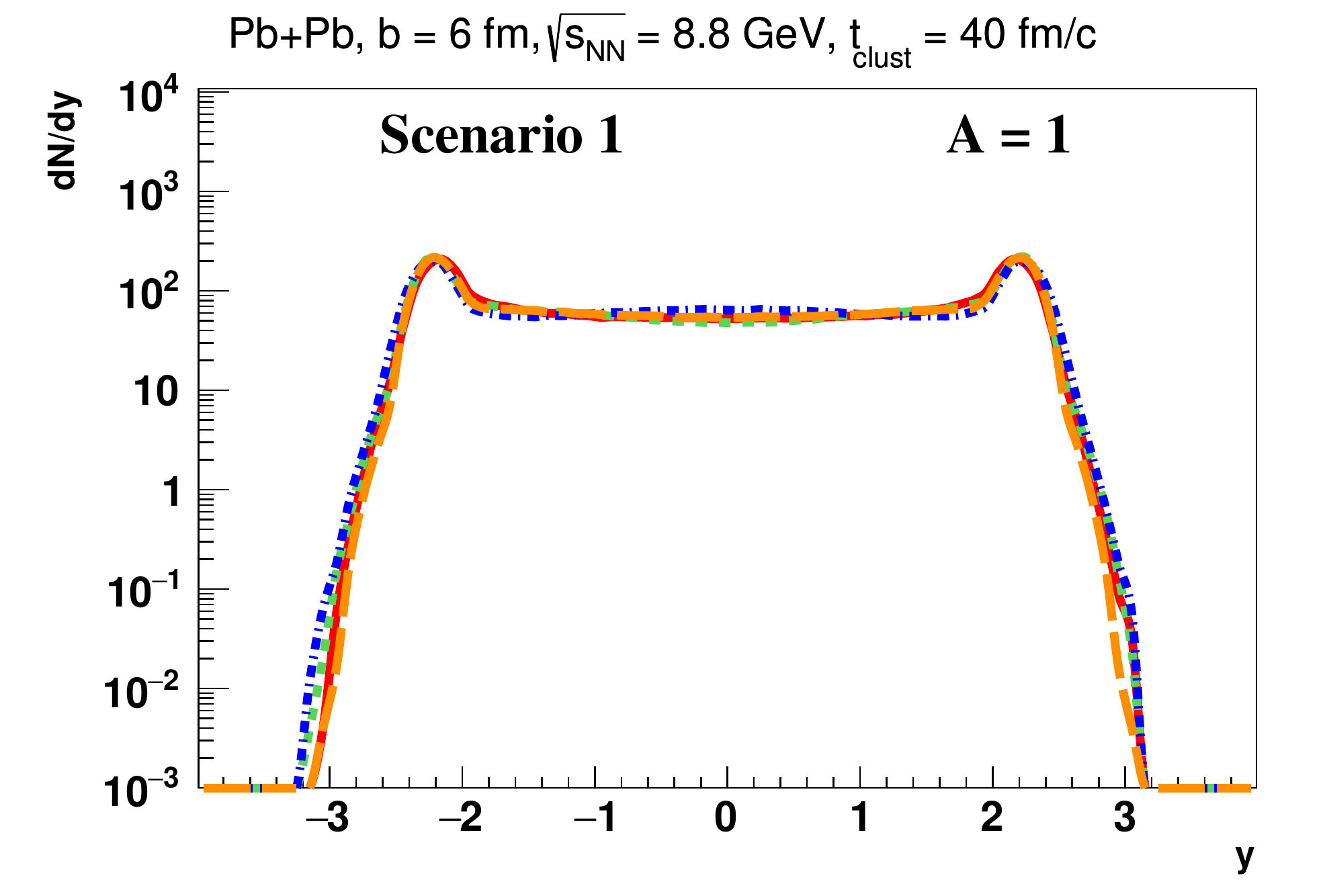} &
          \includegraphics{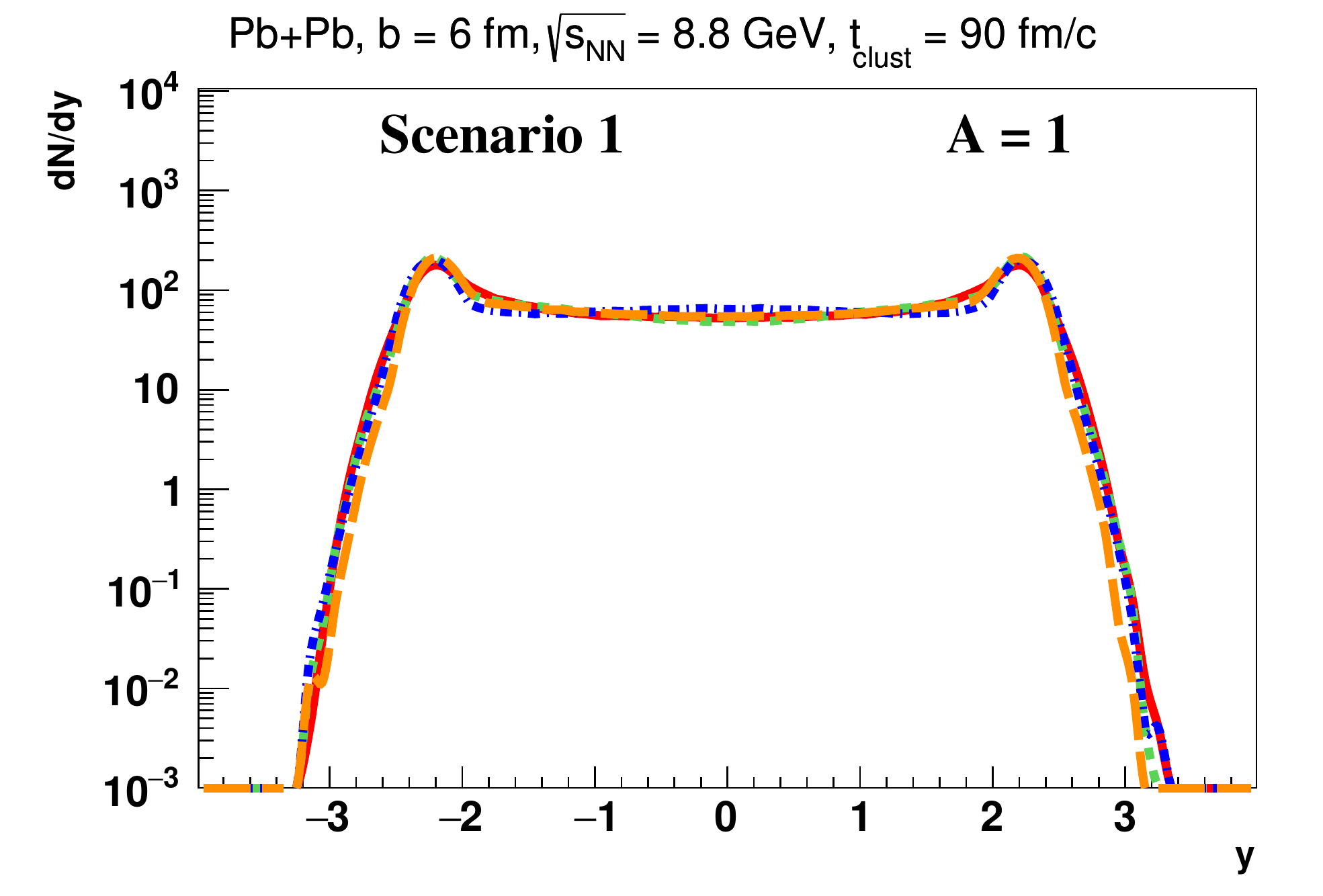} &
          \includegraphics{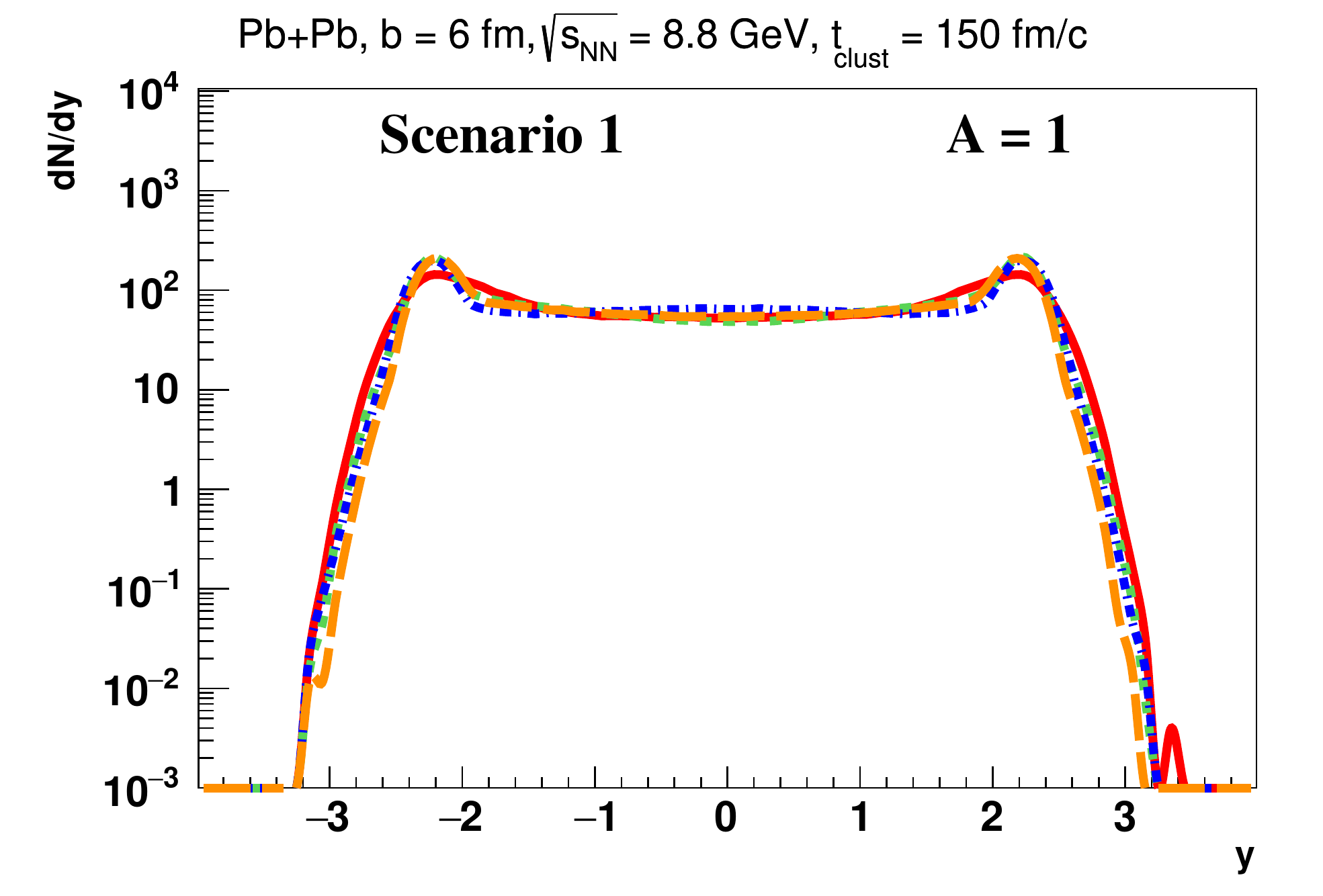} \\
        \end{tabular}
    }
\caption{\label{fig:8.8dndyA1} The rapidity distributions of single baryons ($p$, $n$, $\Lambda$ and $\Sigma^{0}$) in semi-peripheral ($b=6$ fm) $Pb+Pb$ collisions at $\sqrt{s}=8.8$ GeV. The left plot: $y$-distributions at 40 fm/c, the central panel -- at 90 fm/c, the right plot -- at 150 fm/c.
The color coding is the same as in Fig.~\ref{fig:2.52dndyA1}.
}
\end{figure*}

The rapidity spectra for $Pb+Pb$ collisions at $\sqrt{s}=8.8$ GeV are shown in Fig.~\ref{fig:8.8dndyA1}, the colors are the same as in Fig.~\ref{fig:2.52dndyA1}. 
It is very remarkable that all four transport approaches give a very similar rapidity distribution and hence a very similar stopping despite of the complexity of this process. 
That is important for our study of clusters since they are
more sensitive to the spacial and momentum correlations than single particle observables.

\begin{figure*}
    \resizebox{\textwidth}{!}{
        \includegraphics{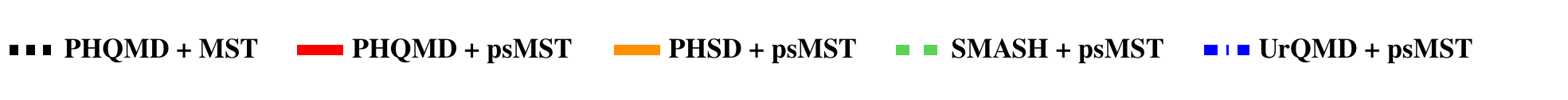} 
    } \\
    \resizebox{\textwidth}{!}{
        \begin{tabular}{ccc}
          \includegraphics{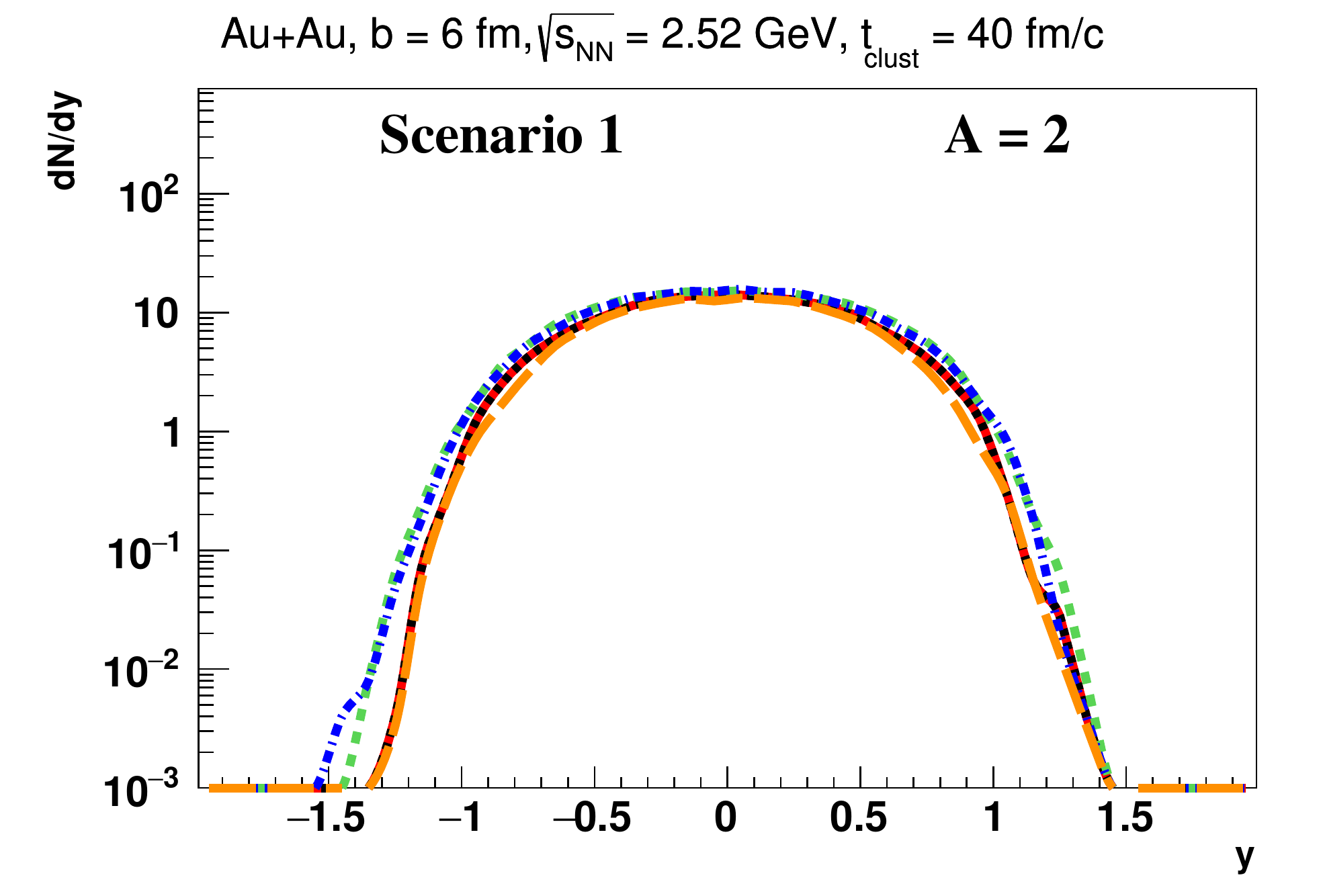} &
          \includegraphics{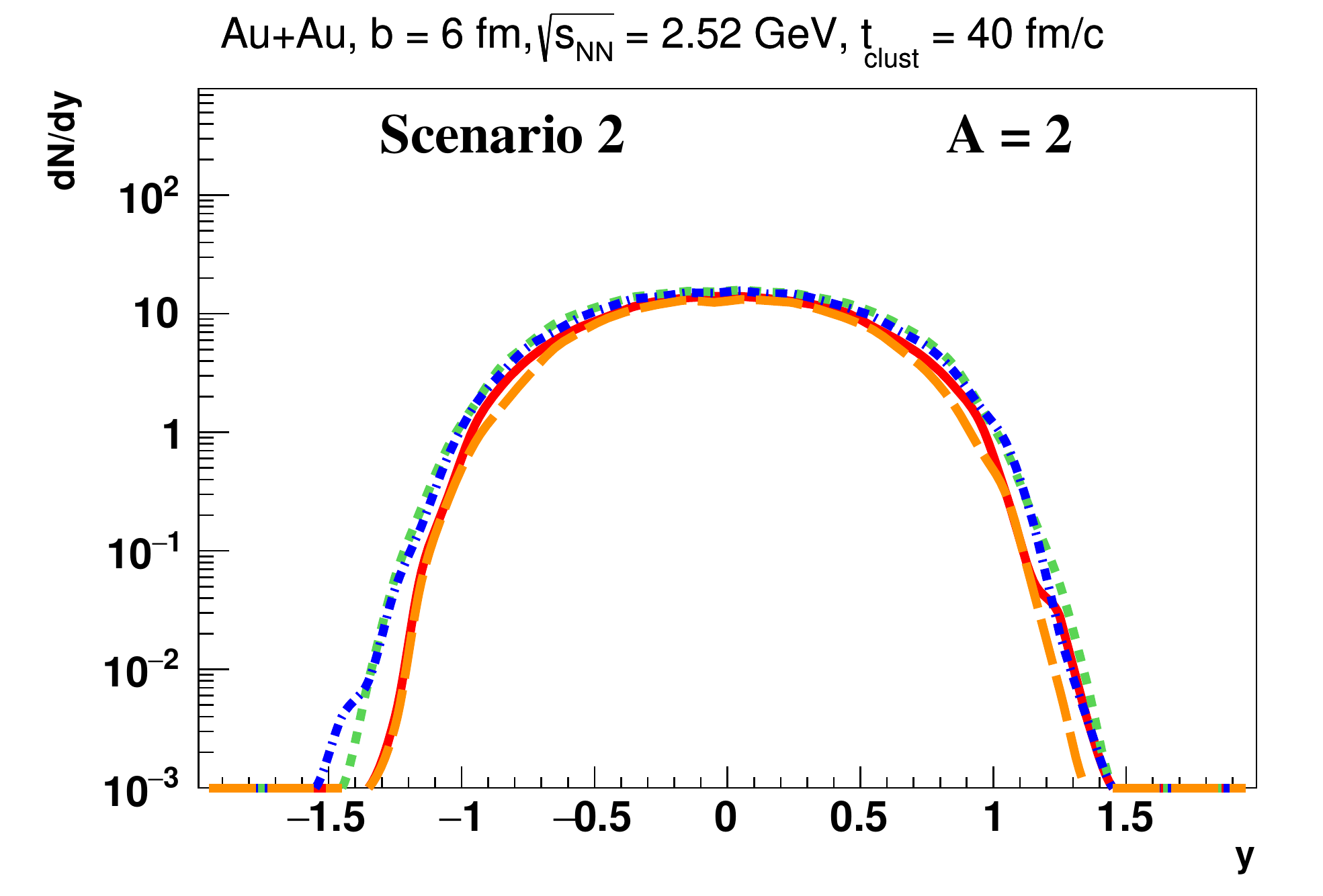} &
          \includegraphics{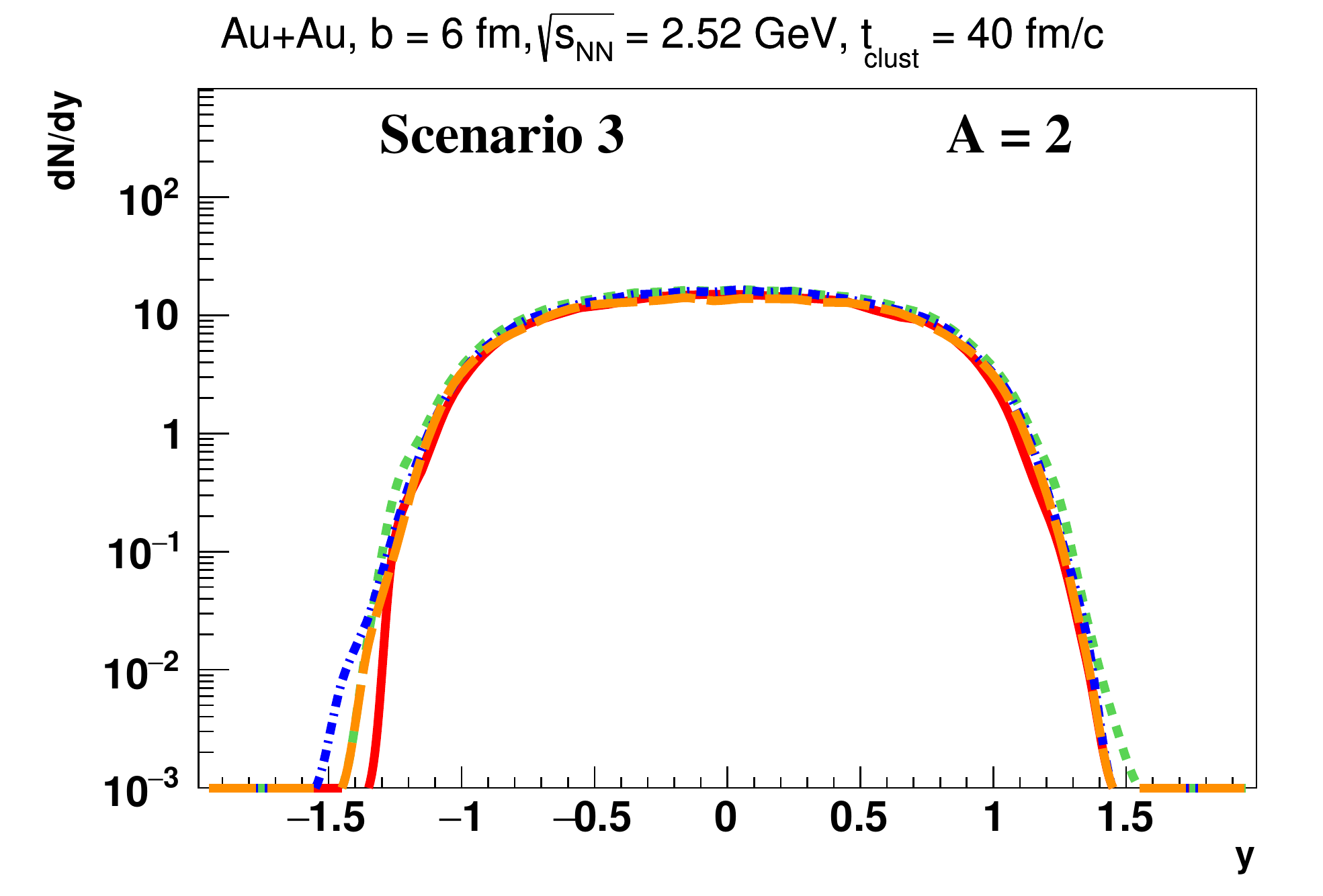} \\
          \includegraphics{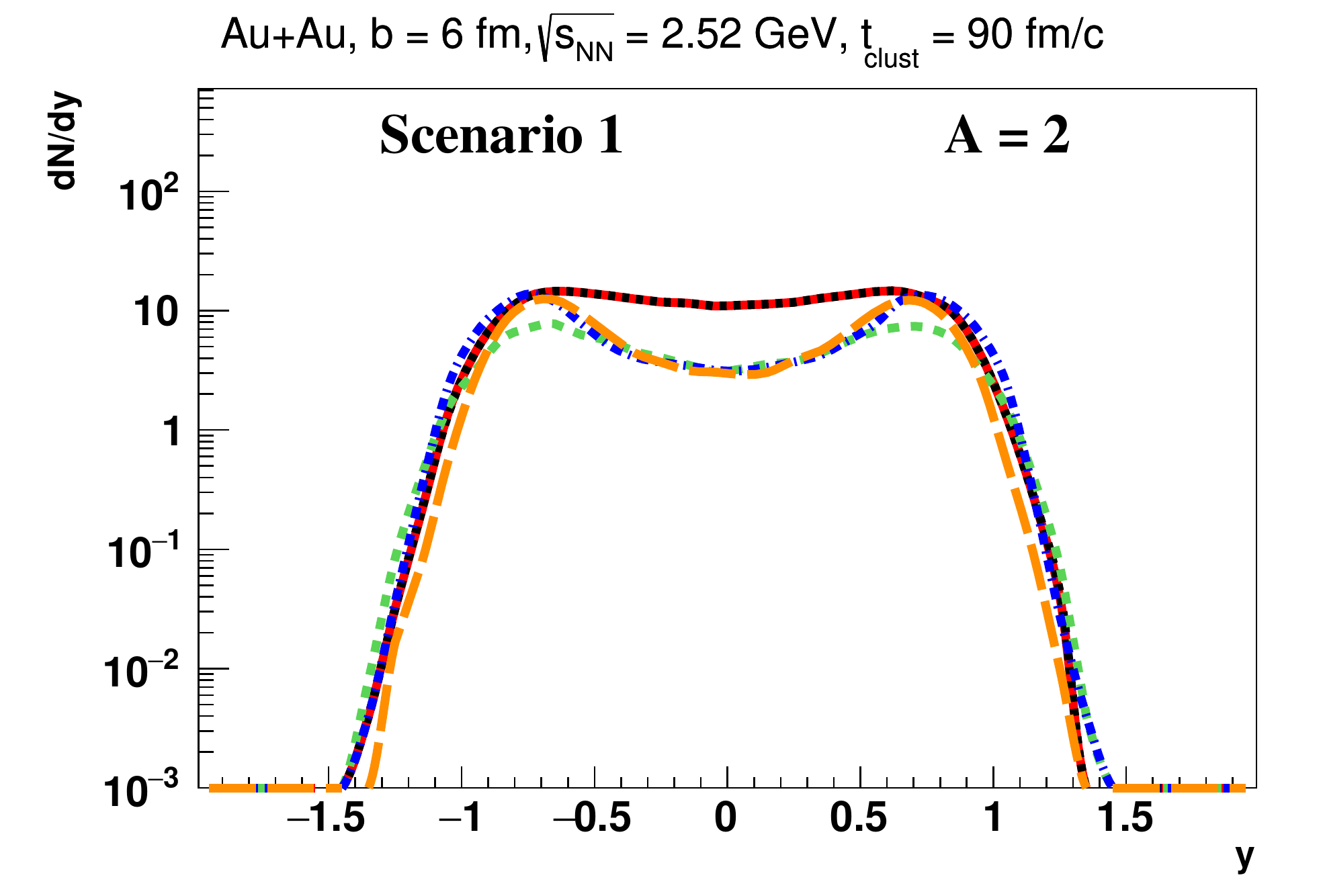} &
          \includegraphics{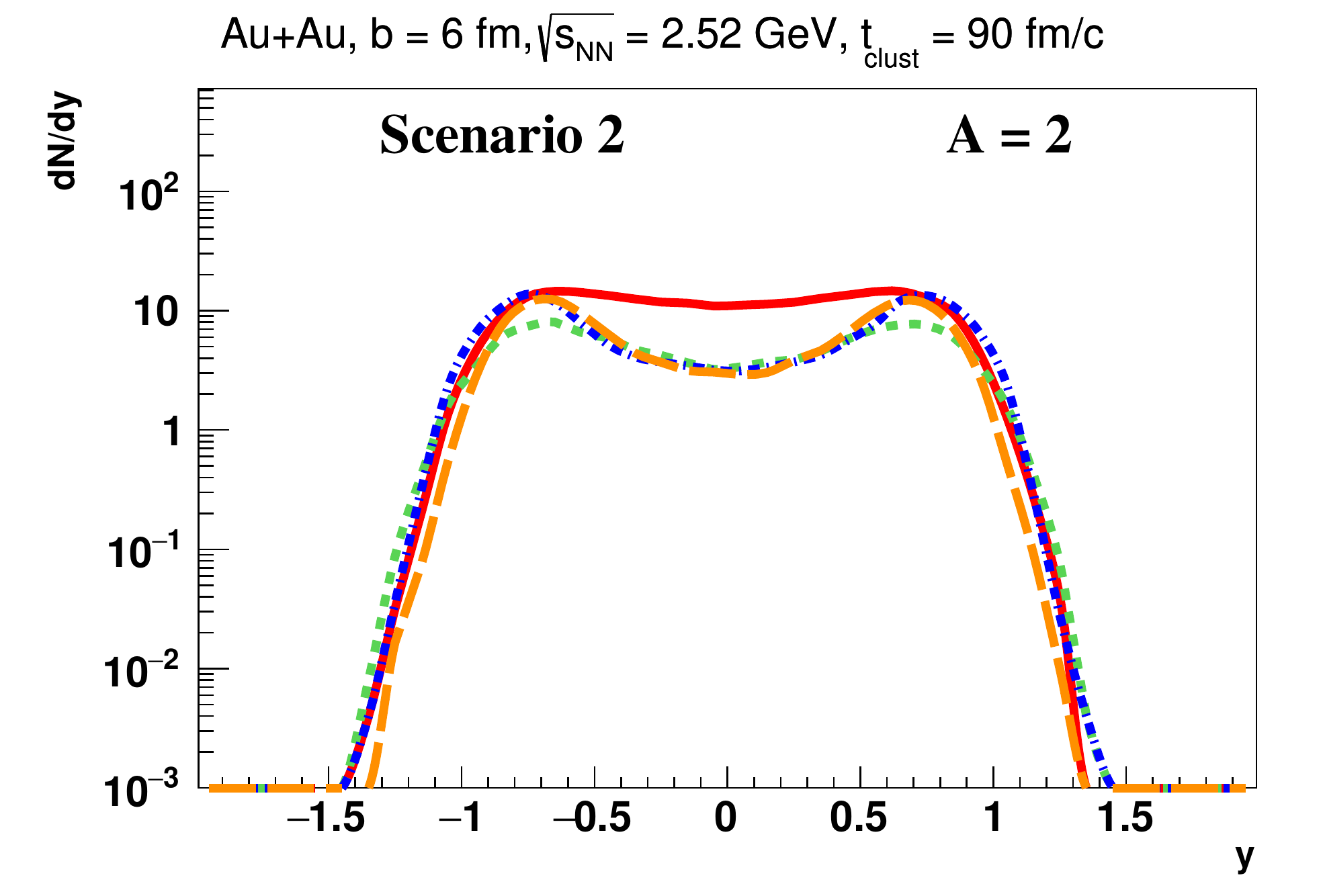} &
          \includegraphics{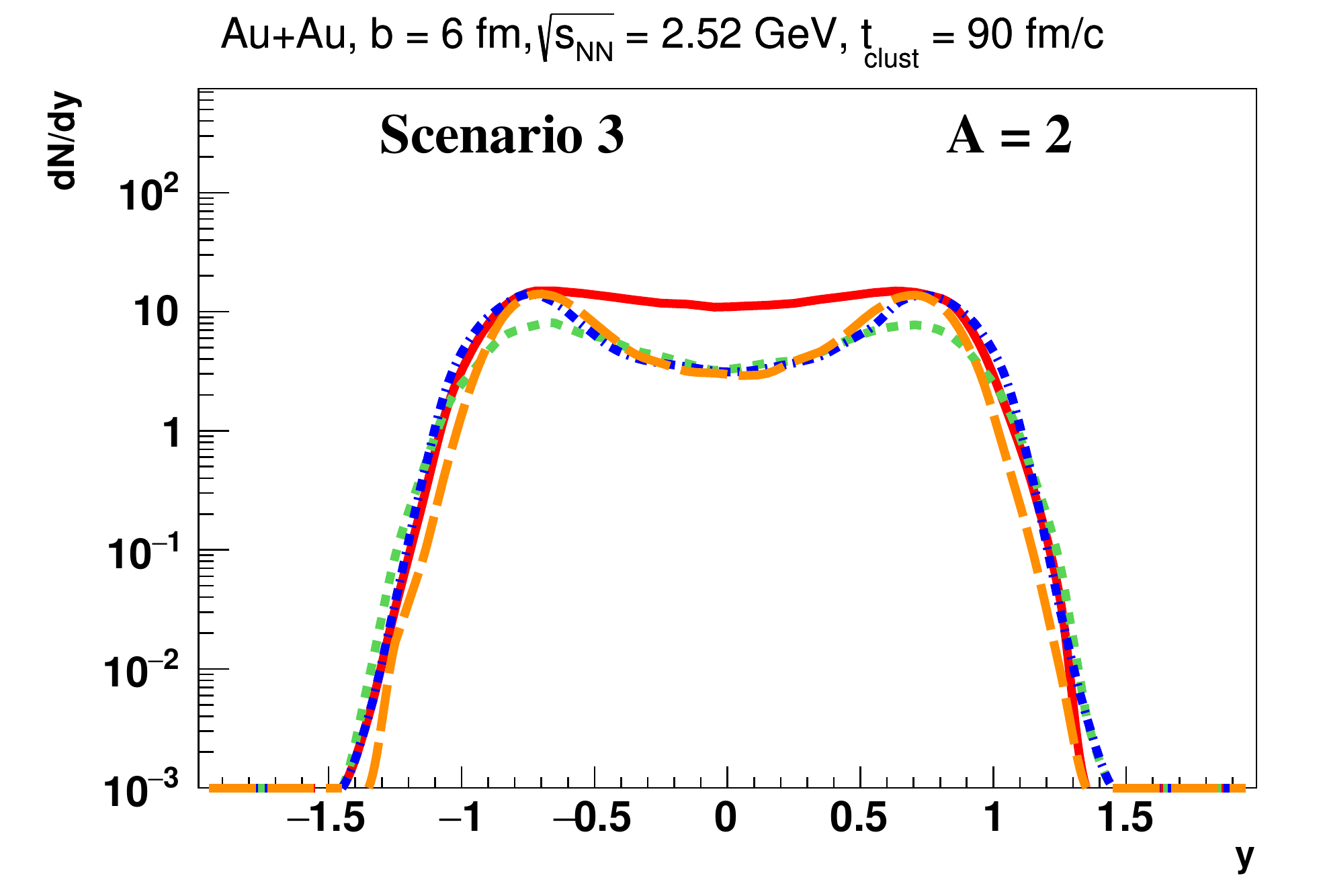} \\
          \includegraphics{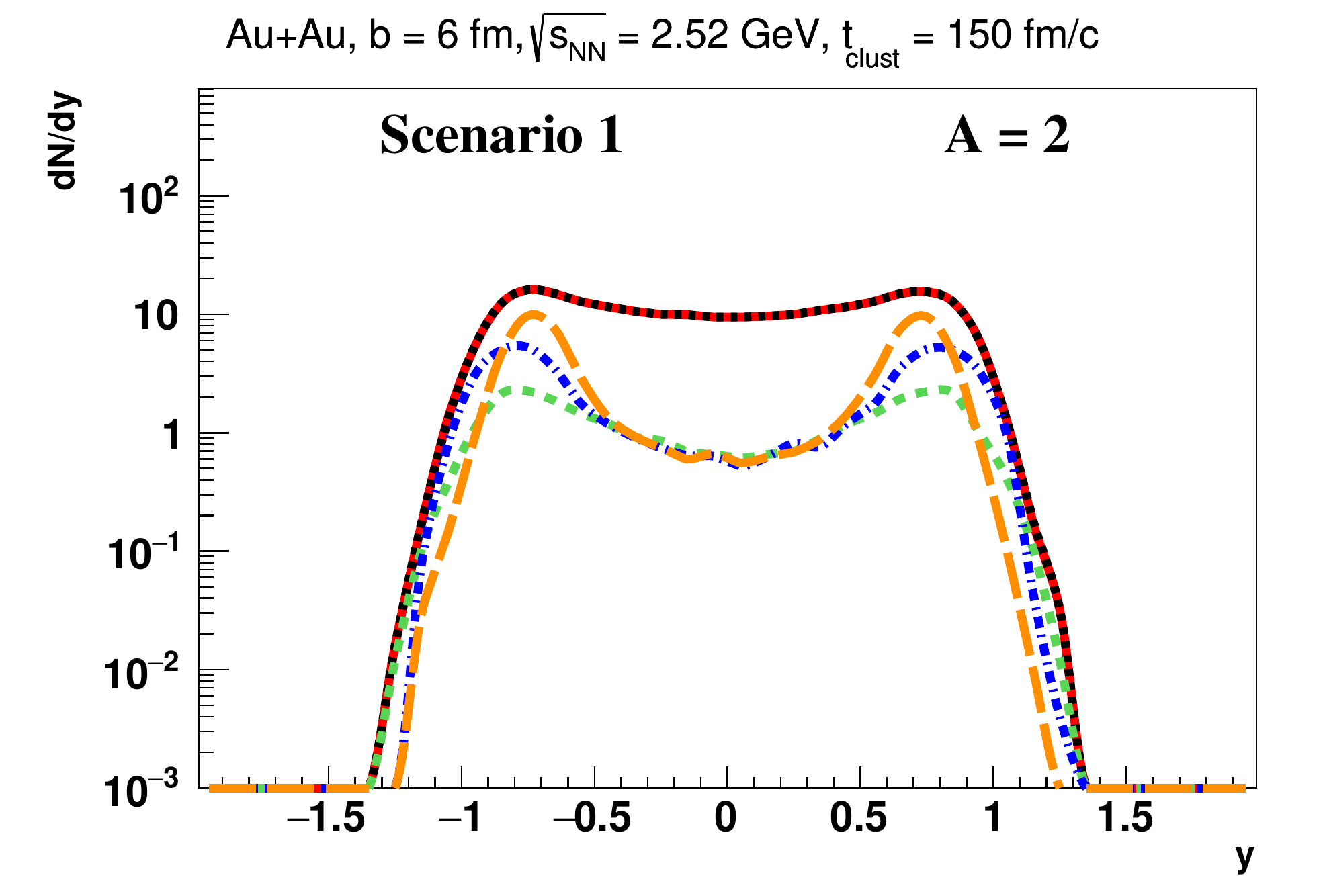} &
          \includegraphics{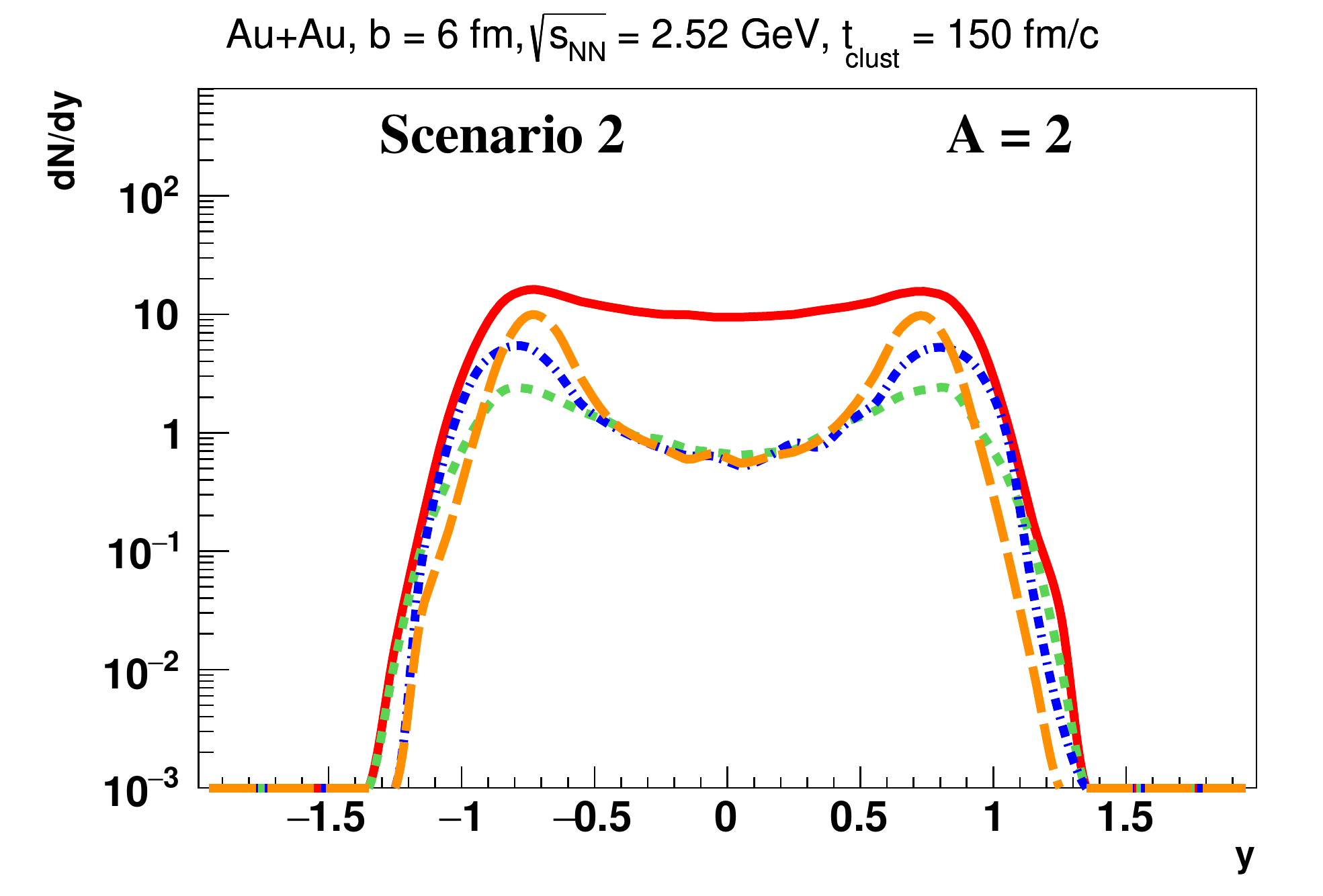} &
          \includegraphics{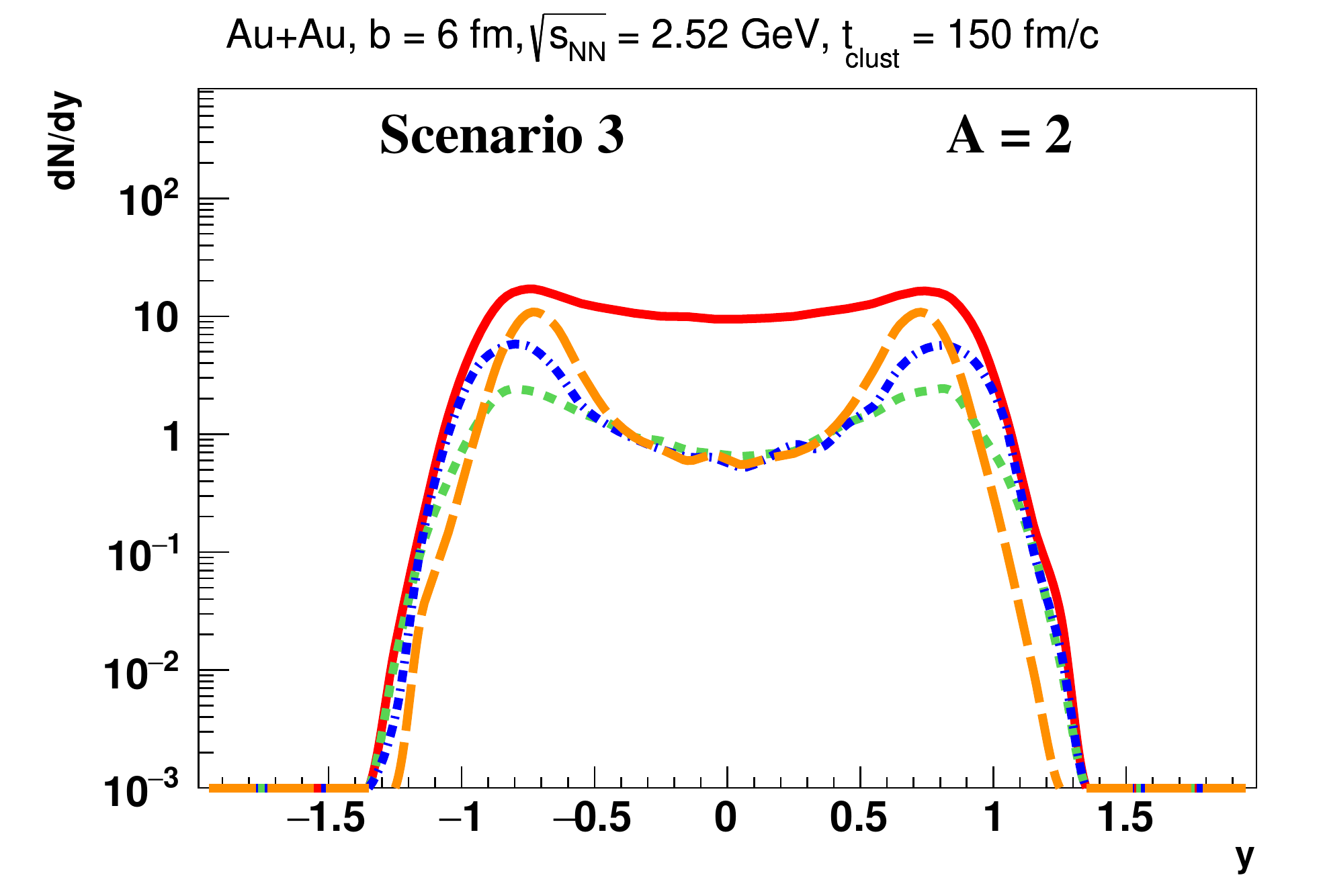} \\
        \end{tabular}
    }
\caption{\label{fig:2.52dndyA2} The rapidity distributions of the clusters with the mass number $A = 2$ in semi-peripheral ($b=6$ fm) $Au+Au$ collisions at $\sqrt{s}=2.52$ GeV. The left column: "Scenario 1", the central column: "Scenario 2", the right column: "Scenario 3" (see text).
The black short dashed lines ("PHQMD+MST") show the PHQMD results at time $t_{clust}$ within the MST cluster recognition model, 
the red solid lines show the PHQMD results within psMST,
the orange long dashed lines indicate the PHSD results within psMST, the green dashed lines correspond to the SMASH results within psMST, the blue dot dashed lines show the UrQMD results within psMST.
The upper row corresponds to the model calculations at $t_{clust} =40$ fm/c,
the middle row -- at $t_{clust} =90$ fm/c,
the lower row  -- at $t_{clust} =150$ fm/c.
}
\end{figure*}

We continue with the rapidity spectra of clusters with mass number $A = 2$ in semi-peripheral ($b = 6$ fm)  
Au+Au collisions at $\sqrt{s}=2.52$ GeV, which are shown in Fig.~\ref{fig:2.52dndyA2} for different Scenarios for the cluster recognition: 
the left column corresponds to "Scenario 1" (see Section ~\ref{sec:2}), the central column -- to  "Scenario 2" and the right column -- to  "Scenario 3".
In the upper row we apply the cluster analysis at $t_{clust} =40$ fm/c, in the middle row -- at $t_{clust} = 90$ fm/c and in the lower row  -- at $t_{clust} = 150$ fm/c.
The red solid lines show PHQMD results with psMST, the orange long dashed lines indicate results of PHSD with psMST, the green dashed lines correspond to SMASH results with psMST and blue dot dashed lines show results of UrQMD with psMST. 
Additionally, we present the PHQMD results with its internal MST cluster recognition algorithm as described in Ref. \cite{Aichelin:2019tnk}  -- cf. the back short dashed lines ("PHQMD+MST") in the left column of "Scenario 1".  
For "scenario 1" the results for PHQMD with psMST and that obtained by applying the internal MST algorithm are identical - as expected.

At an early time, $t_{clust} = 40$ fm/c, all models show a similar rapidity distributions, at the later times  $t_{clust} = 90,\ 150$ fm/c  the PHQMD  predicts more clusters in the  mid-rapidity region than the MF-based  
PHSD approach (with potential) or the cascade approaches, SMASH and  UrQMD, (without potential).   This shows the importance of the n-body dynamics (realized by 2-body potential interactions between the baryons in PHQMD) for the cluster formation.

\begin{figure*}
    \resizebox{\textwidth}{!}{
        \begin{tabular}{ccc}
          \includegraphics{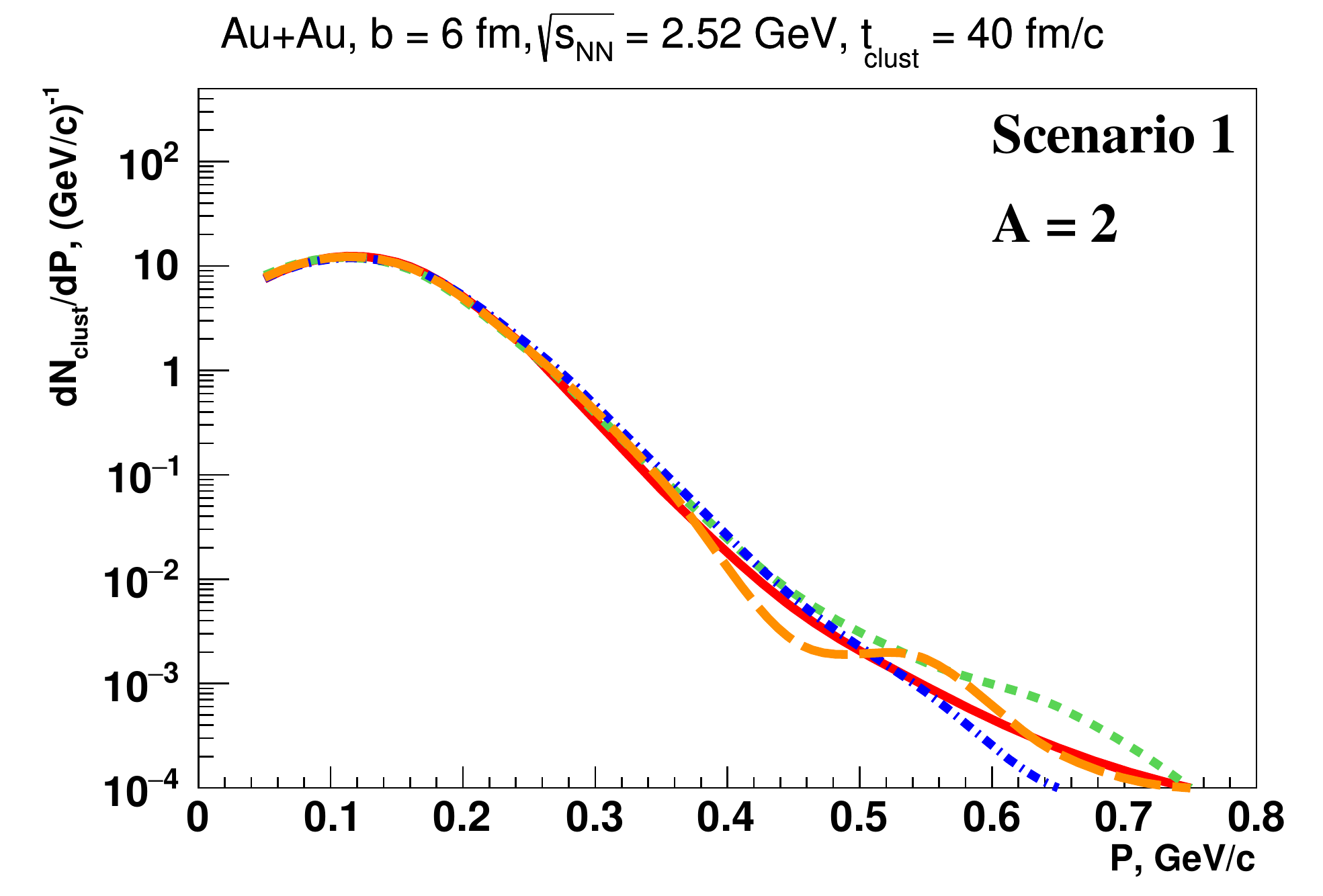} &
          \includegraphics{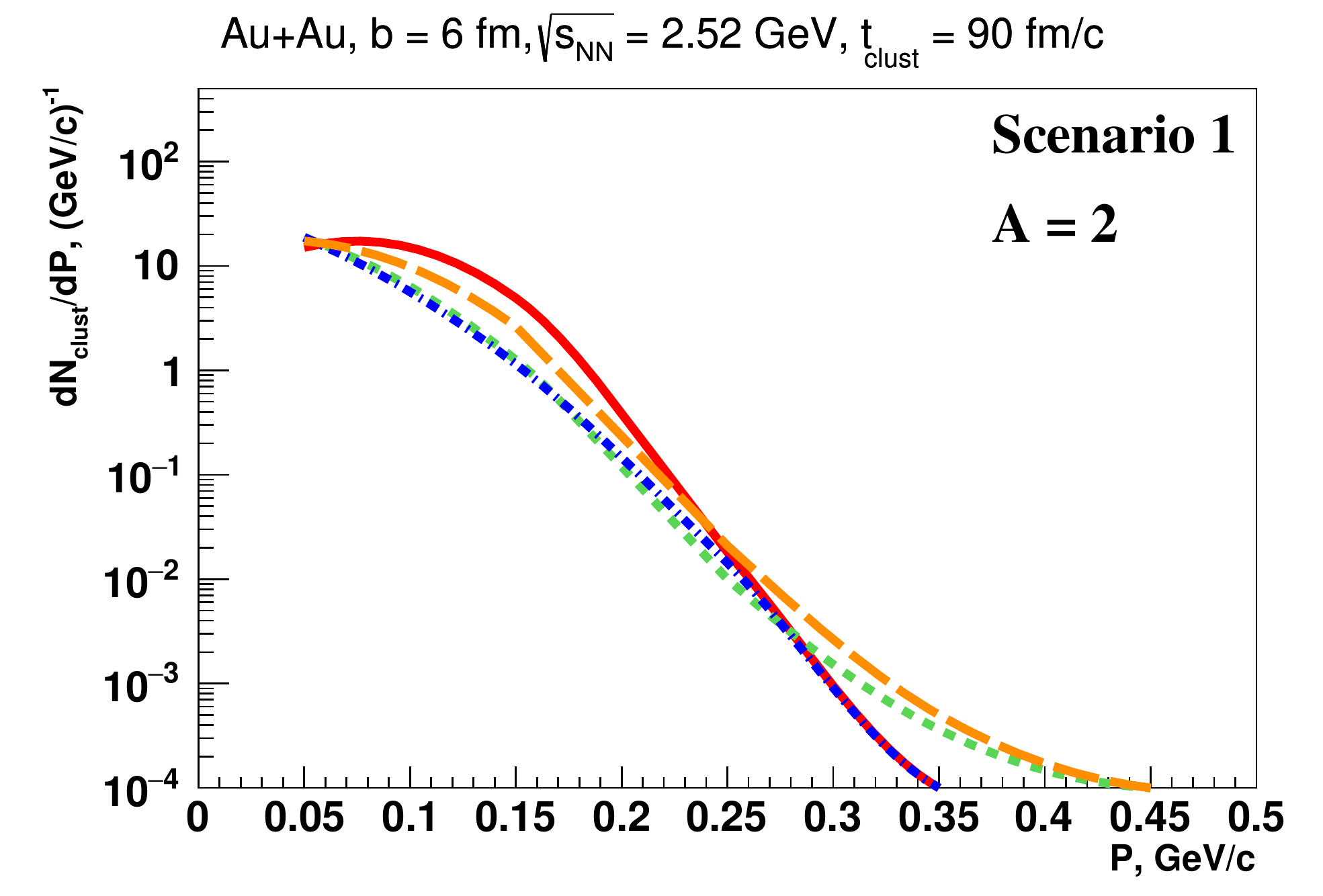} &
          \includegraphics{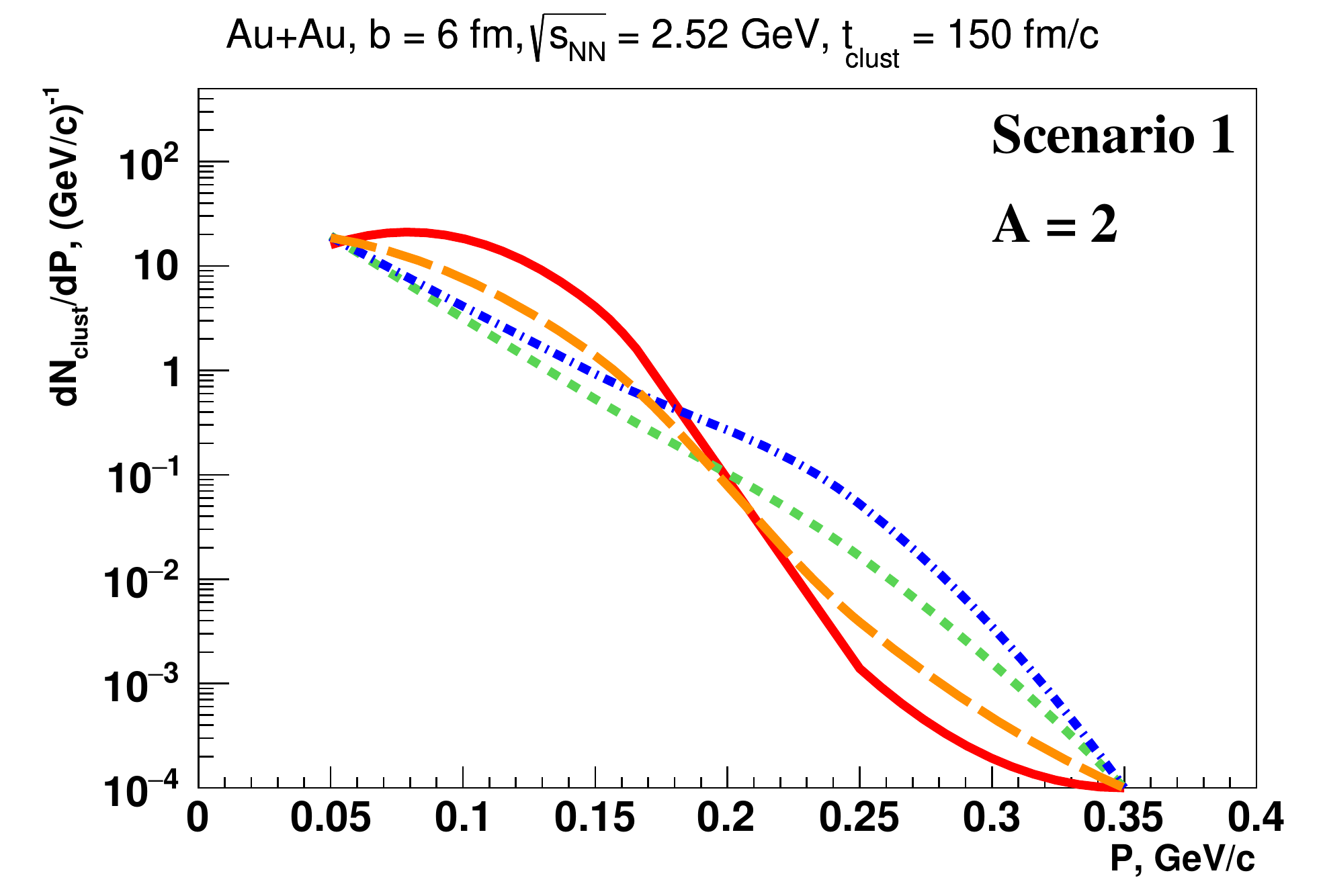} \\
        \end{tabular}
    }
\caption{\label{fig:2.52dndpA2} The momentum spectra of baryons ($p$, $n$, $\Lambda$ and $\Sigma^{0}$) from $A = 2$ clusters in semi-peripheral ($b=6$ fm) $Au+Au$ collisions at $\sqrt{s}=2.52$ GeV (integrated over all rapidity range). The momentum is calculated in the cluster center of mass frame. The left column: $t_{clust} = 40$ fm/c, the central column: $t_{clust} = 90$ fm/c, the right column: $t_{clust} = 150$ fm/c.
The color coding is the same as in Fig.~\ref{fig:2.52dndyA2}.
}
\end{figure*}

The momentum spectra of baryons in $A = 2$ clusters (integrated over all rapidities) found by psMST within "Scenario 1"
in semi-peripheral ($b = 6$ fm) Au+Au collisions at $\sqrt{s}=2.52$ GeV are shown in Fig.~\ref{fig:2.52dndpA2}. 
We remind that in this scenario only coordinate space information is used to determine the cluster size and that the nucleon momenta are calculated in the cluster center of mass frame.

Similar to the rapidity distributions, the momentum
distributions at the early time  $t_{clust} = 40$ fm/c are very close to each other, however, different 
dynamics of hadrons leads to a spreading of momentum distributions at later times.
The Fig.~\ref{fig:2.52dndpA2} also gives an indication of the fraction of clusters which can be discarded by the momentum condition of the "Scenario 2" and "Scenario 3".

\begin{figure*}
    \resizebox{\textwidth}{!}{
        \includegraphics{plots/scenario1/header.pdf} 
    } \\
    \resizebox{\textwidth}{!}{
        \begin{tabular}{ccc}
          \includegraphics{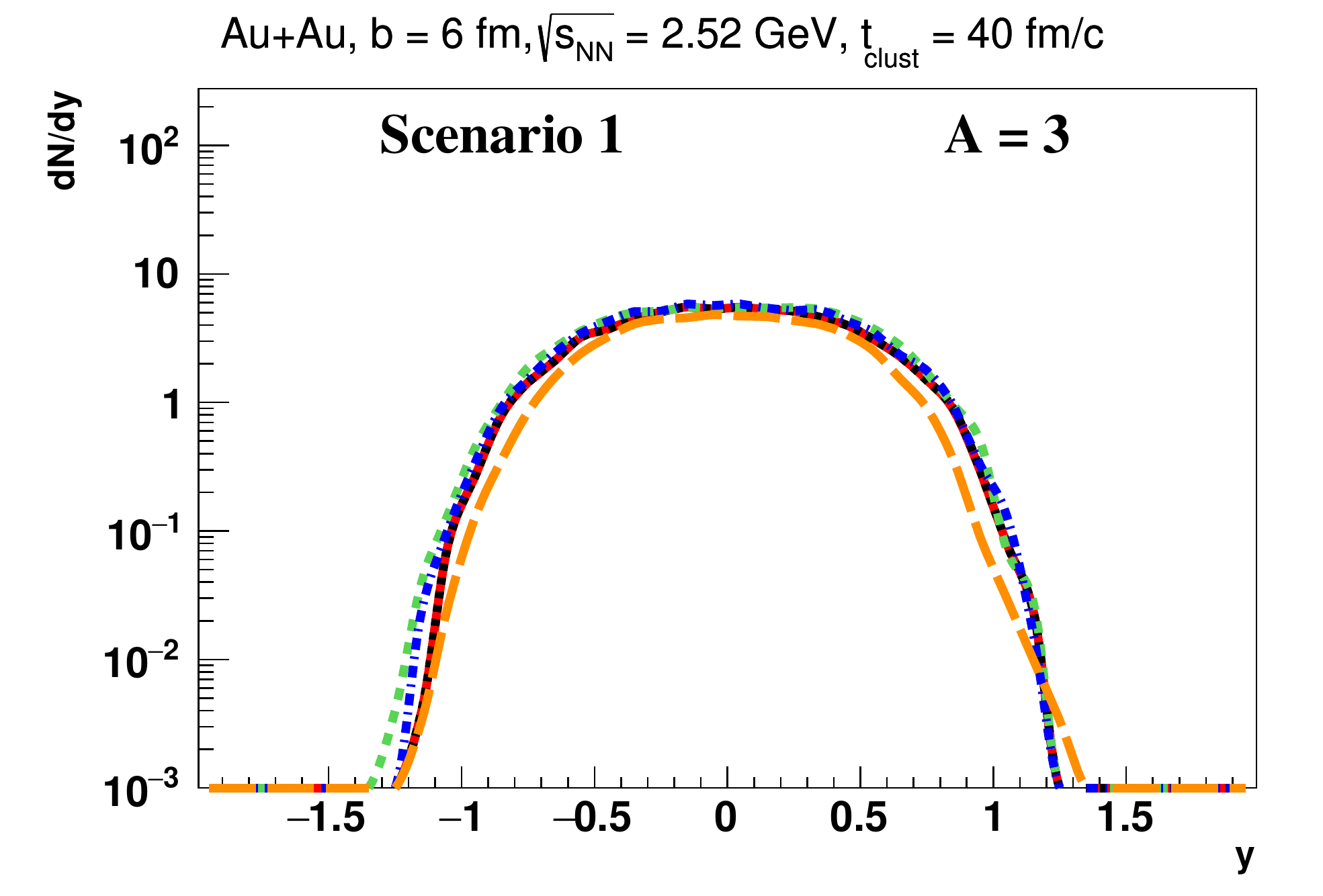} &
          \includegraphics{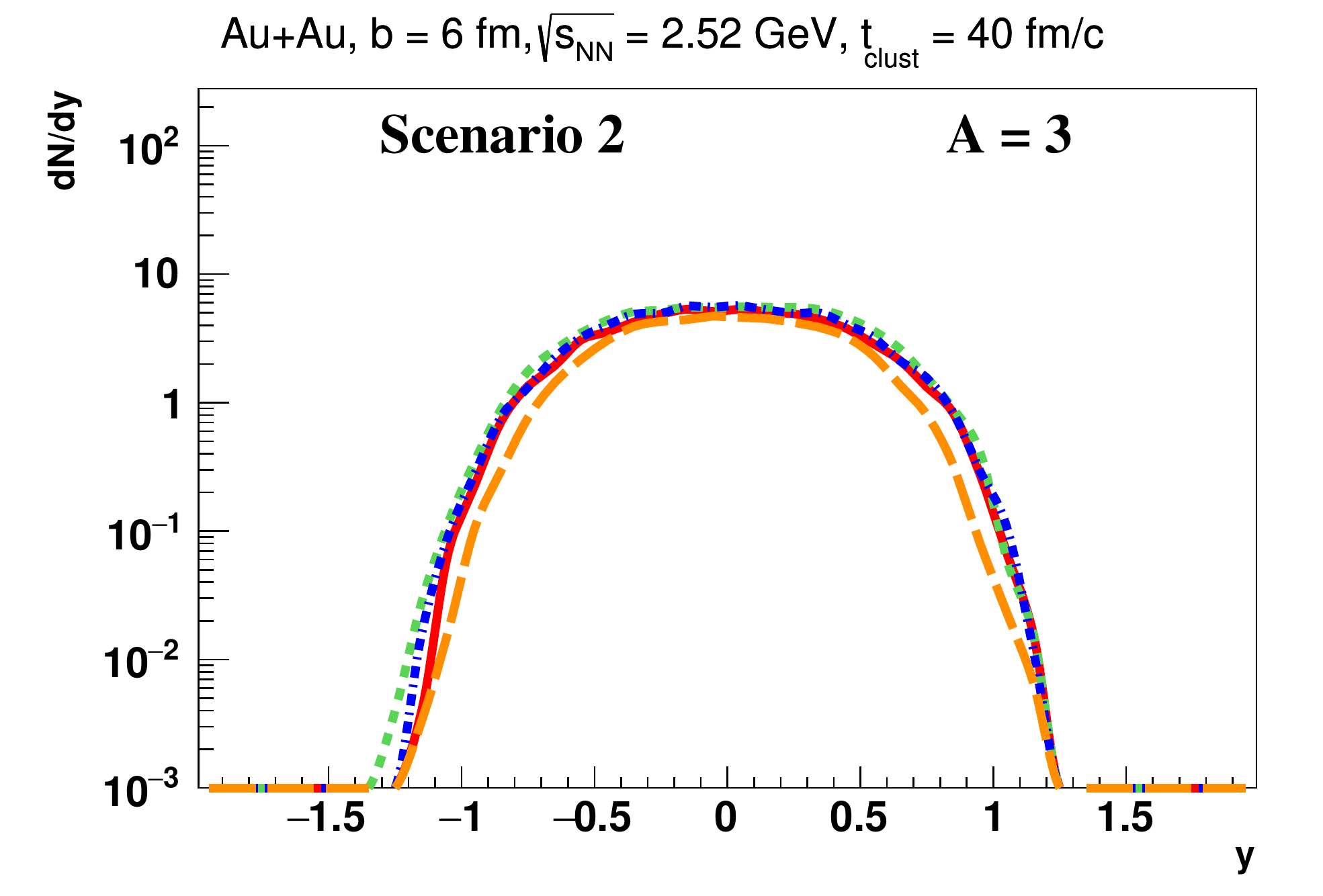} &
          \includegraphics{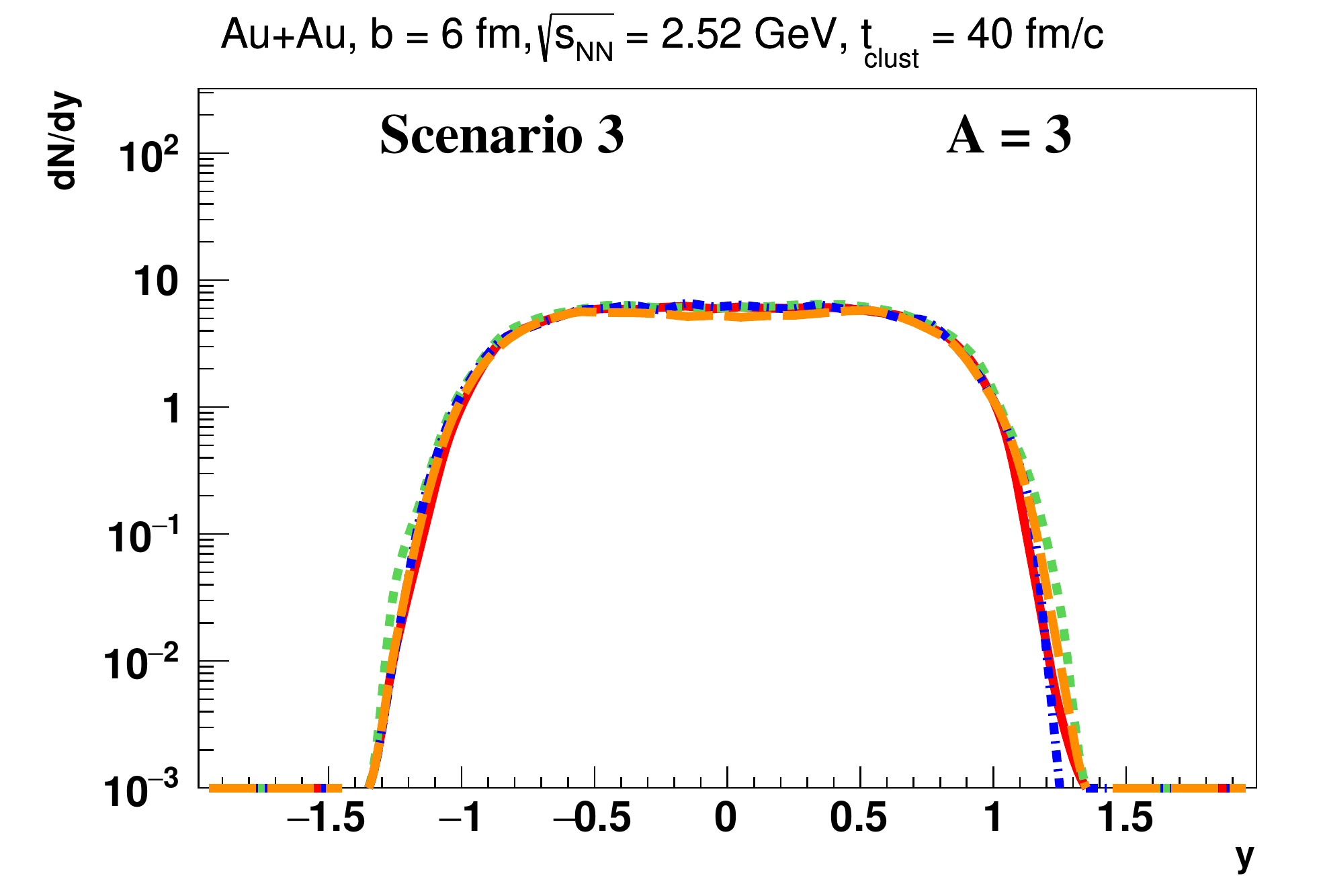} \\
          \includegraphics{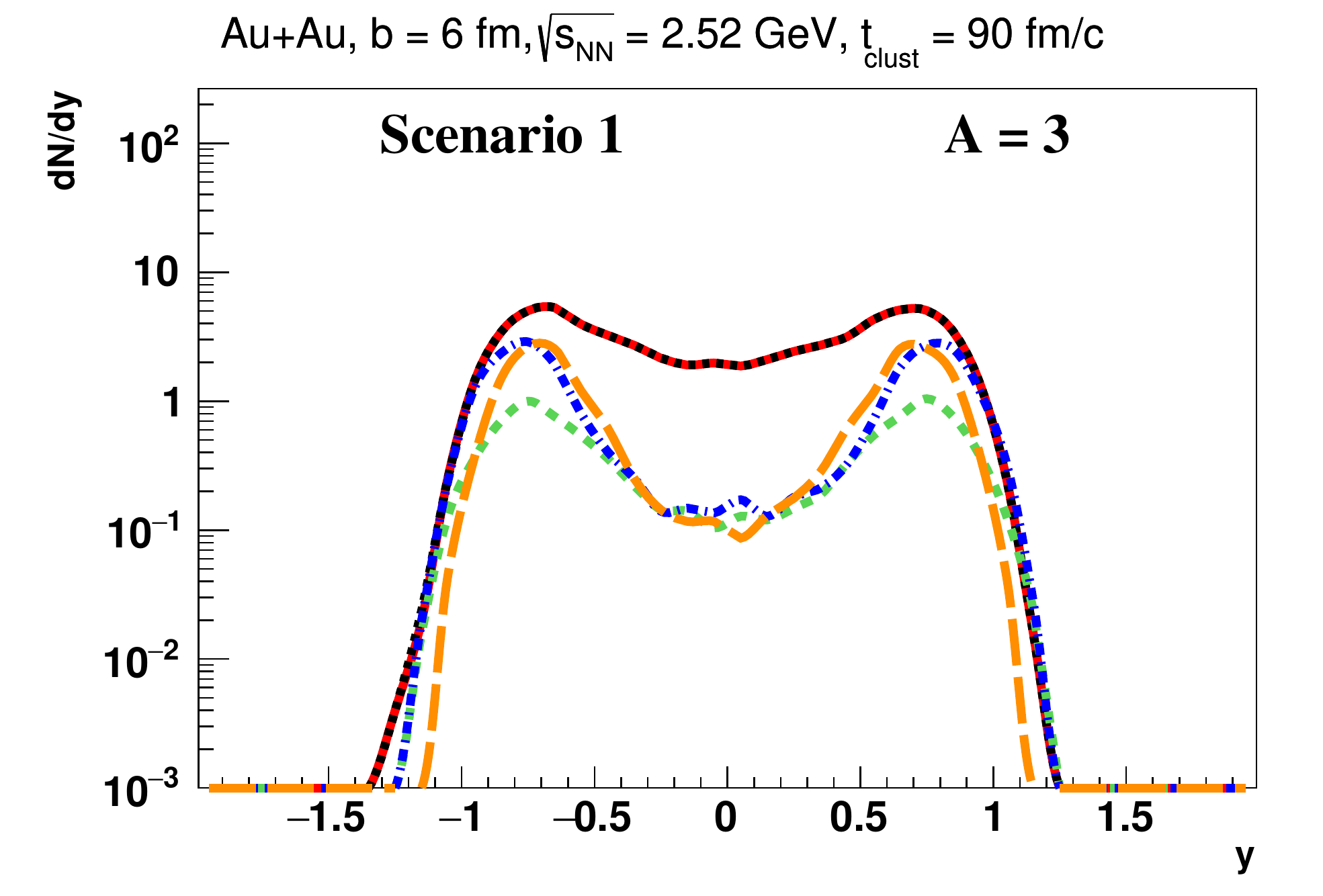} &
          \includegraphics{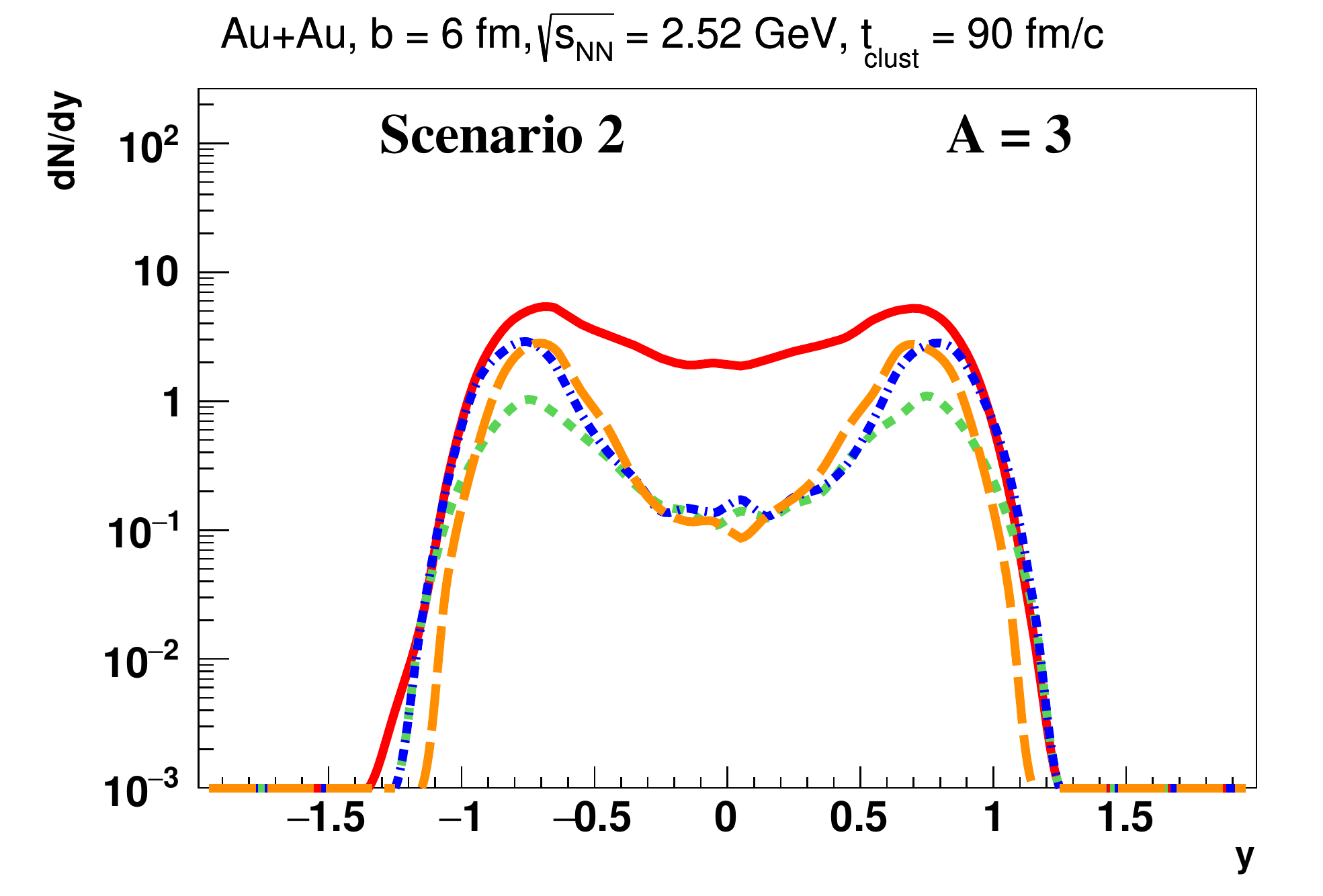} &
          \includegraphics{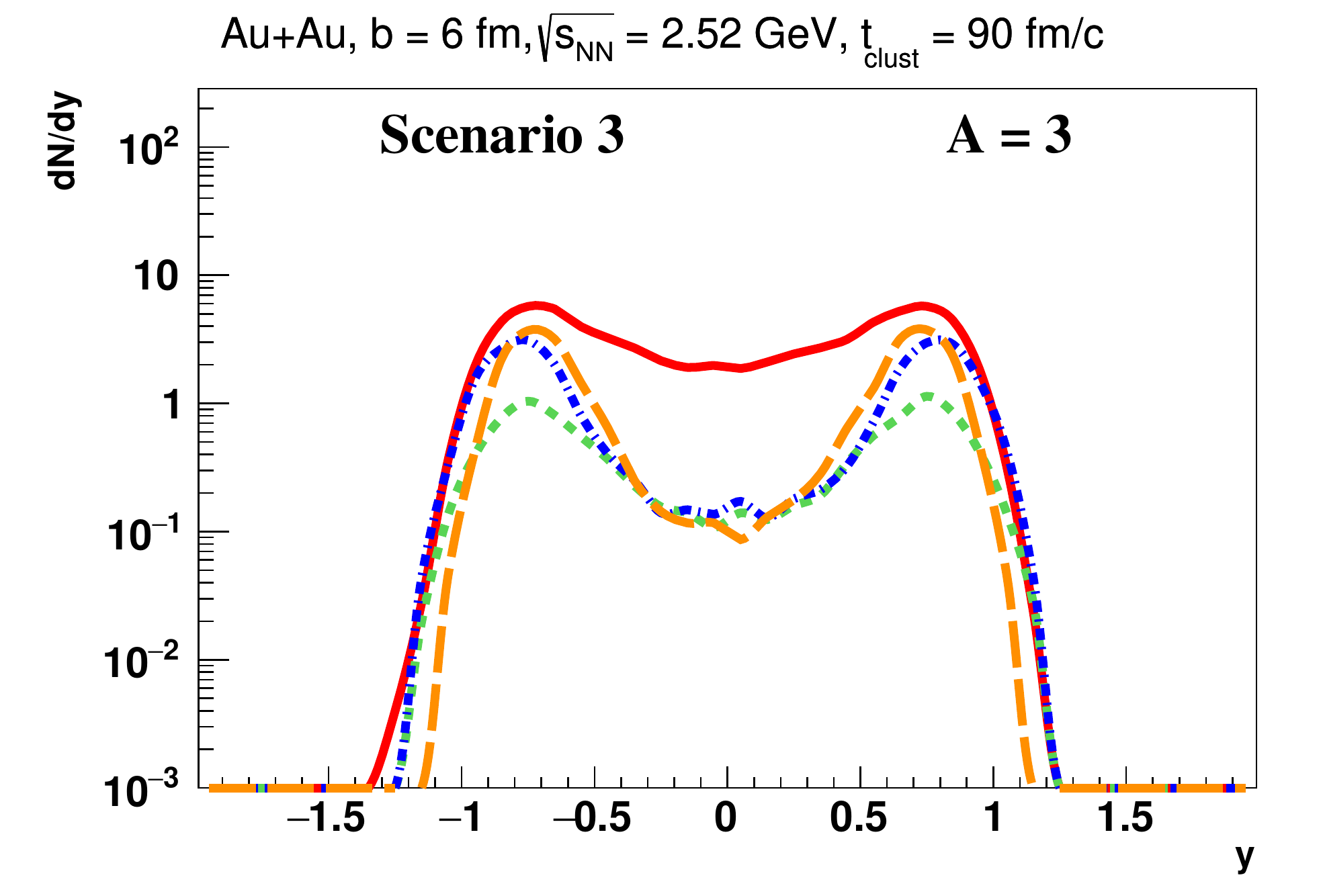} \\
          \includegraphics{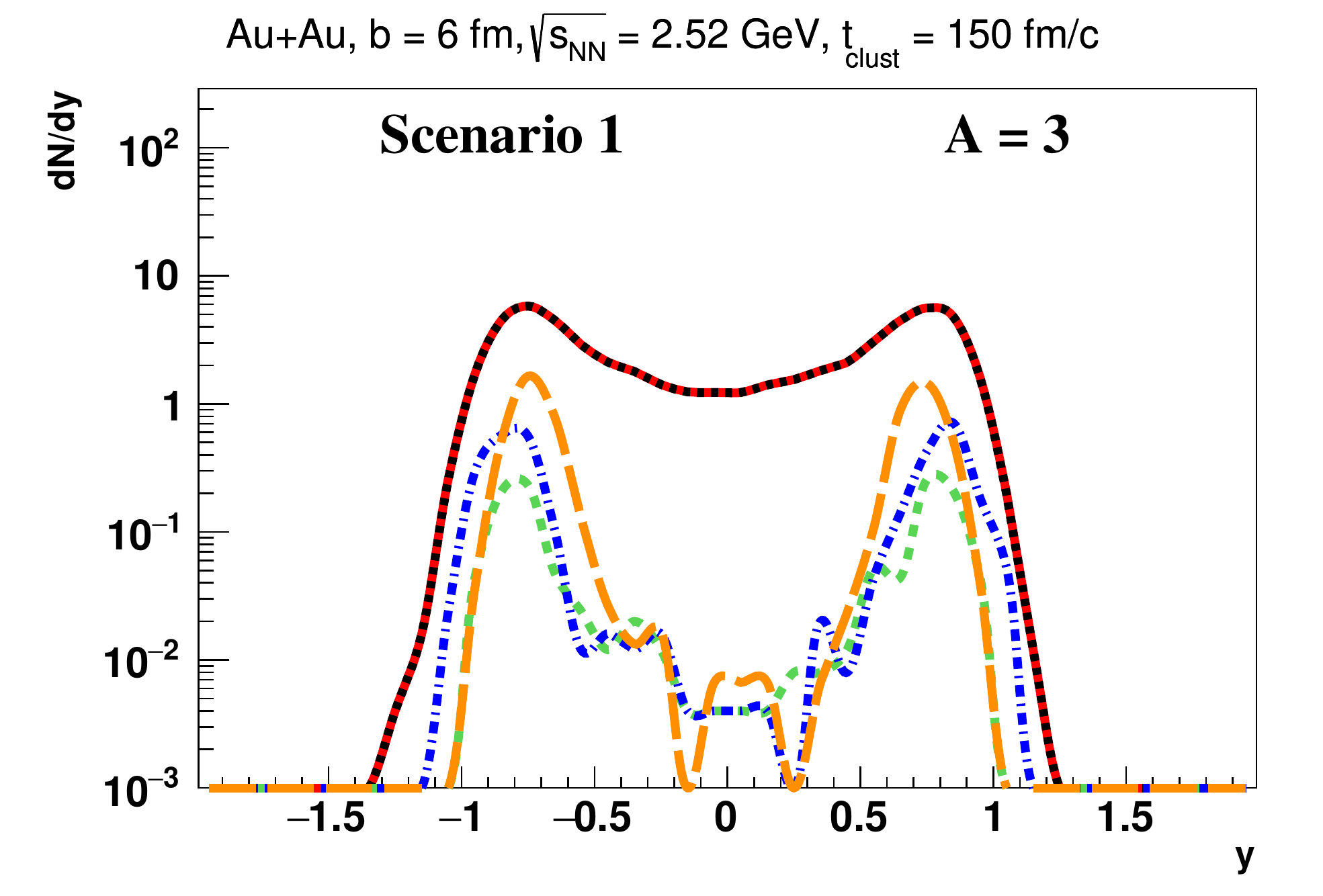} &
          \includegraphics{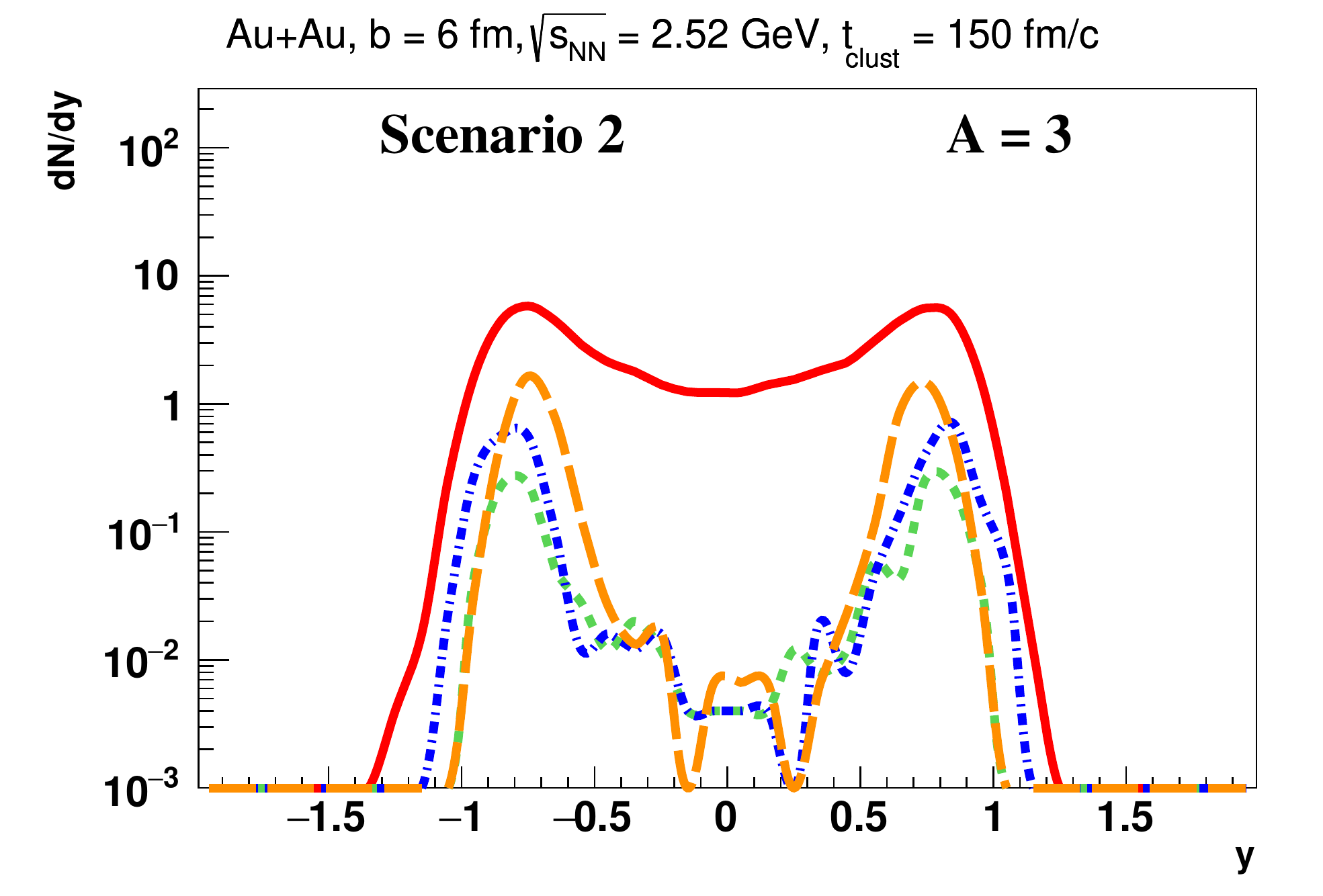} &
          \includegraphics{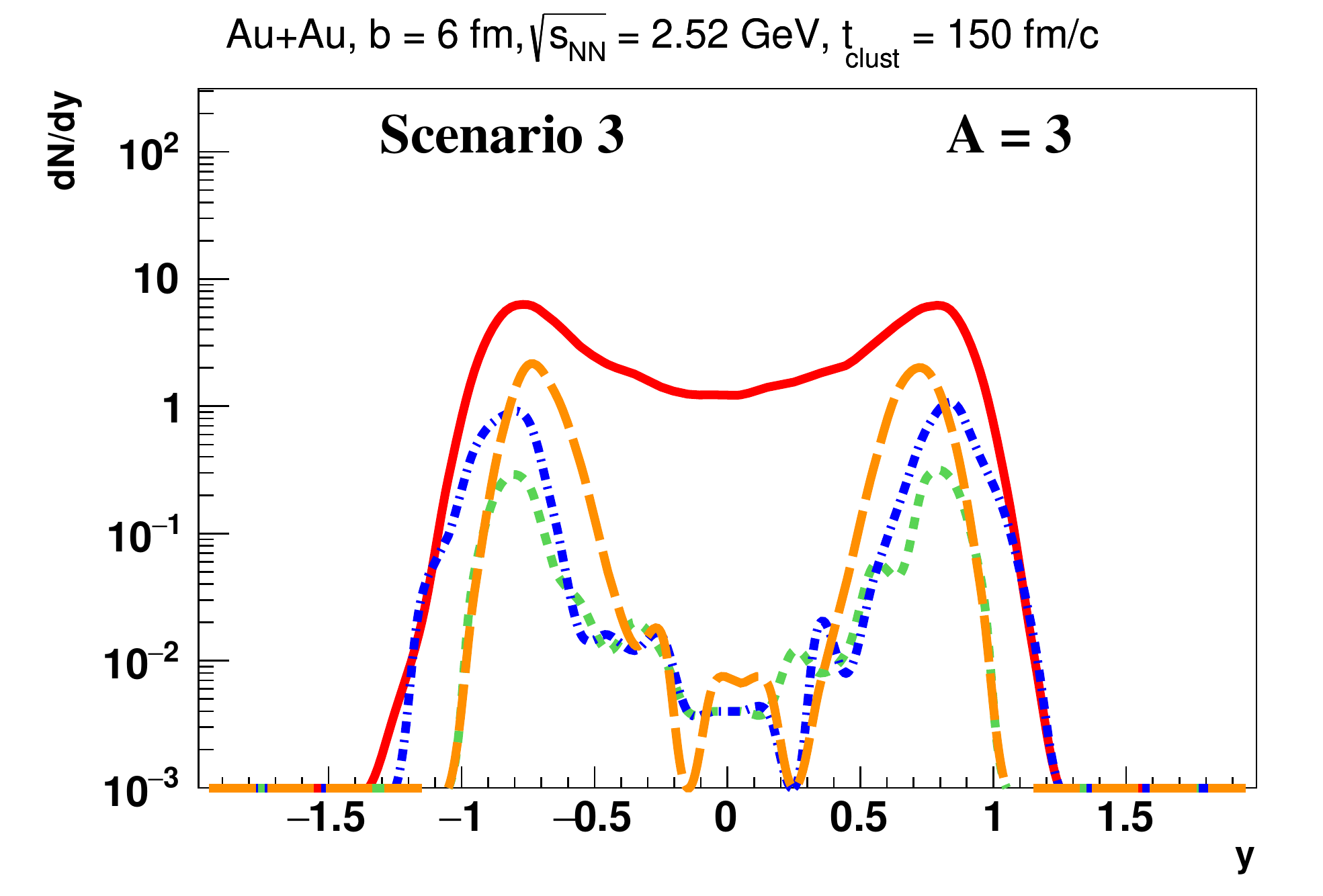} \\
        \end{tabular}
    }
\caption{\label{fig:2.52dndyA3} The rapidity distributions of clusters with the mass number $A = 3$ at $t_{clust} = 40, 90, 150$ fm/c in semi-peripheral ($b=6$ fm) $Au+Au$ collisions at $\sqrt{s}=2.52$ GeV. The left column: "Scenario 1", the center column: "Scenario 2", the right column: "Scenario 3". The color coding is the same as in Fig.~\ref{fig:2.52dndyA2}.}
\end{figure*}

\begin{figure*}
    \resizebox{\textwidth}{!}{
        \begin{tabular}{ccc}
          \includegraphics{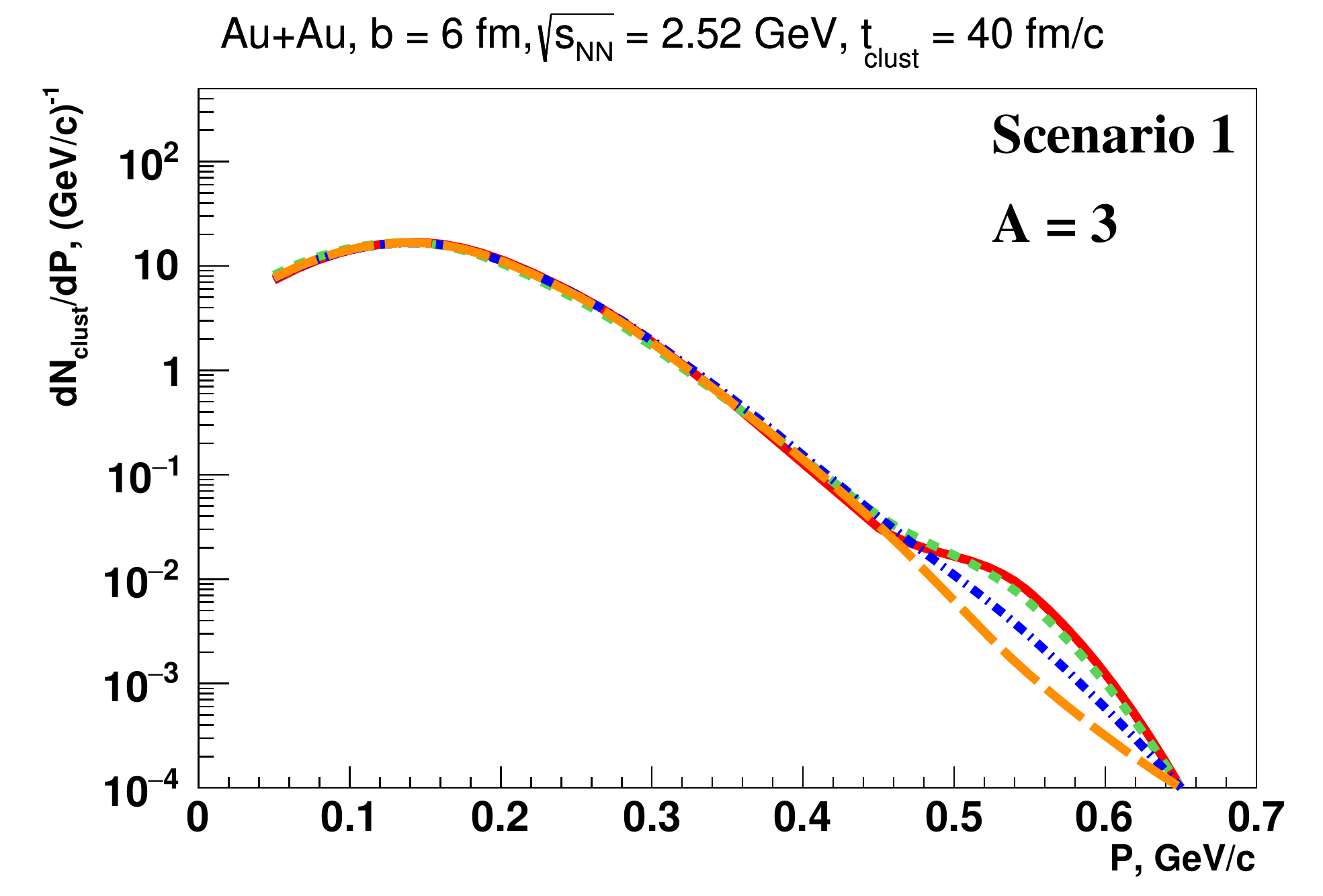} &
          \includegraphics{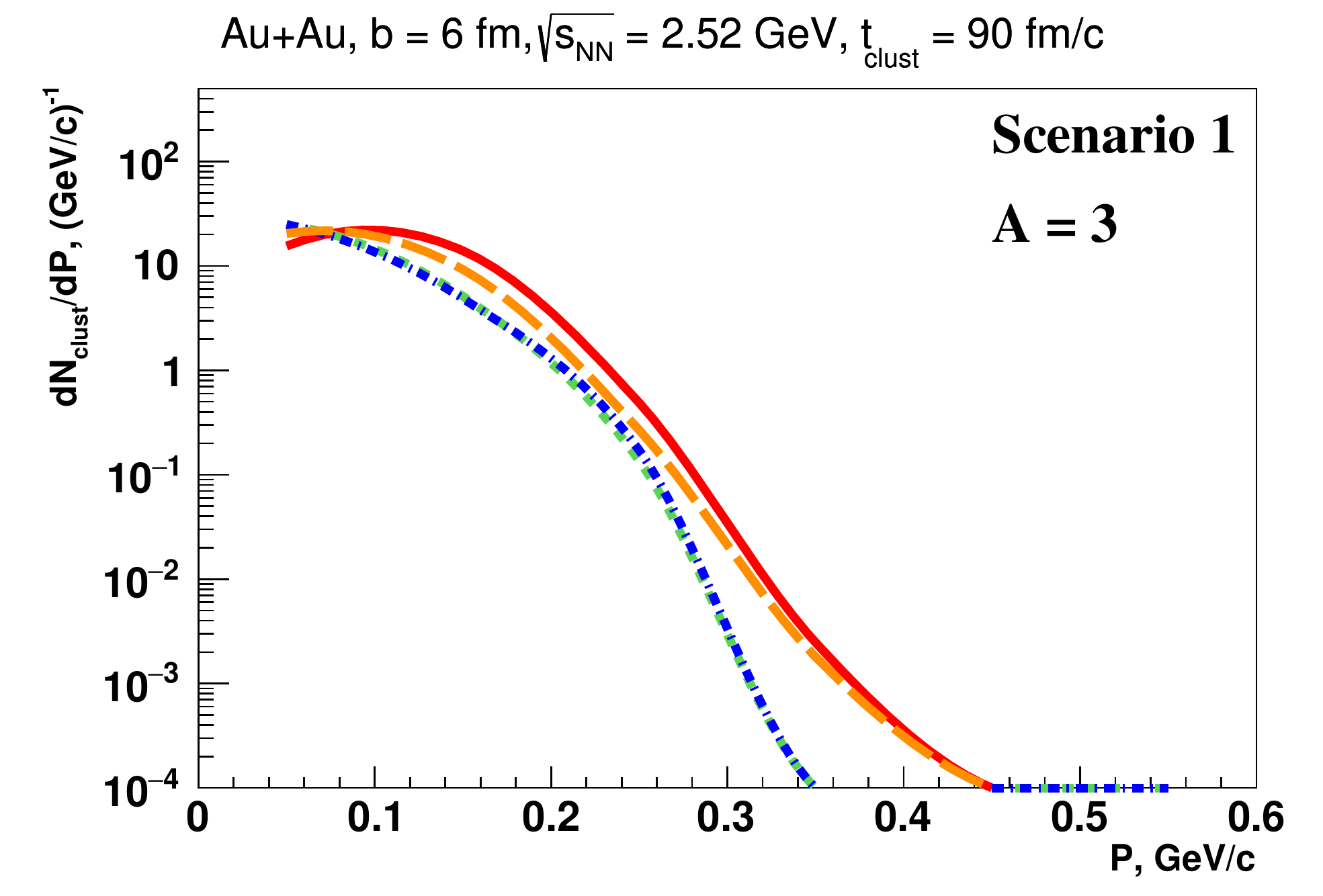} &
          \includegraphics{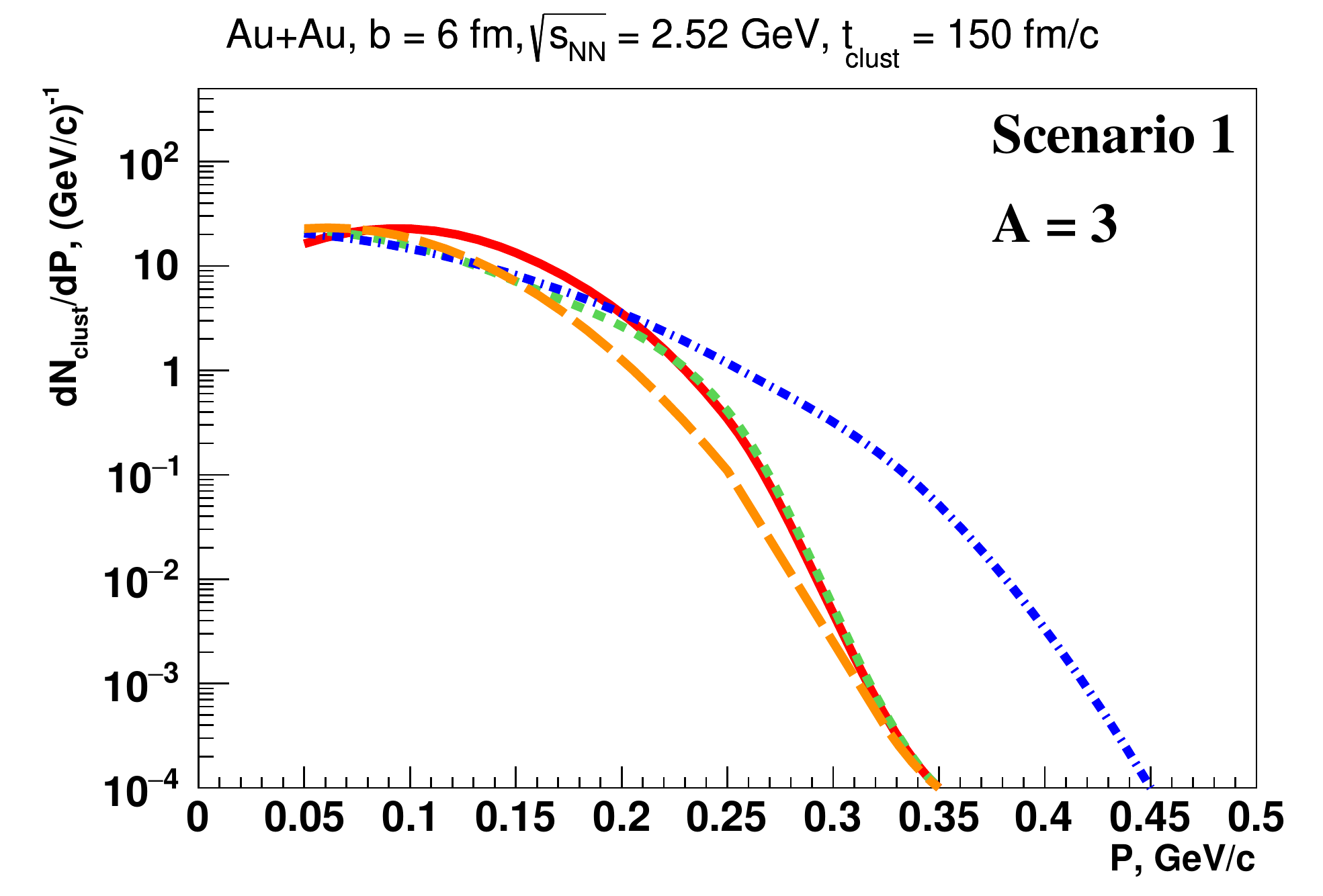} \\
        \end{tabular}
    }
\caption{\label{fig:2.52dndpA3} The momentum spectra of baryons ($p$, $n$, $\Lambda$ and $\Sigma^{0}$) from $A = 3$ clusters in semi-peripheral ($b=6$ fm) $Au+Au$ collisions at $\sqrt{s}=2.52$ GeV (integrated over all rapidity range). The momentum is calculated in the cluster center of mass frame. The left column: $t_{clust} = 40$ fm/c, the center column: $t_{clust} = 90$ fm/c, the right column: $t_{clust} = 150$ fm/c. The color coding is the same as in Fig.~\ref{fig:2.52dndyA2}.
}
\end{figure*}

\begin{figure*}
    \resizebox{\textwidth}{!}{
        \includegraphics{plots/scenario1/header.pdf} 
    } \\
    \resizebox{\textwidth}{!}{
        \begin{tabular}{ccc}
          \includegraphics{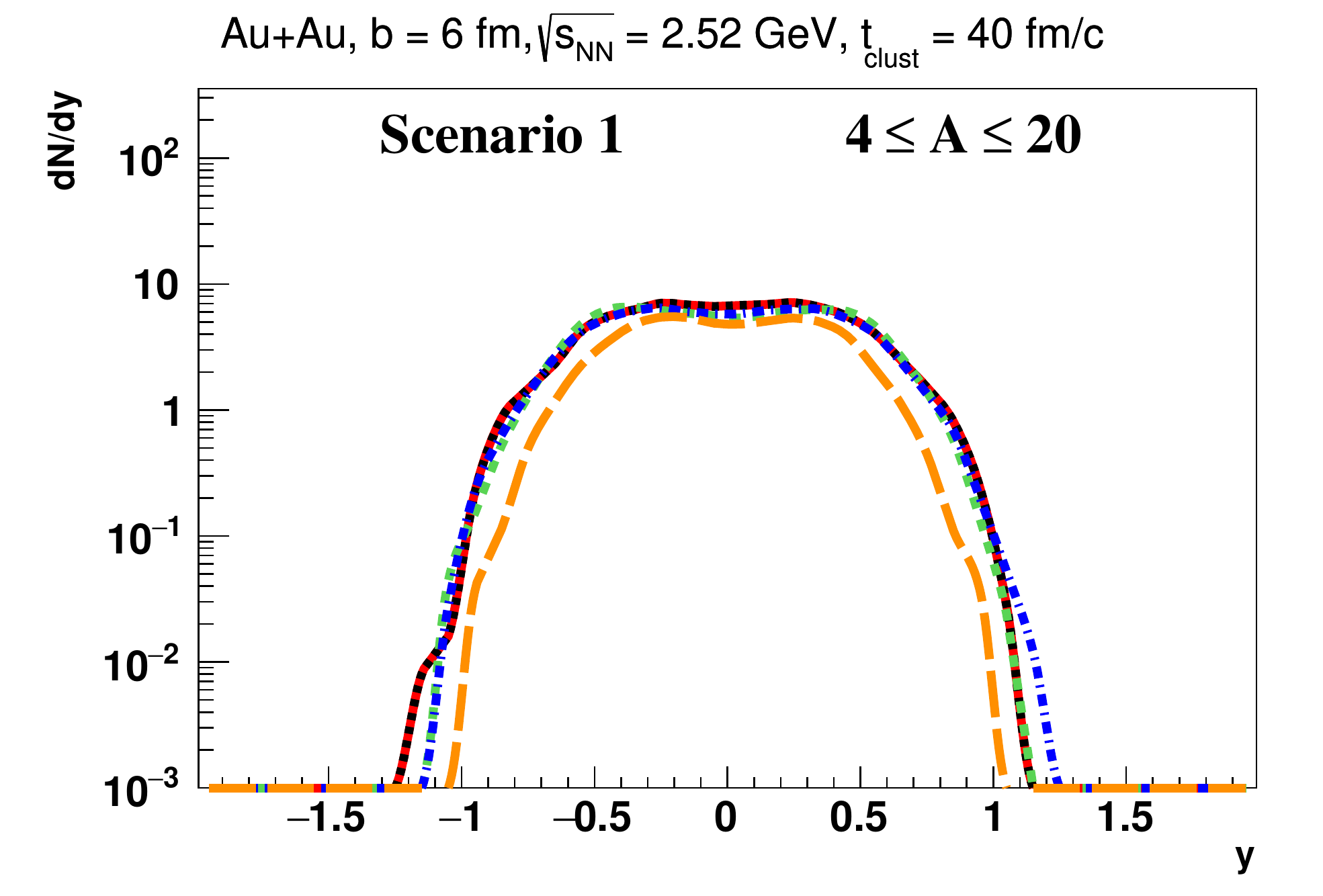} &
          \includegraphics{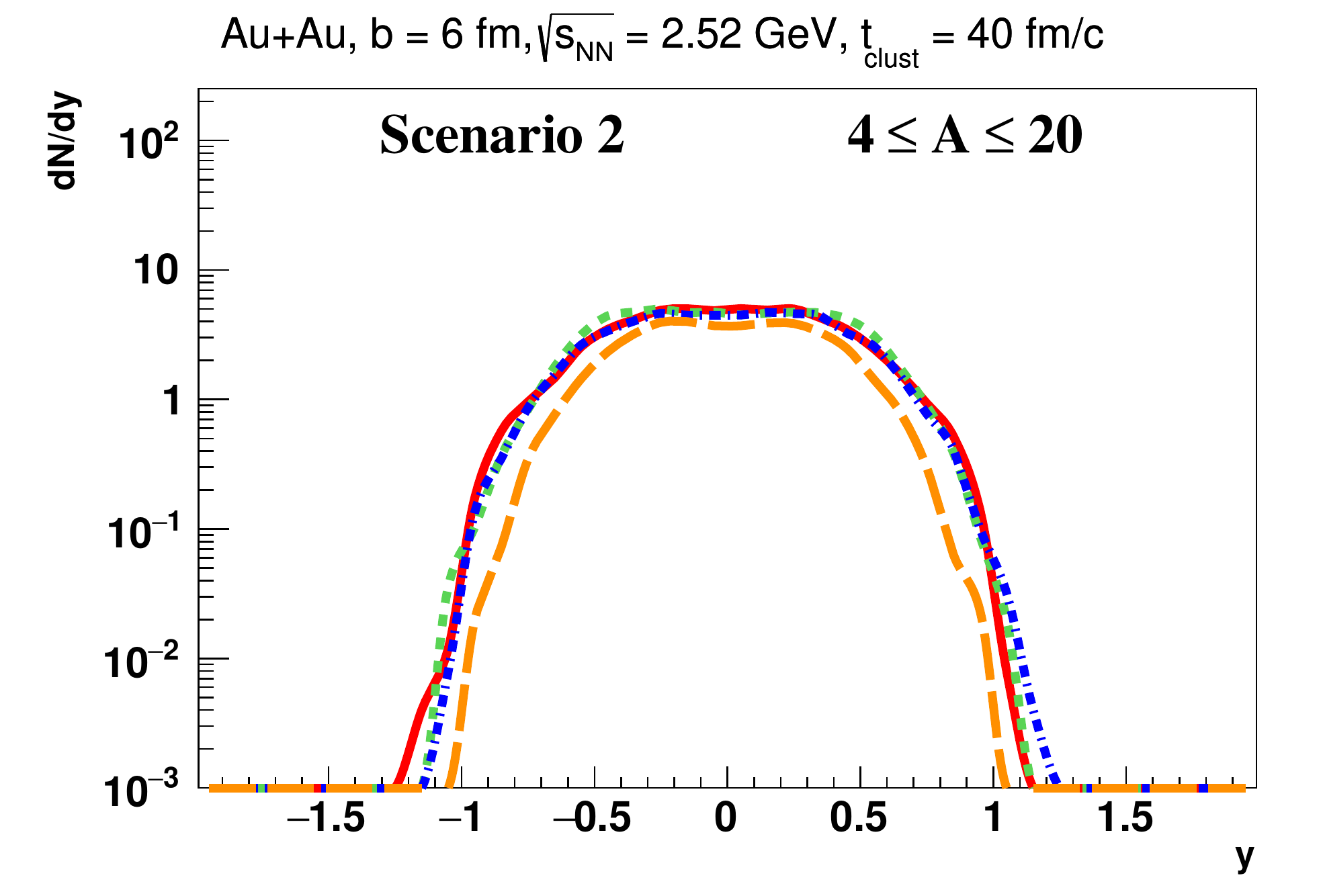} &
          \includegraphics{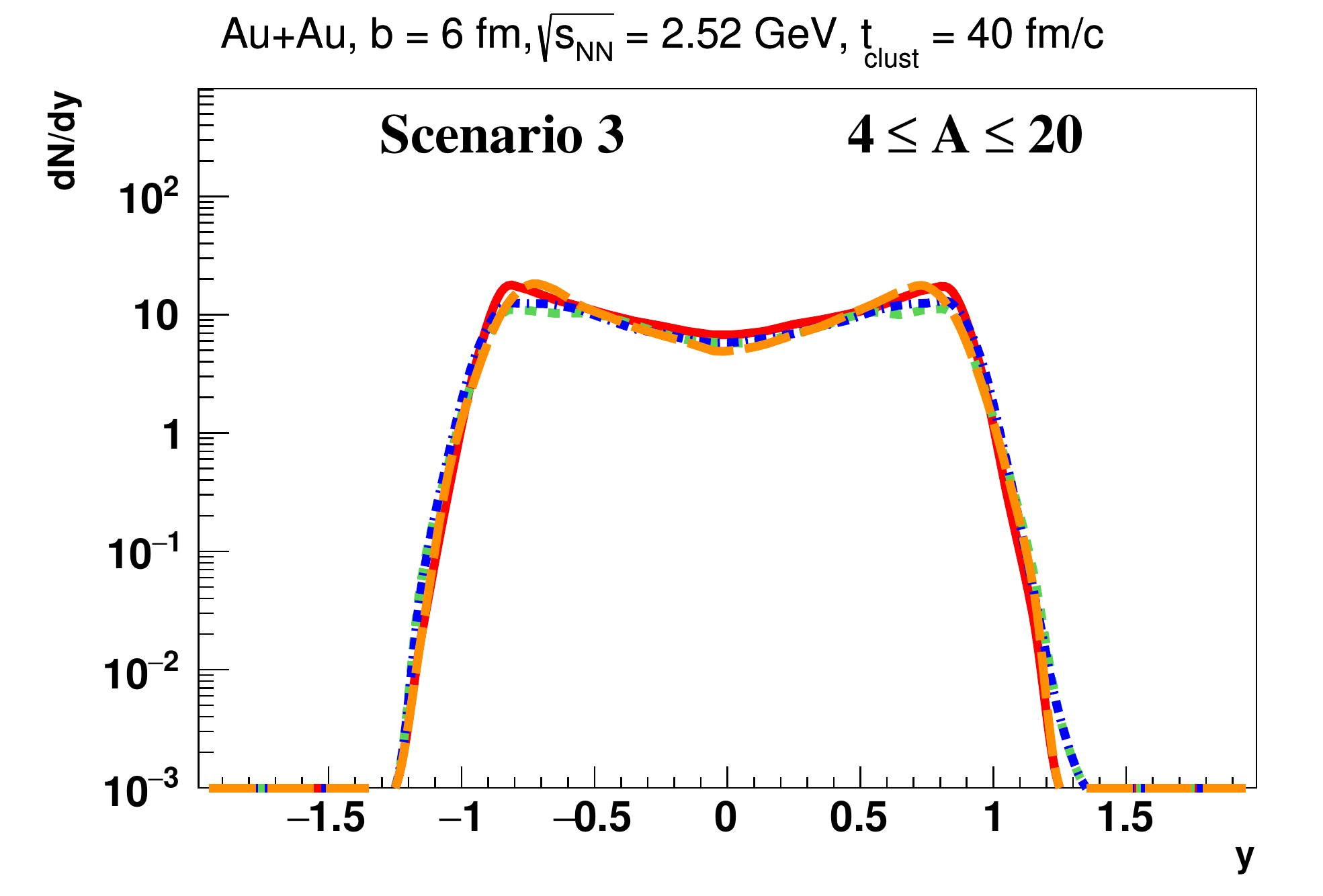} \\
          \includegraphics{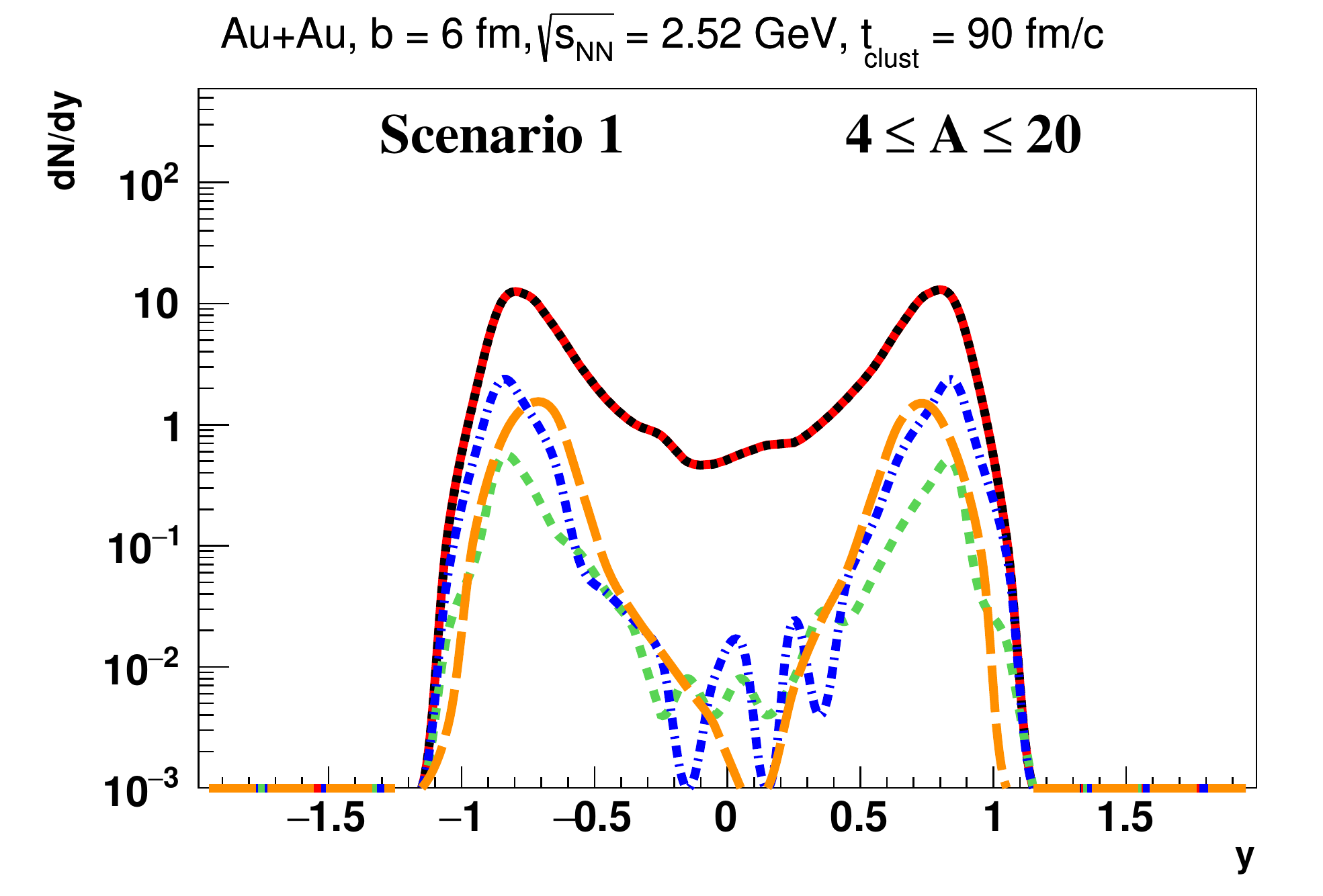} &
          \includegraphics{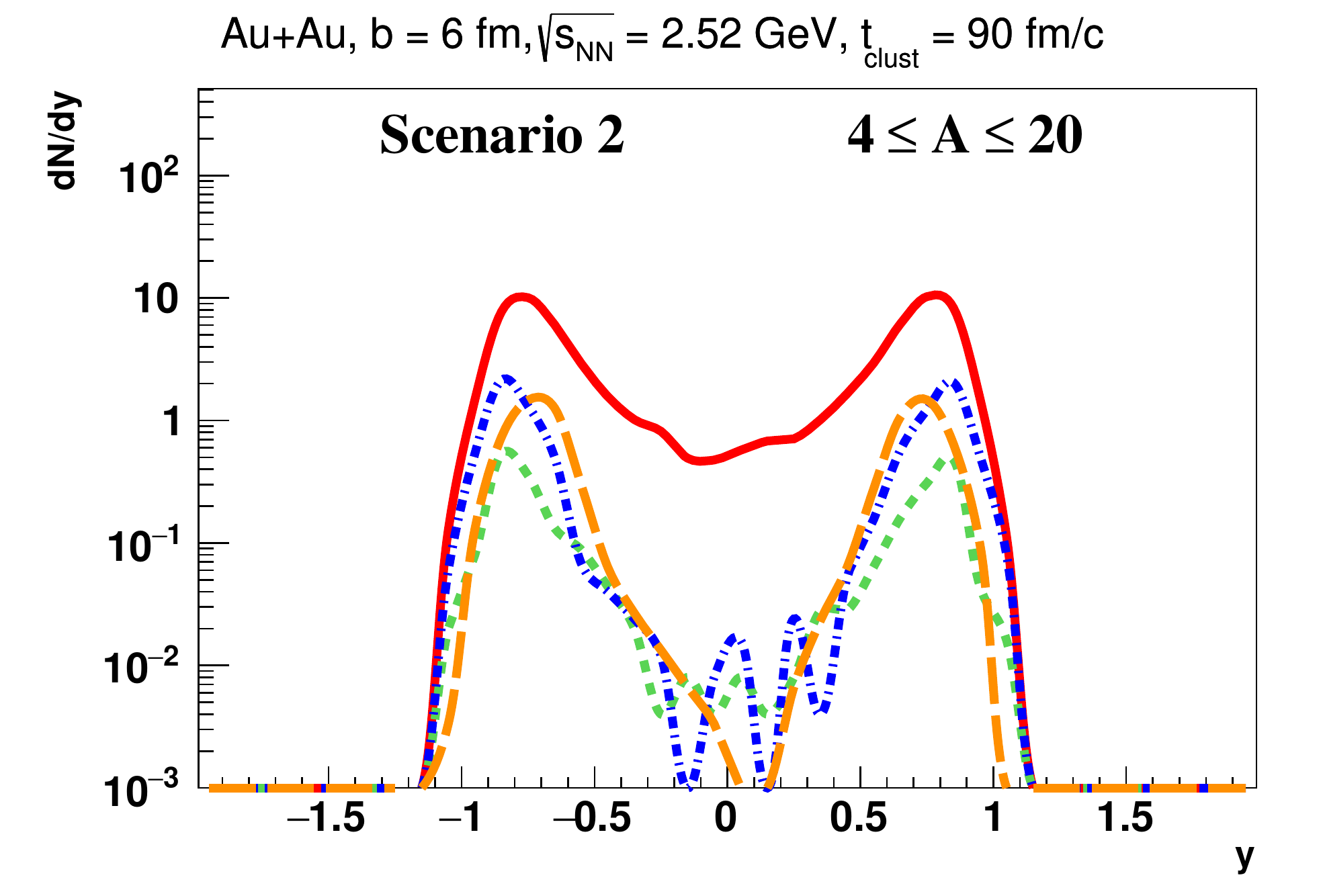} &
          \includegraphics{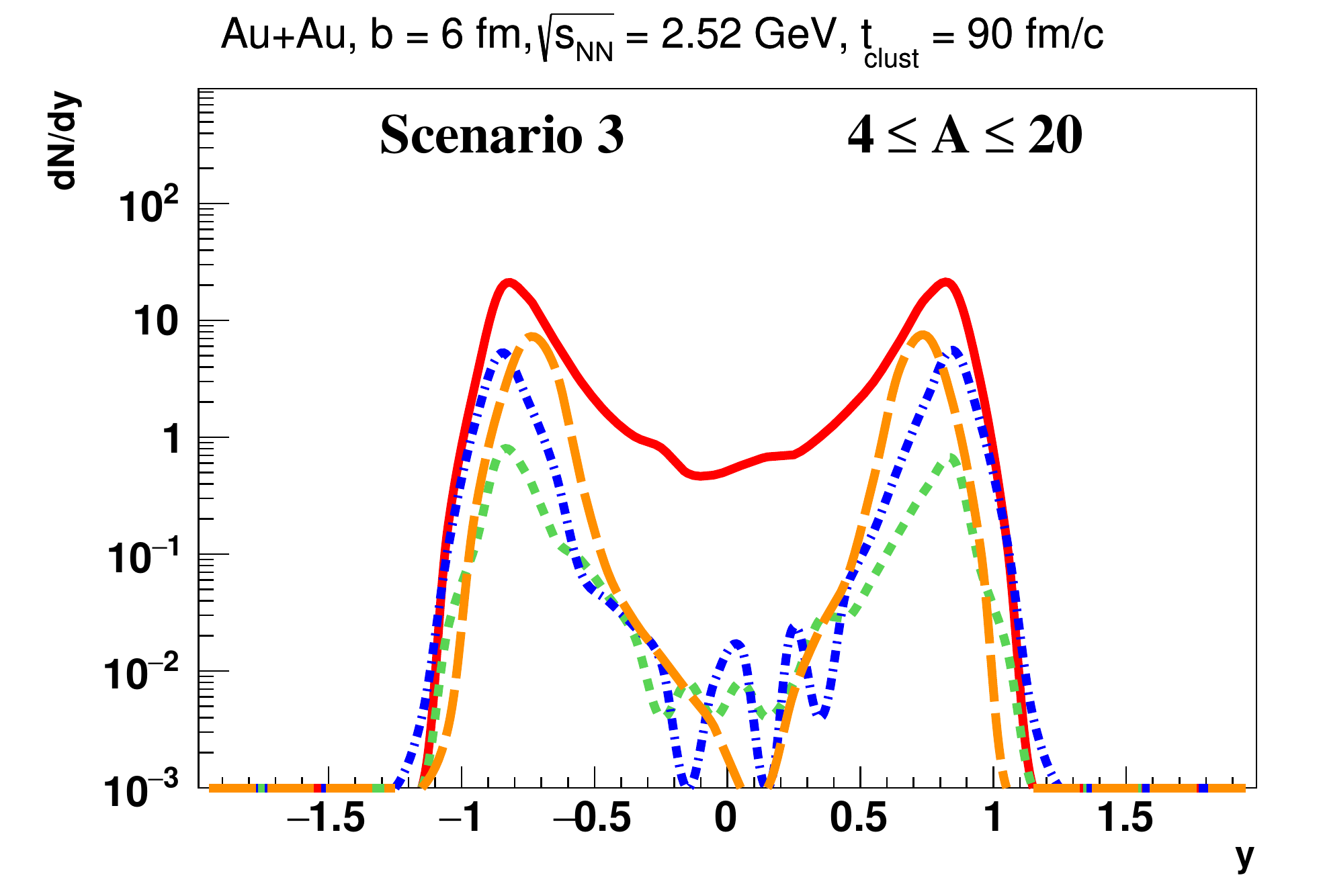} \\
          \includegraphics{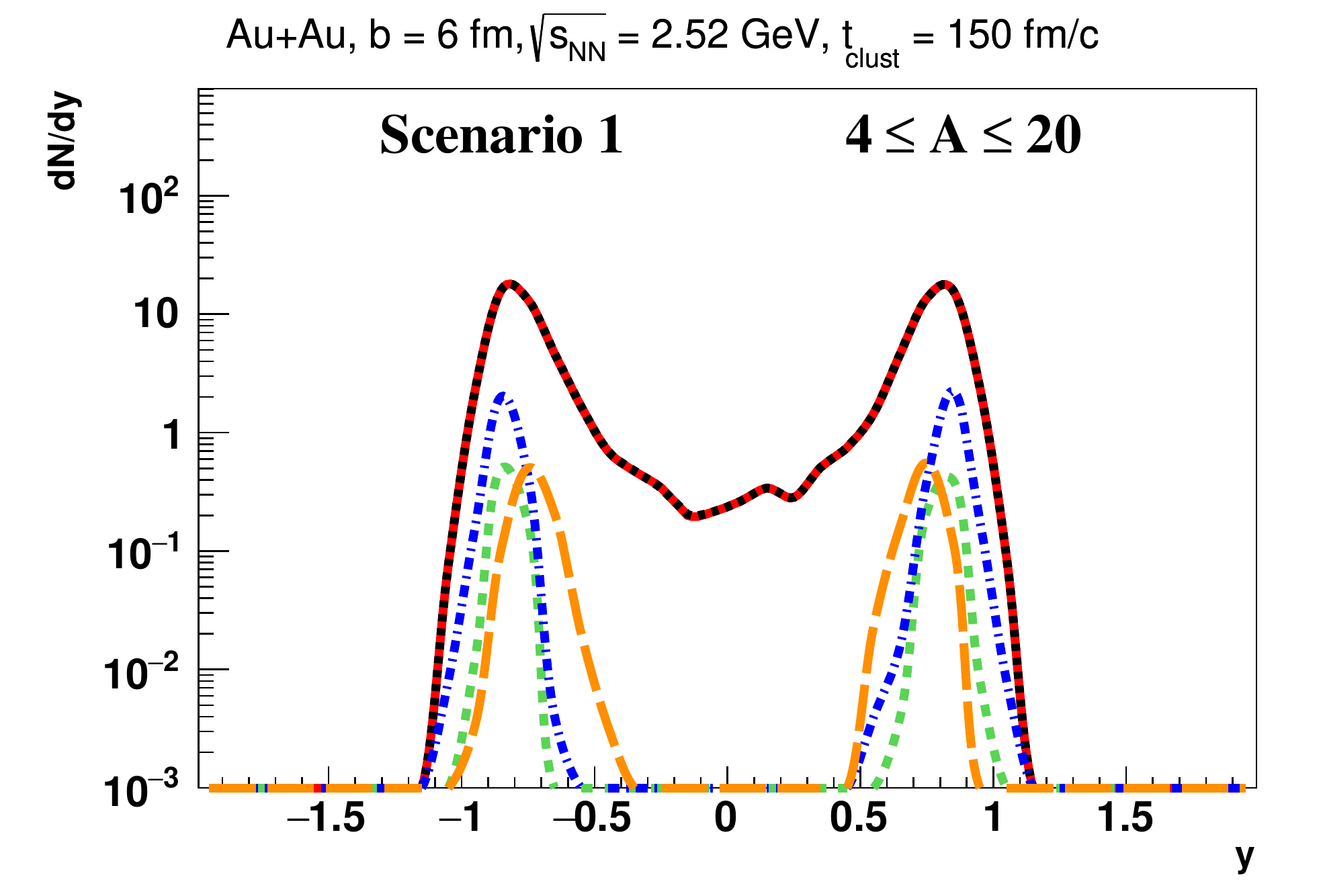} &
          \includegraphics{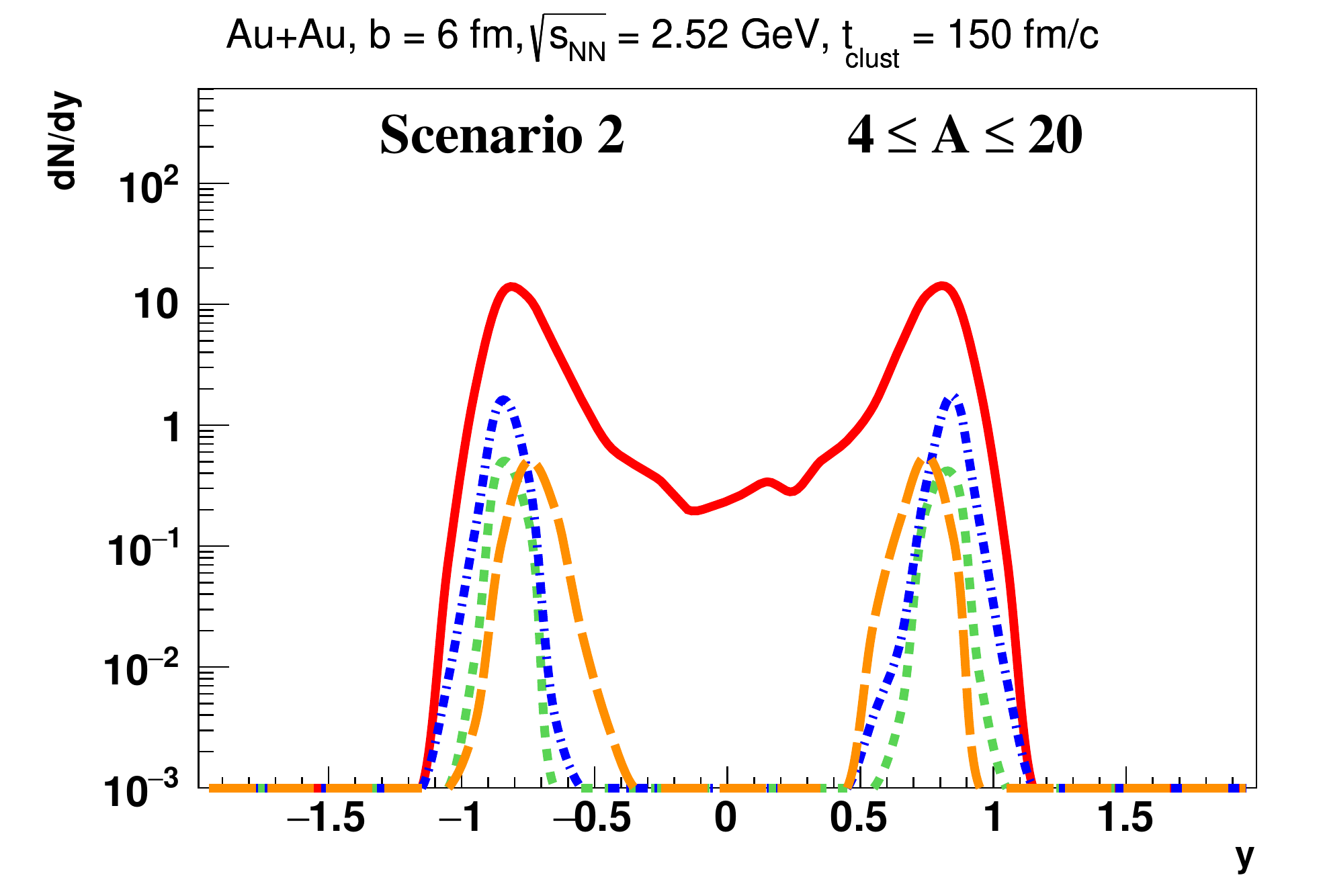} &
          \includegraphics{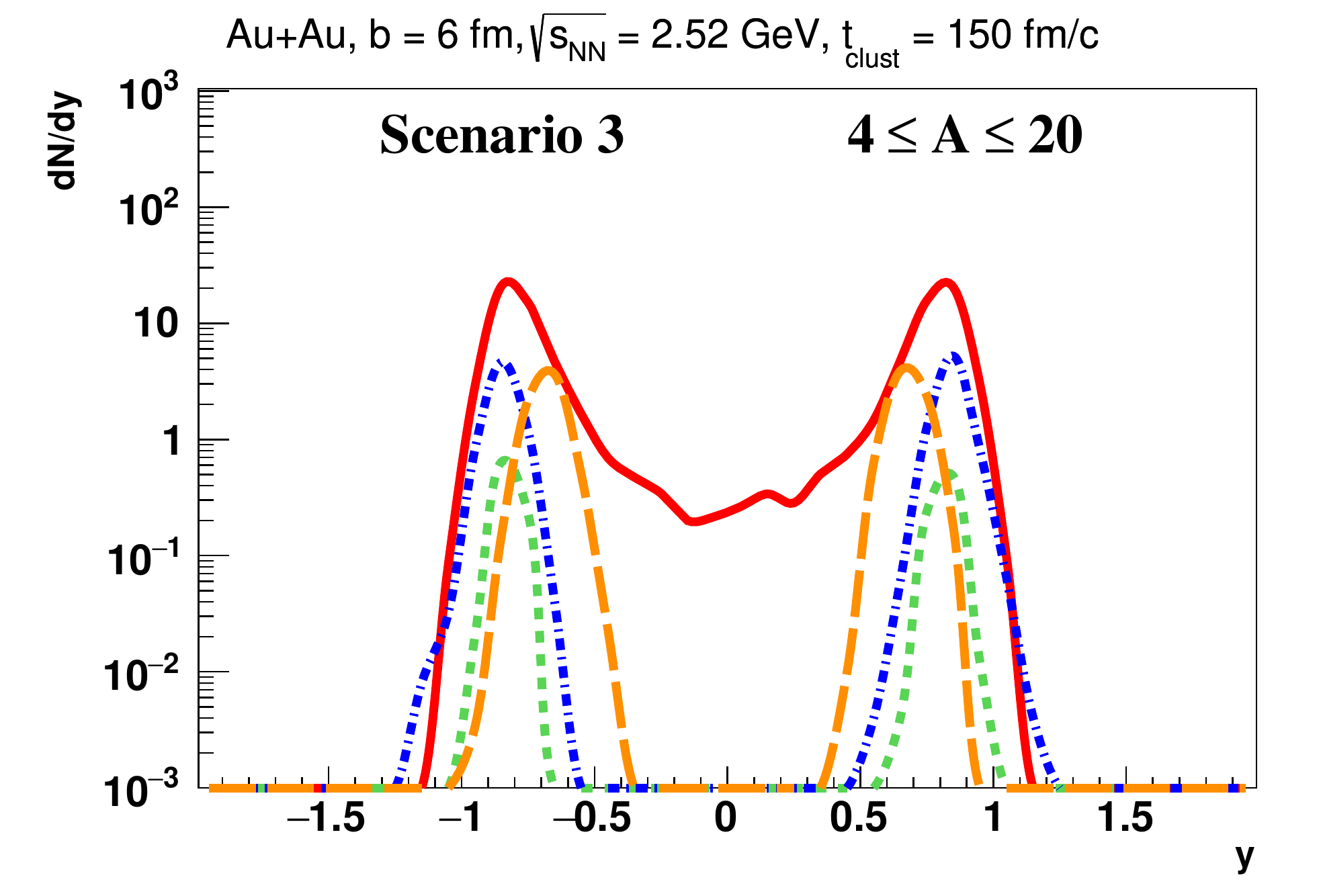} \\
        \end{tabular}
    }
\caption{\label{fig:2.52dndyA4_20} The rapidity distributions of clusters with the mass number $4 \leq A \leq 20$ at $t_{clust} = 40, 90, 150$ fm/c in semi-peripheral ($b=6$ fm) $Au+Au$ collisions at $\sqrt{s}=2.52$ GeV. The left column: "Scenario 1", the center column: "Scenario 2", the right column: "Scenario 3". The color coding is the same as in Fig.~\ref{fig:2.52dndyA2}.}
\end{figure*}

\begin{figure*}
    \resizebox{\textwidth}{!}{
        \begin{tabular}{ccc}
          \includegraphics{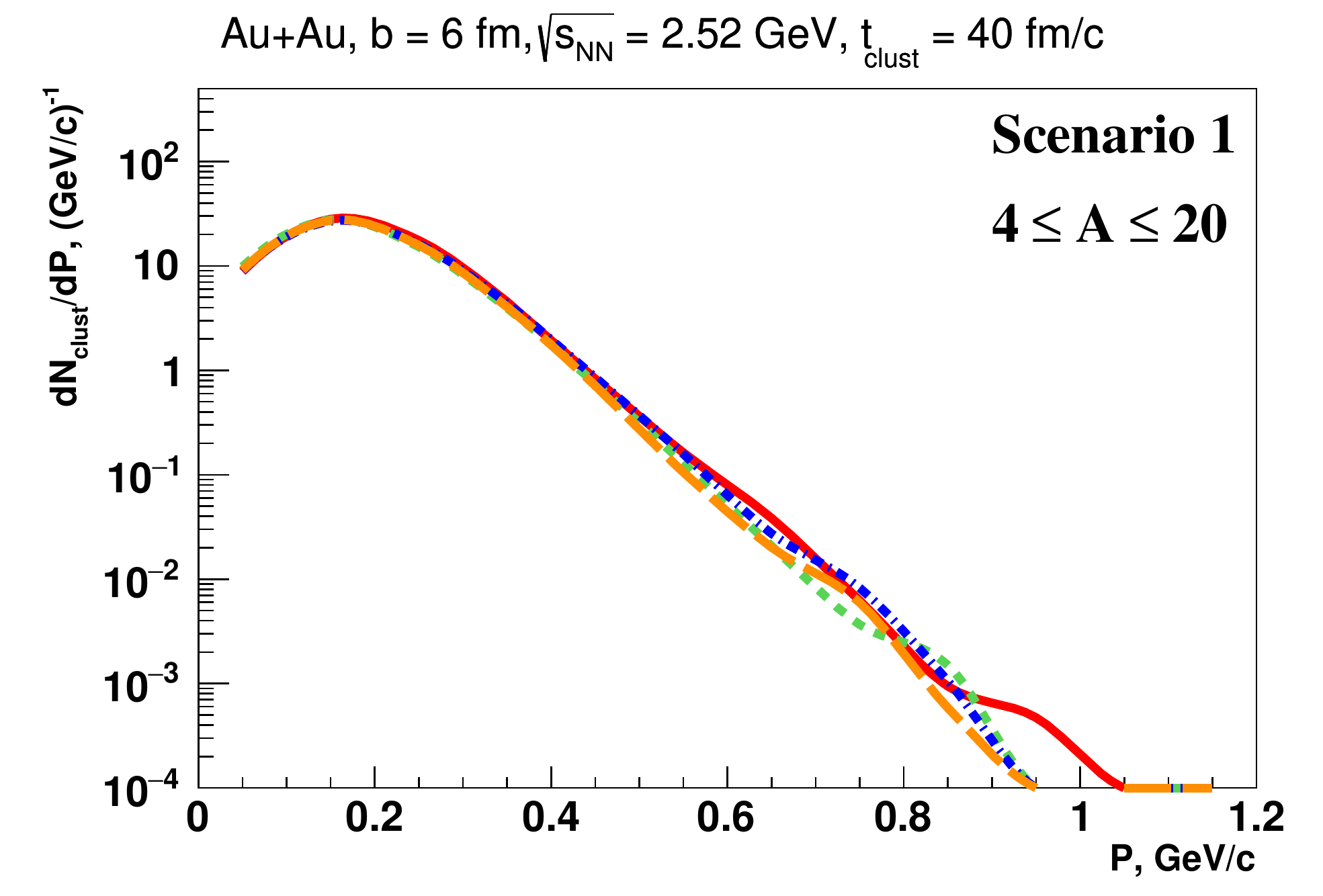} &
          \includegraphics{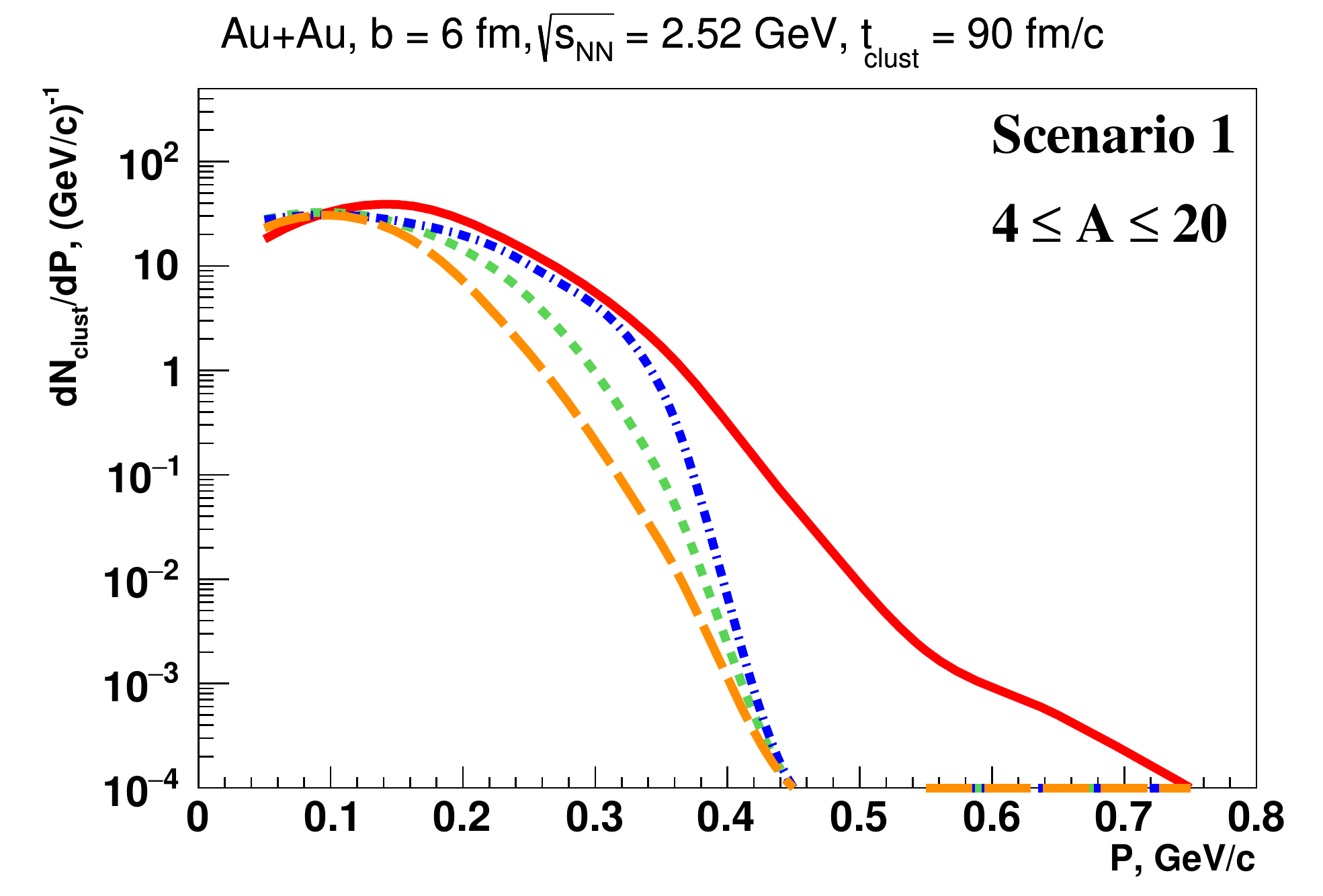} &
          \includegraphics{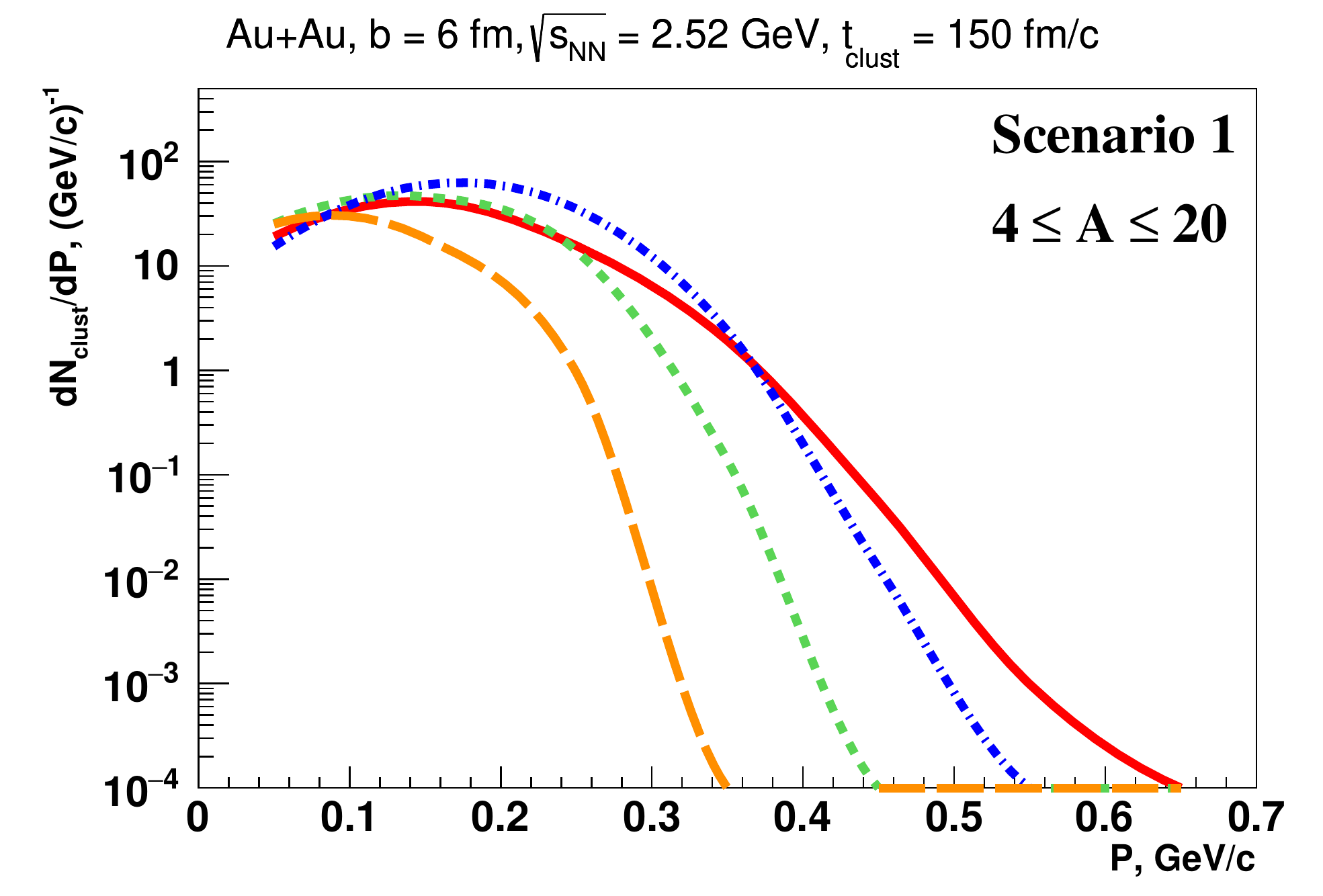} \\
        \end{tabular}
    }
\caption{\label{fig:2.52dndpA4_20} The momentum spectra of baryons ($p$, $n$, $\Lambda$ and $\Sigma^{0}$) from $4 \leq A \leq 20$ clusters in semi-peripheral ($b=6$ fm) $Au+Au$ collisions at $\sqrt{s}=2.52$ GeV (integrated over all rapidity range). The momentum is calculated in the cluster center of mass frame. The left column: $t_{clust} = 40$ fm/c, the center column: $t_{clust} = 90$ fm/c, the right column: $t_{clust} = 150$ fm/c. The color coding is the same as in Fig.~\ref{fig:2.52dndyA2}.}
\end{figure*}

A similar trend is found for clusters with the mass number $A = 3$, shown in Fig.~\ref{fig:2.52dndyA3} and Fig.~\ref{fig:2.52dndpA3}, and for the intermediate mass clusters with $4 \leq A \leq 20$, presented in Fig.~\ref{fig:2.52dndyA4_20} and Fig.~\ref{fig:2.52dndpA4_20}: at the lower energy ($\sqrt{s_{NN}} = 2.52$ GeV) 
PHQMD predicts several orders of magnitude more clusters at mid-rapidity than the other models of this study.
This shows the importance of spacial-momentum correlations which are kept in the QMD dynamics in contradiction to the MF dynamics where the forces between test particles are reduced by Num, the number of test particles per physical particle.

\begin{figure*}
    \resizebox{\textwidth}{!}{
        \includegraphics{plots/scenario1/header.pdf} 
    } \\
    \resizebox{\textwidth}{!}{
        \begin{tabular}{ccc}
          \includegraphics{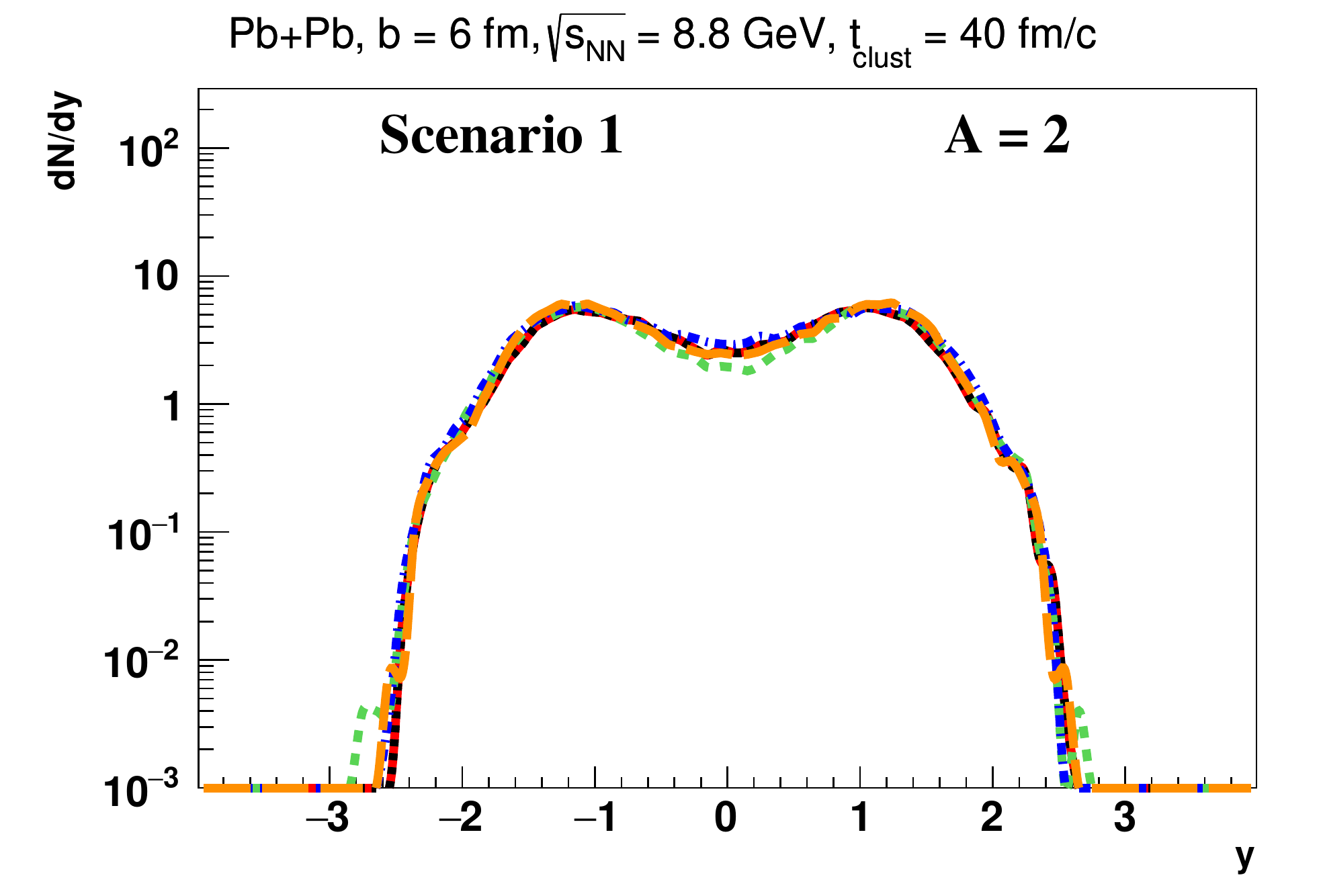} &
          \includegraphics{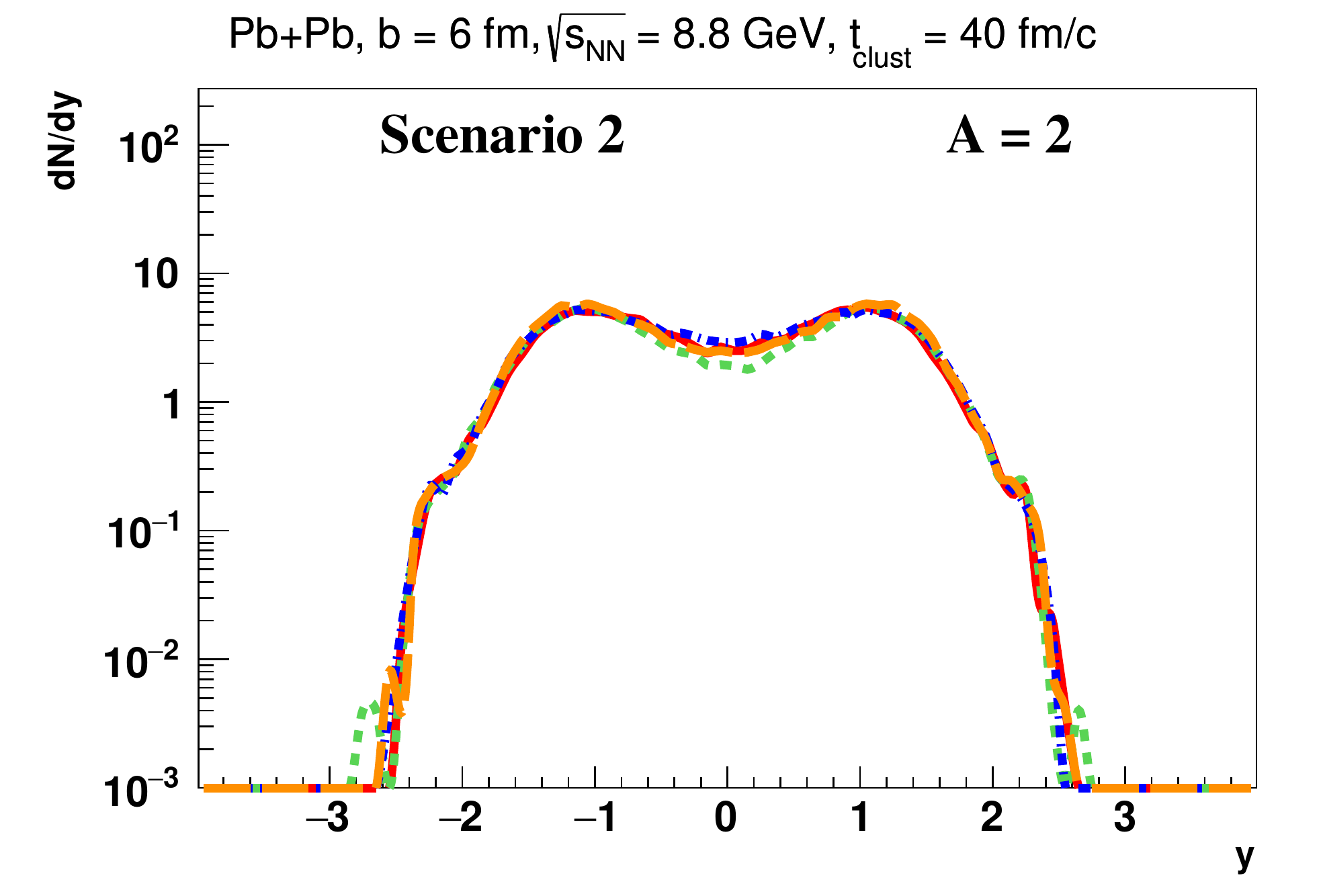} &
          \includegraphics{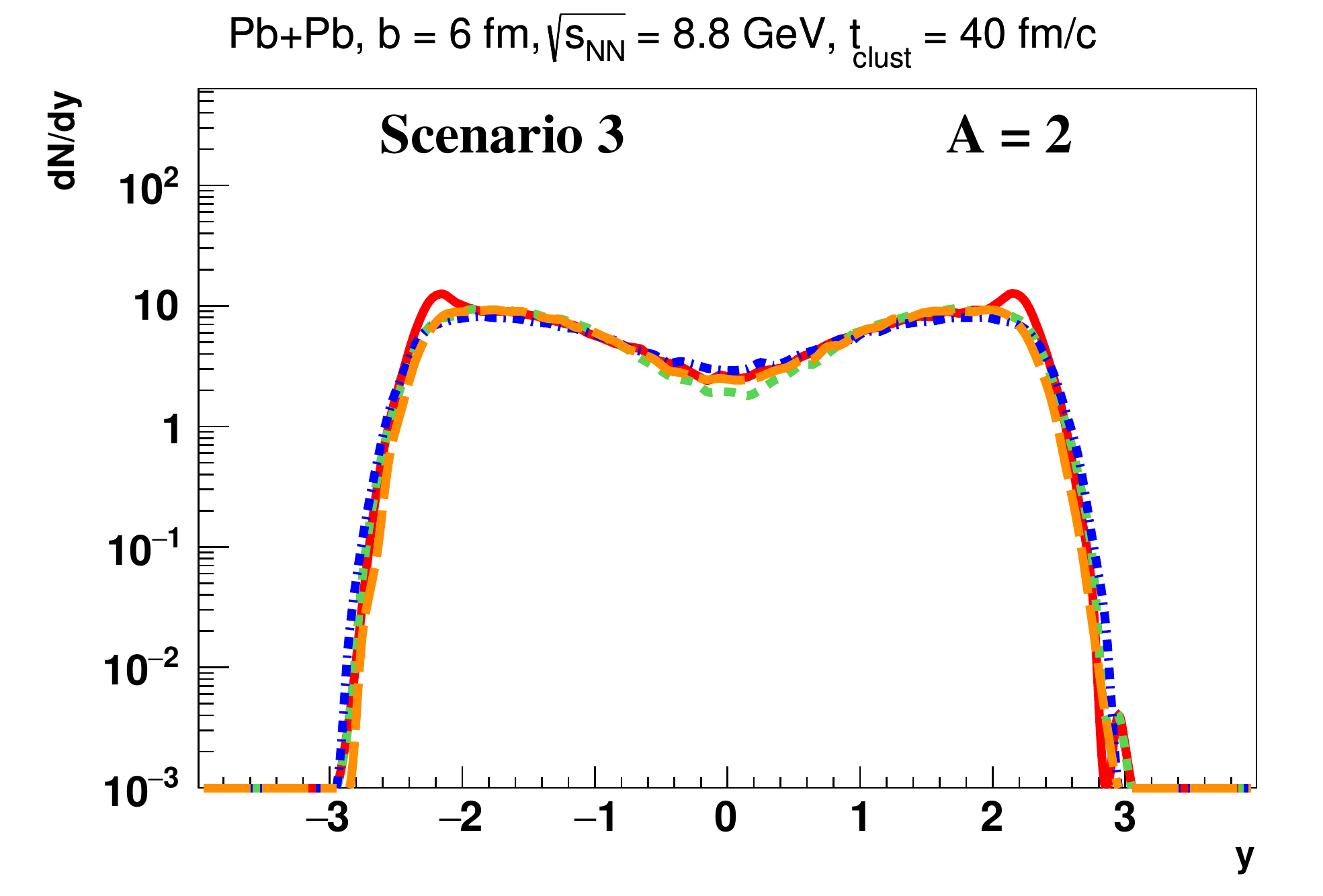} \\
          \includegraphics{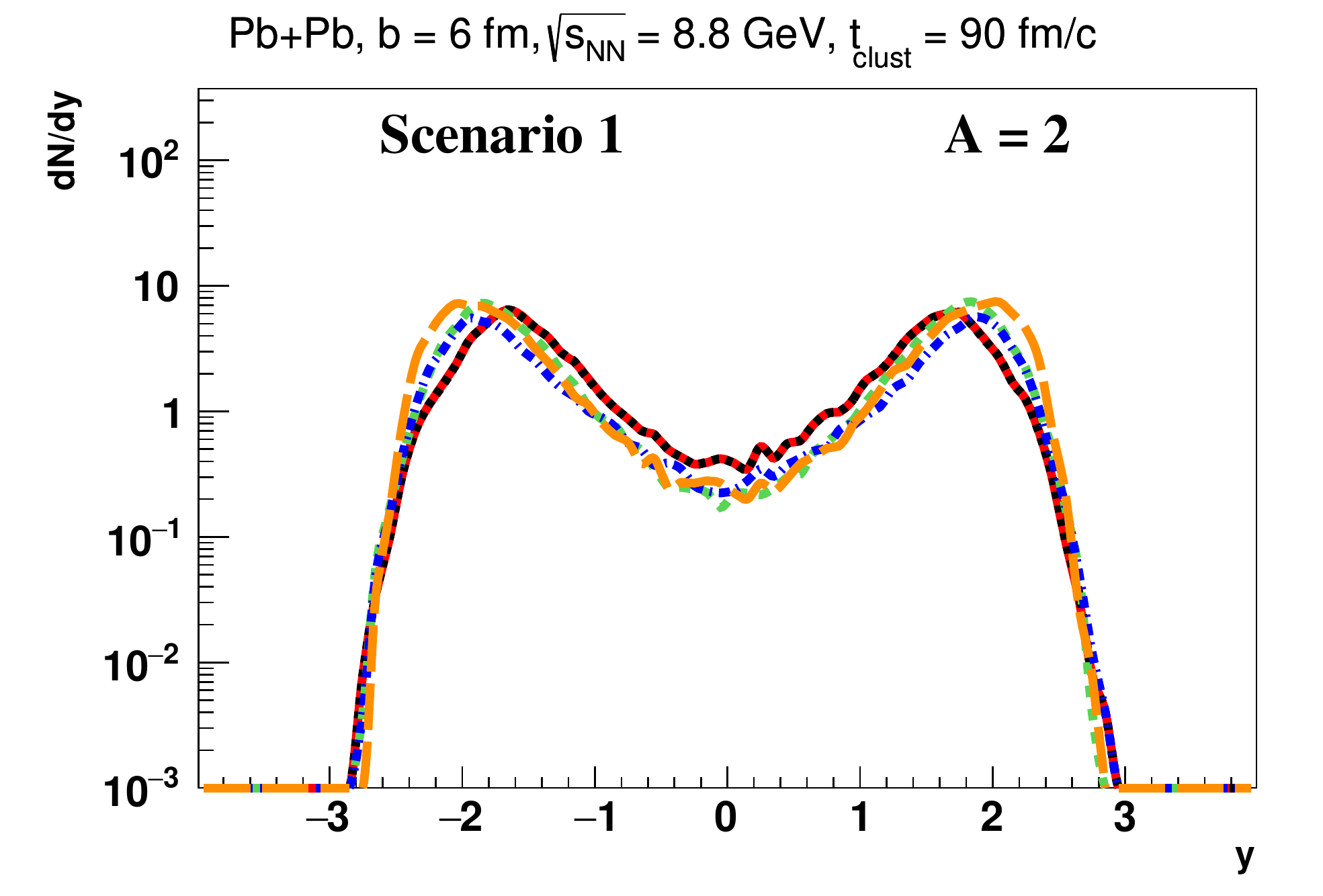} &
          \includegraphics{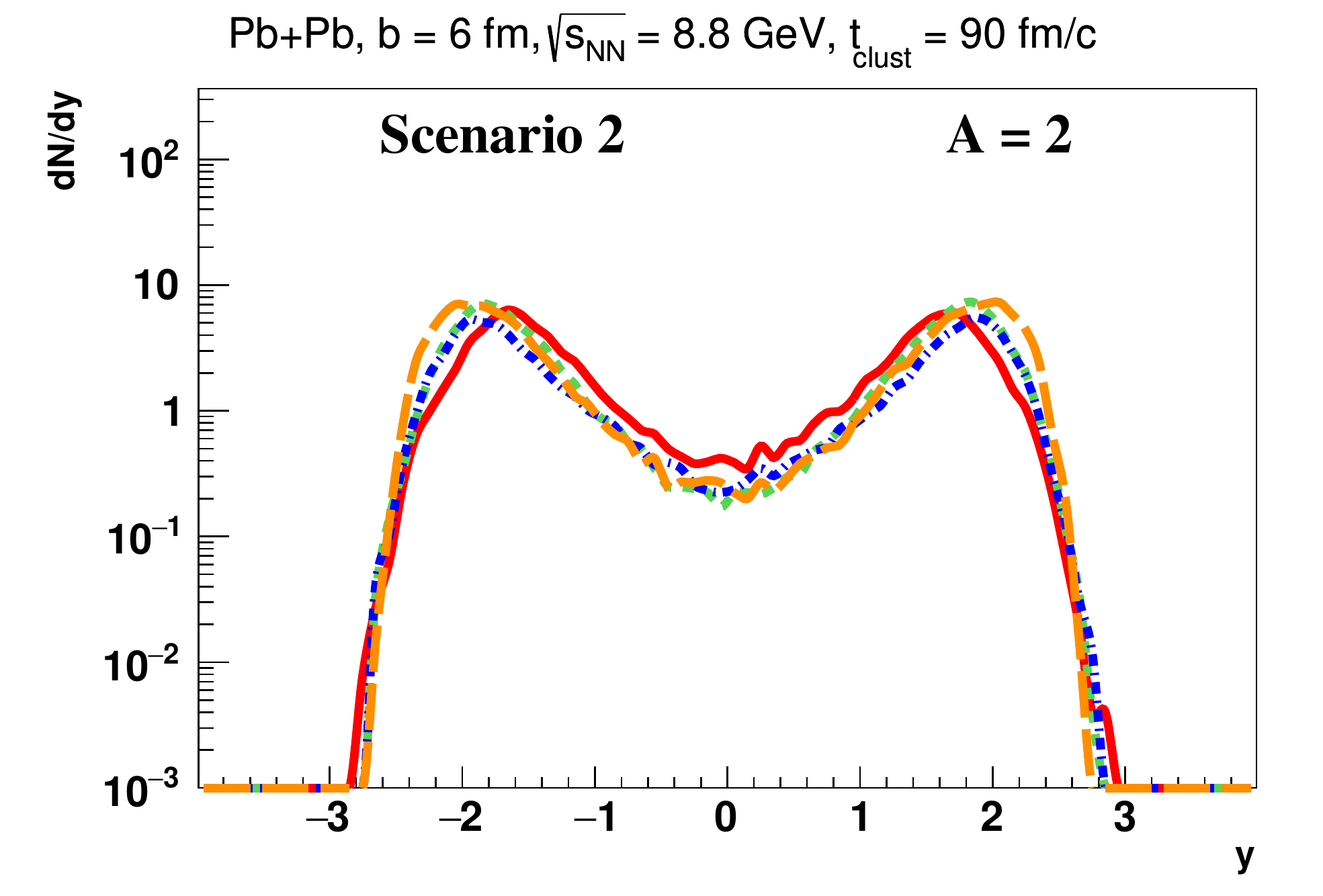} &
          \includegraphics{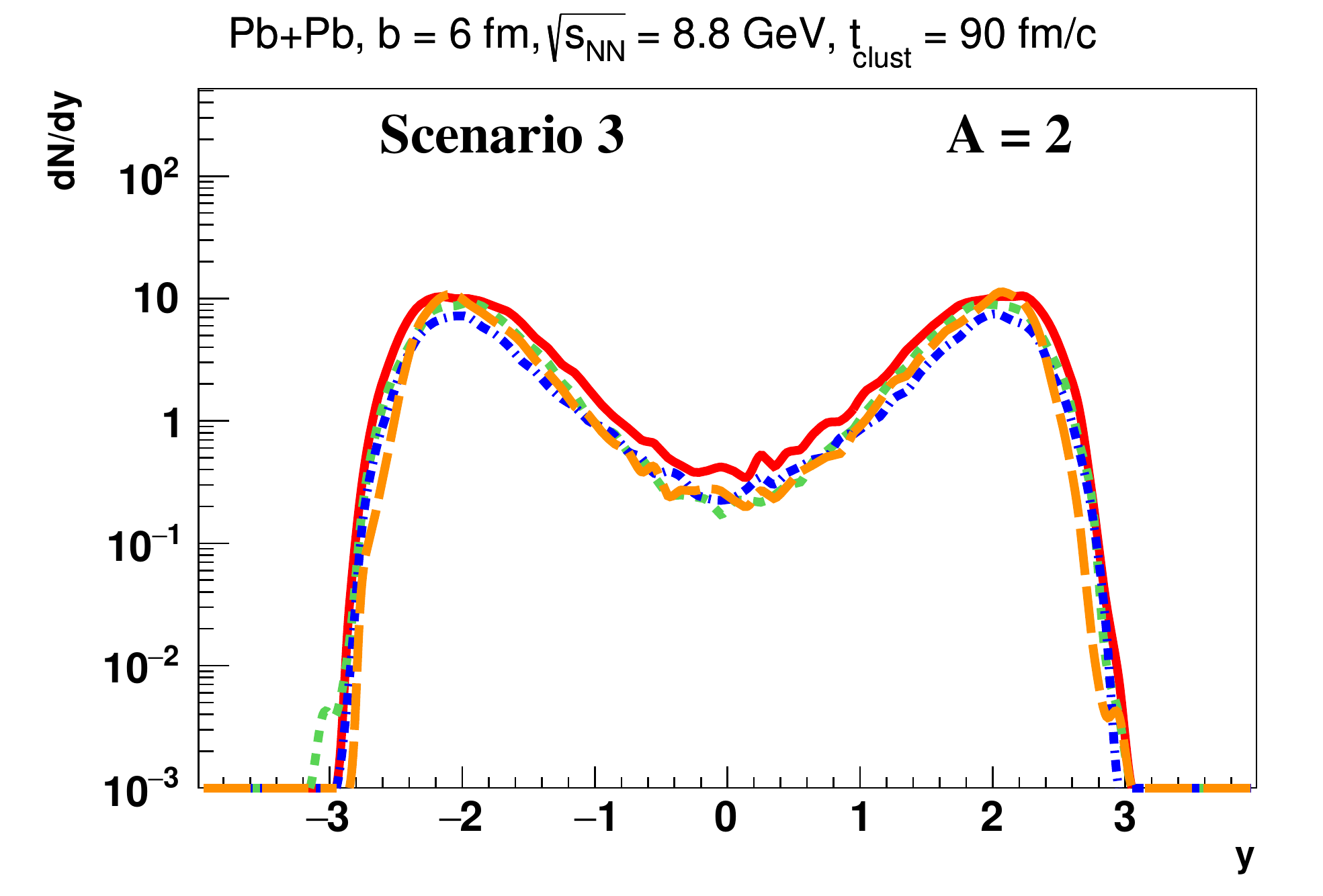} \\
          \includegraphics{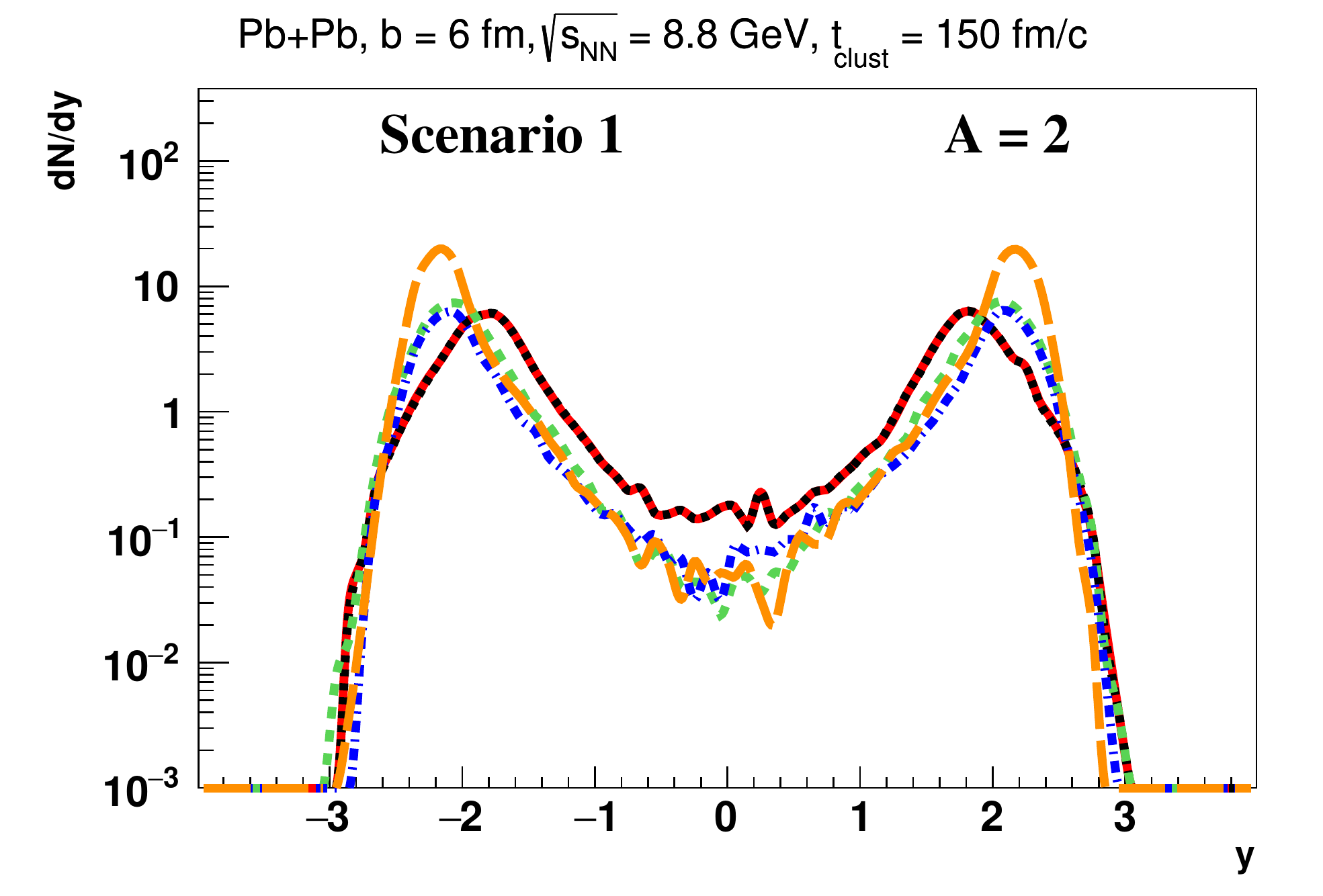} &
          \includegraphics{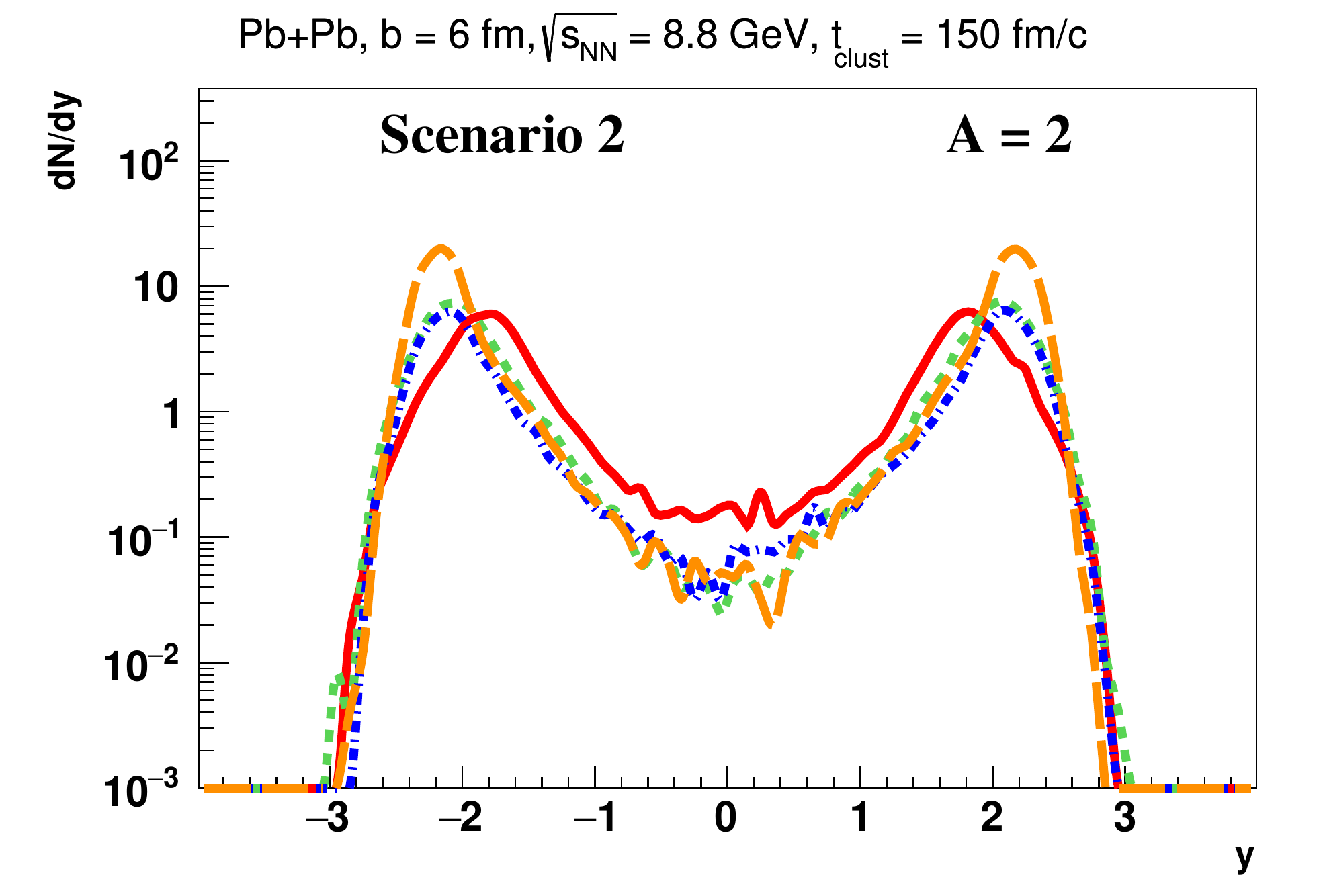} &
          \includegraphics{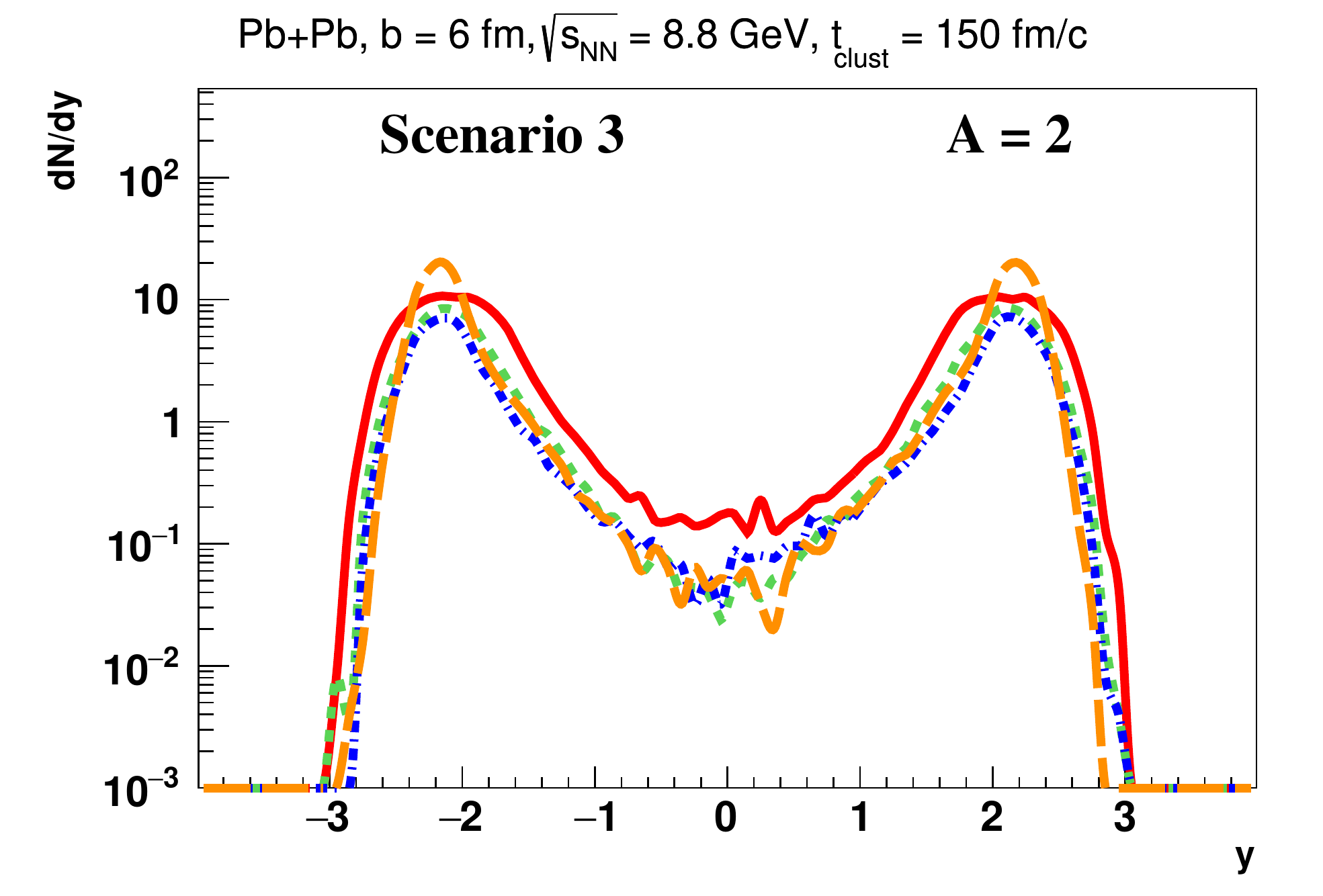} \\
        \end{tabular}
    }
\caption{\label{fig:8.8dndyA2} The rapidity distributions of clusters with the mass number $A = 2$ in semi-peripheral ($b=6$ fm) $Pb+Pb$ collisions at $\sqrt{s}=8.8$ GeV at $t_{clust} = 40, 90, 150$ fm/c. The left column: "Scenario 1", the center column: "Scenario 2", the right column: "Scenario 3". The color coding is the same as in Fig.~\ref{fig:2.52dndyA2}.}
\end{figure*}

\begin{figure*}
    \resizebox{\textwidth}{!}{
        \begin{tabular}{ccc}
          \includegraphics{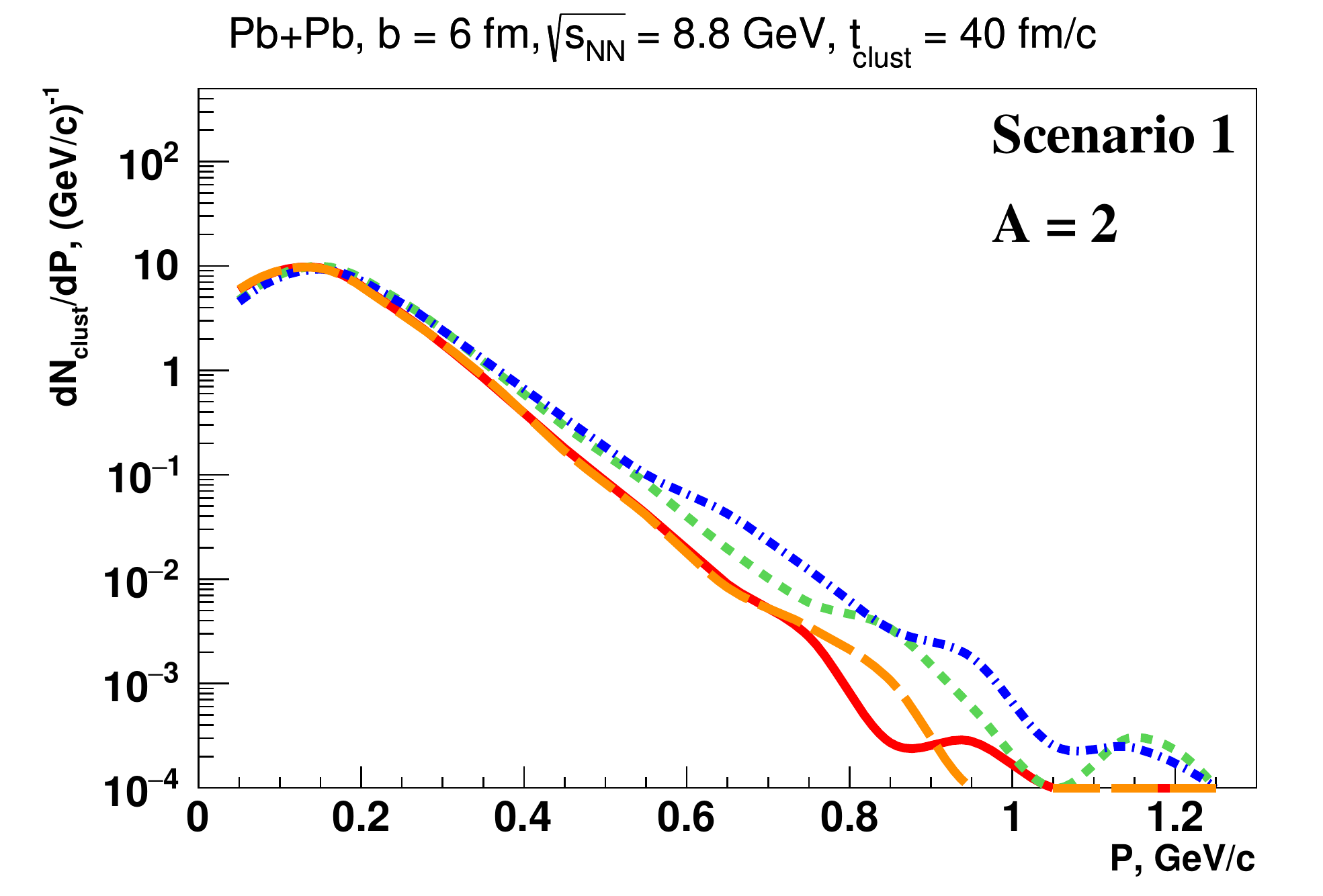} &
          \includegraphics{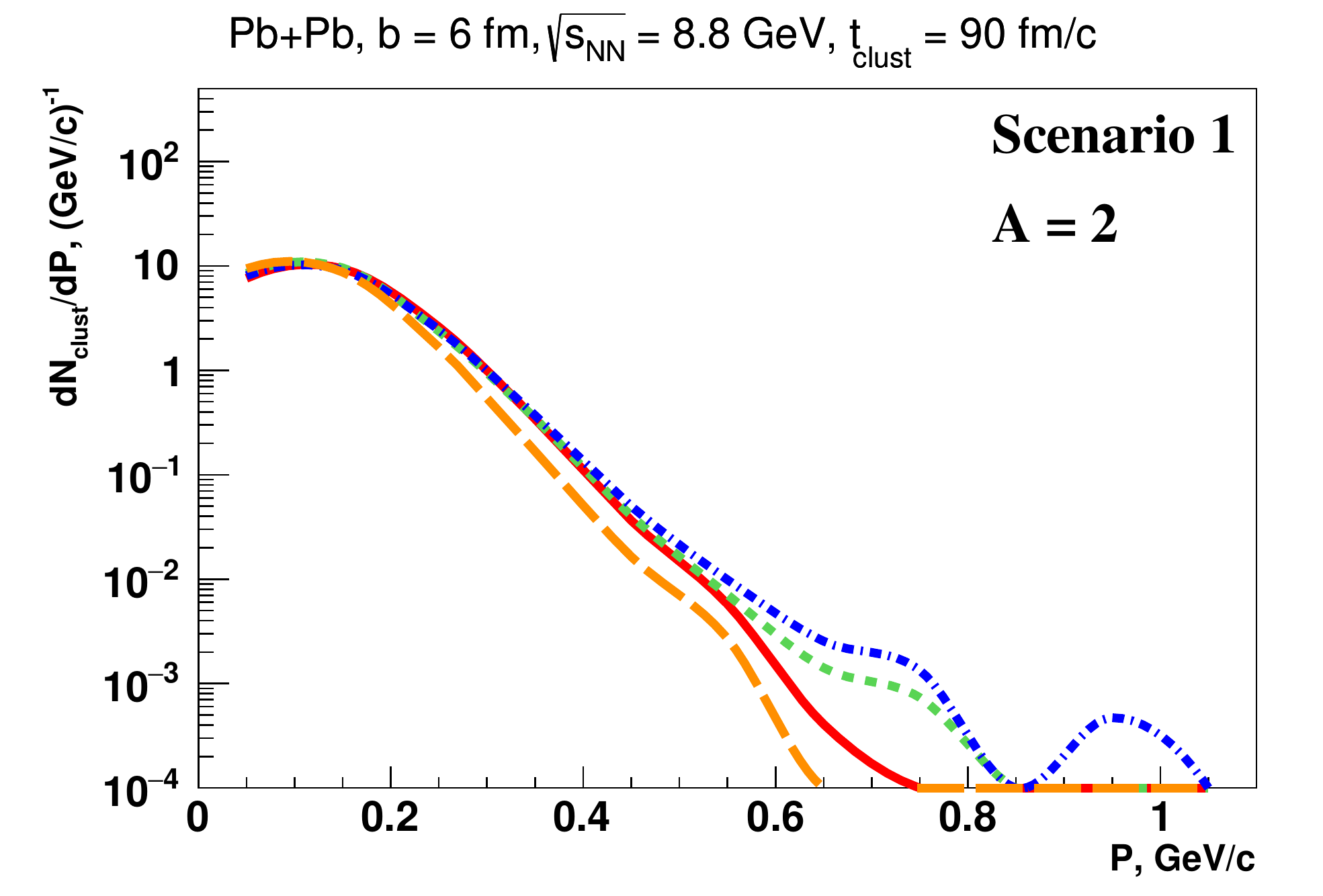} &
          \includegraphics{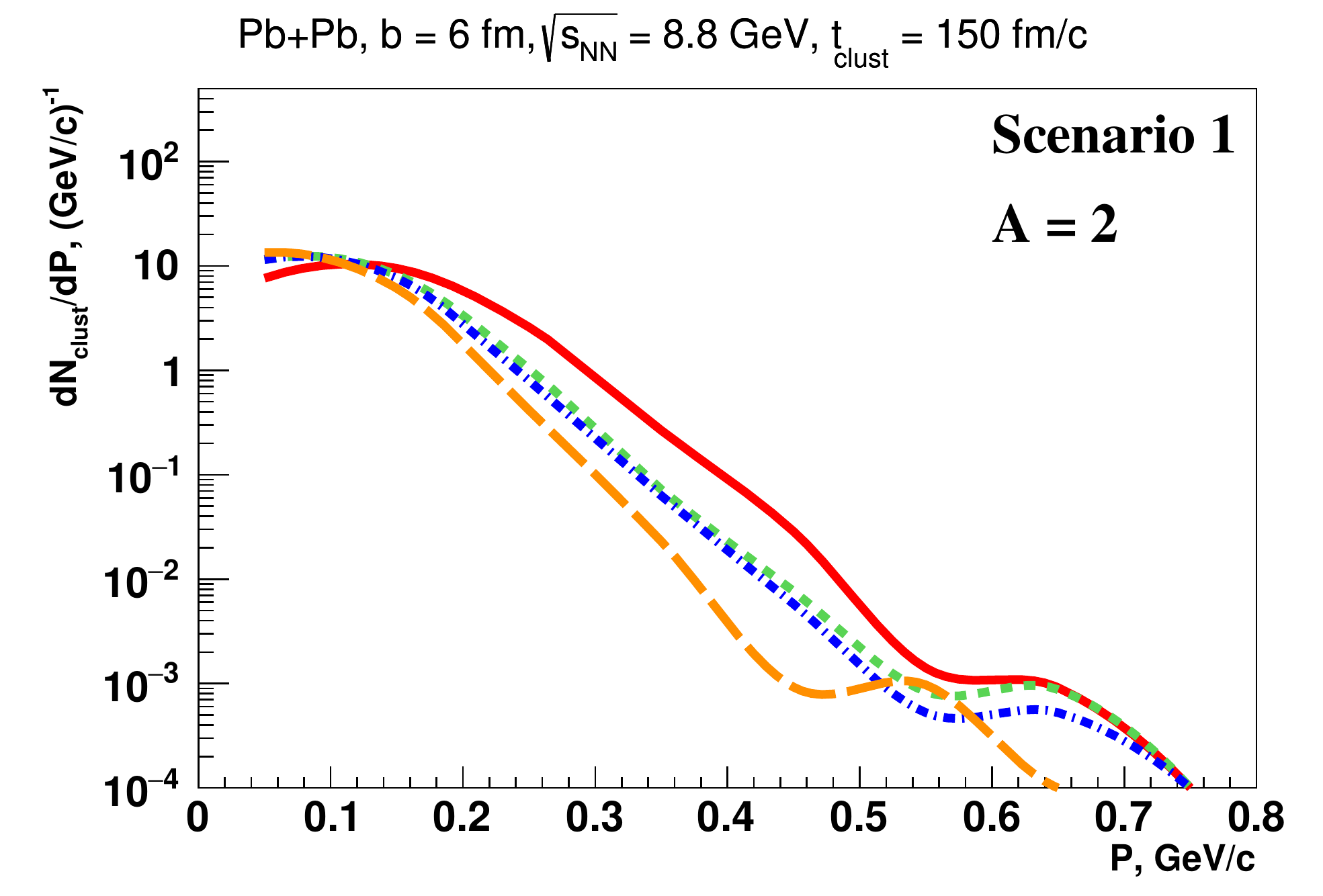} \\
        \end{tabular}
    }
\caption{\label{fig:8.8dndpA2} The momentum spectra of baryons ($p$, $n$, $\Lambda$ and $\Sigma^{0}$) in $A = 2$ clusters in semi-peripheral ($b=6$ fm) $Pb+Pb$ collisions at $\sqrt{s}=8.8$ GeV (integrated over all rapidity range). The momentum is calculated in the cluster center of mass frame. The left column: $t_{clust} = 40$ fm/c, the center column: $t_{clust} = 90$ fm/c, the right column: $t_{clust} = 150$ fm/c. The color coding is the same as in Fig.~\ref{fig:2.52dndyA2}.}
\end{figure*}

\begin{figure*}
    \resizebox{\textwidth}{!}{
        \includegraphics{plots/scenario1/header.pdf} 
    } \\
    \resizebox{\textwidth}{!}{
        \begin{tabular}{ccc}
          \includegraphics{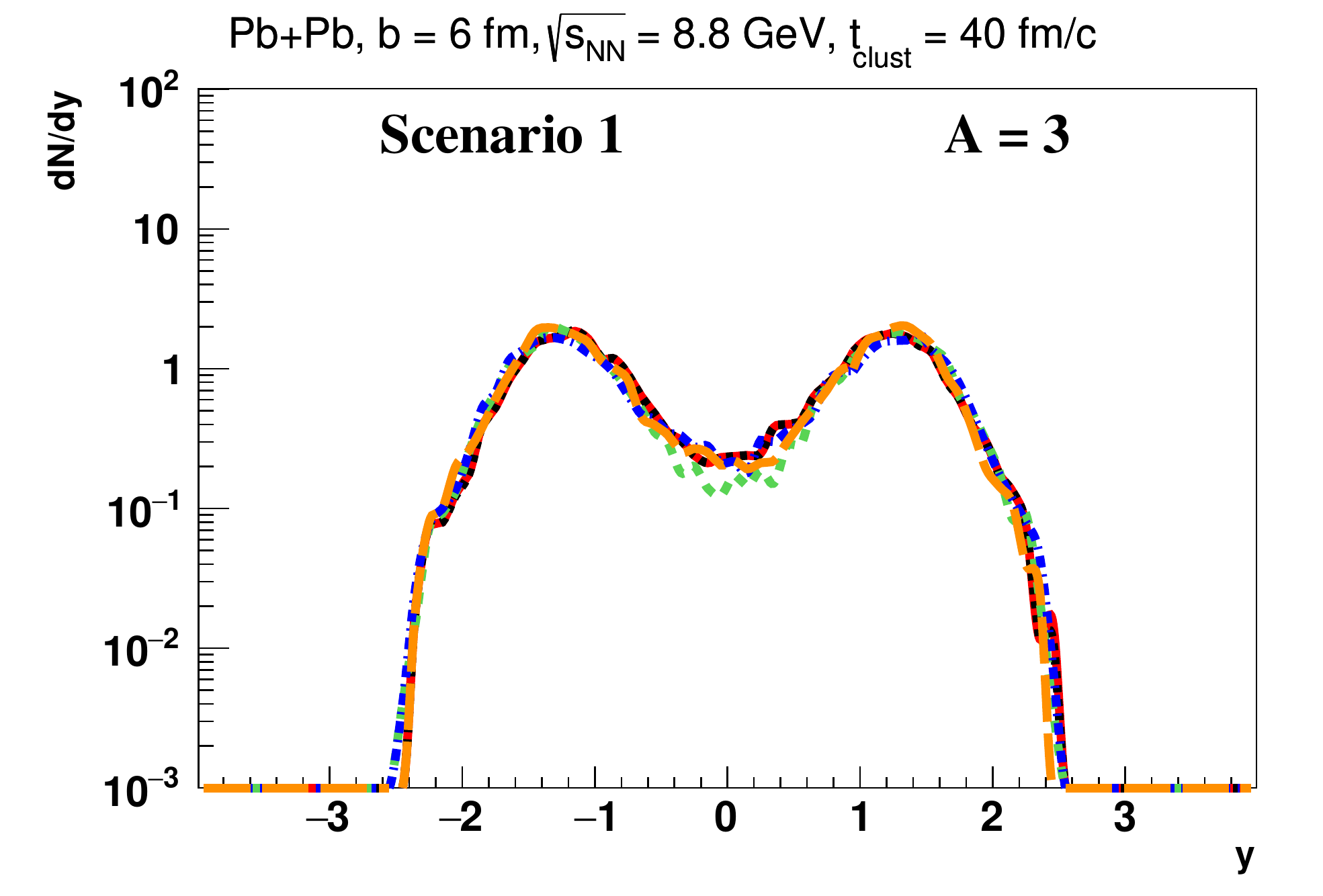} &
          \includegraphics{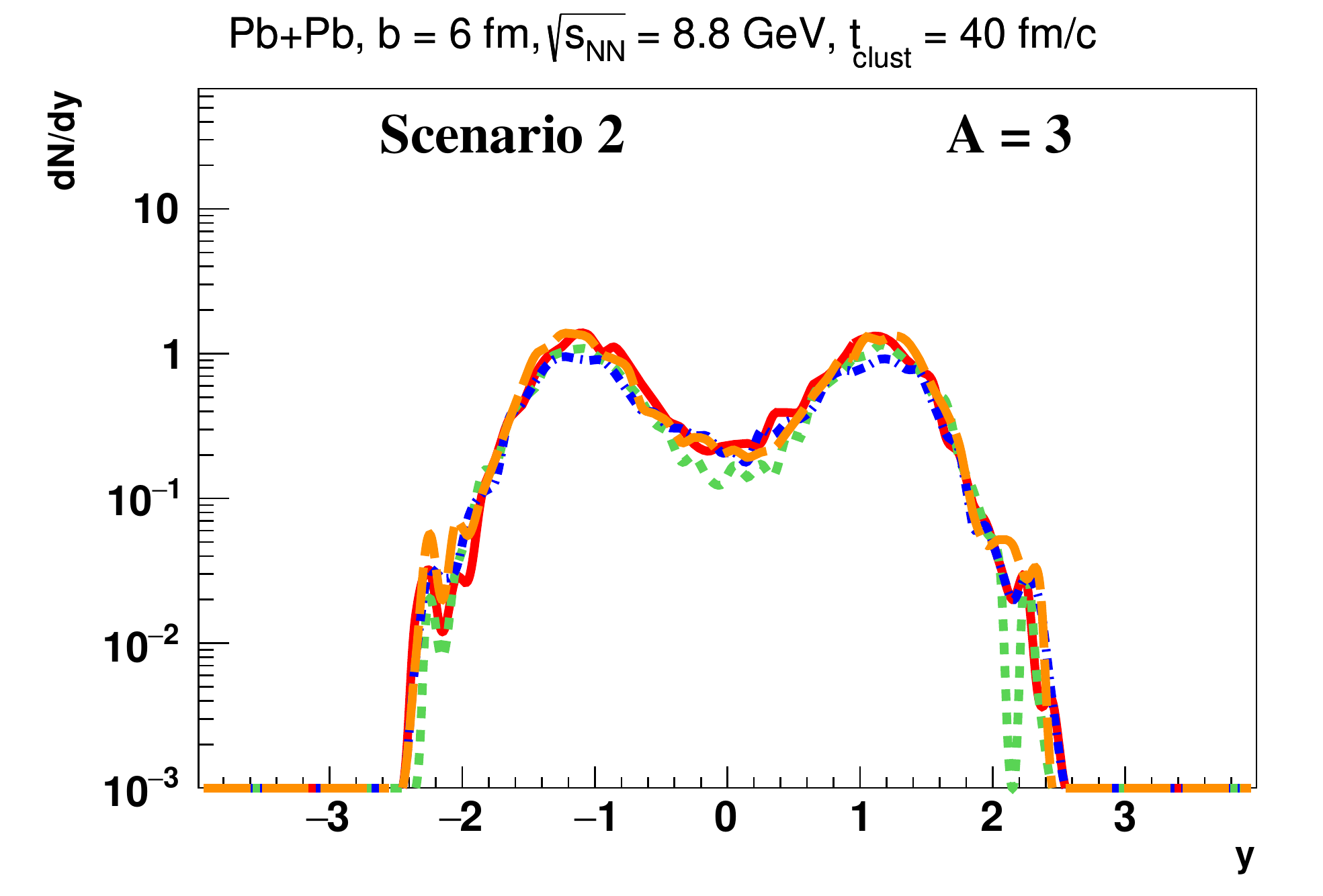} &
          \includegraphics{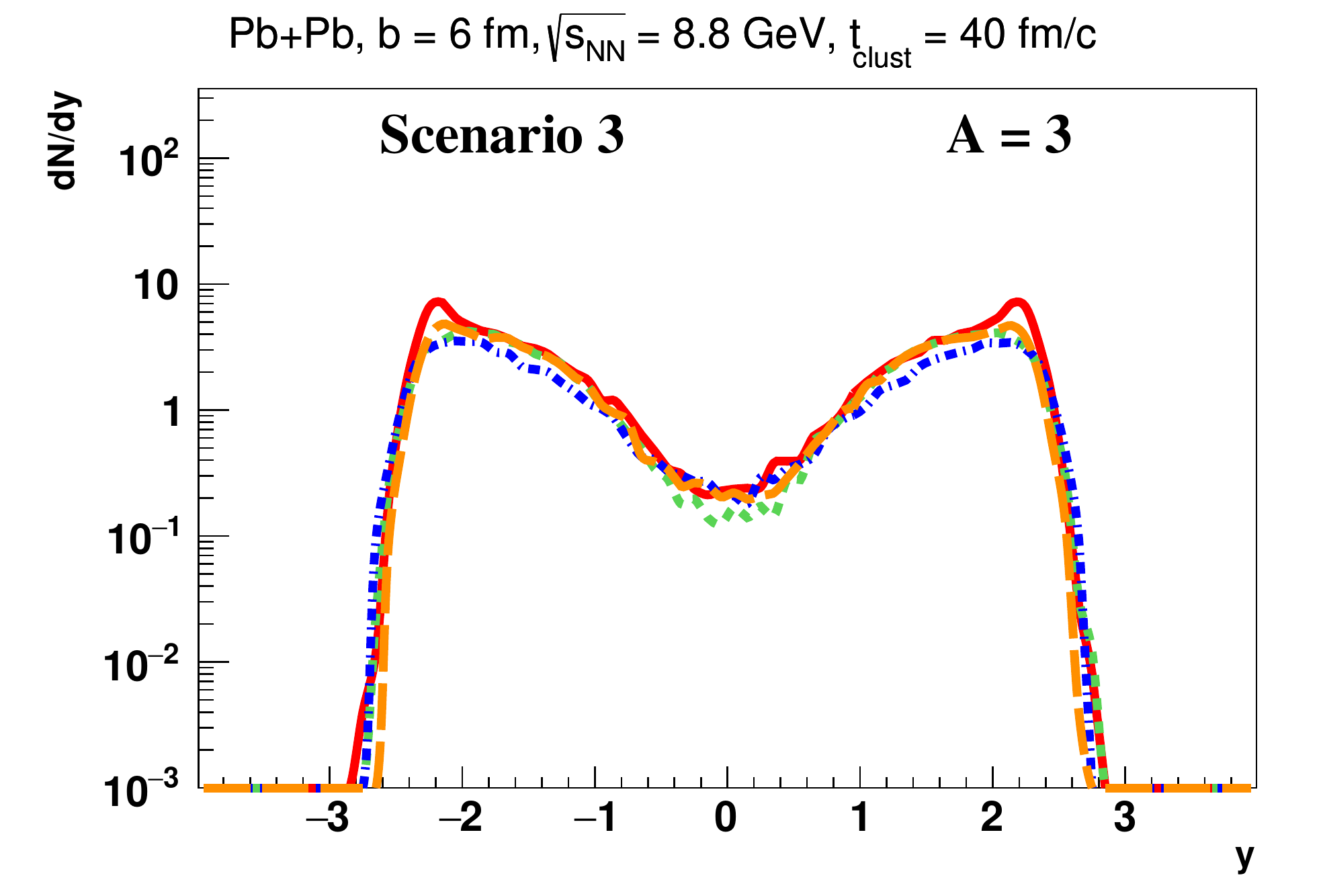} \\
          \includegraphics{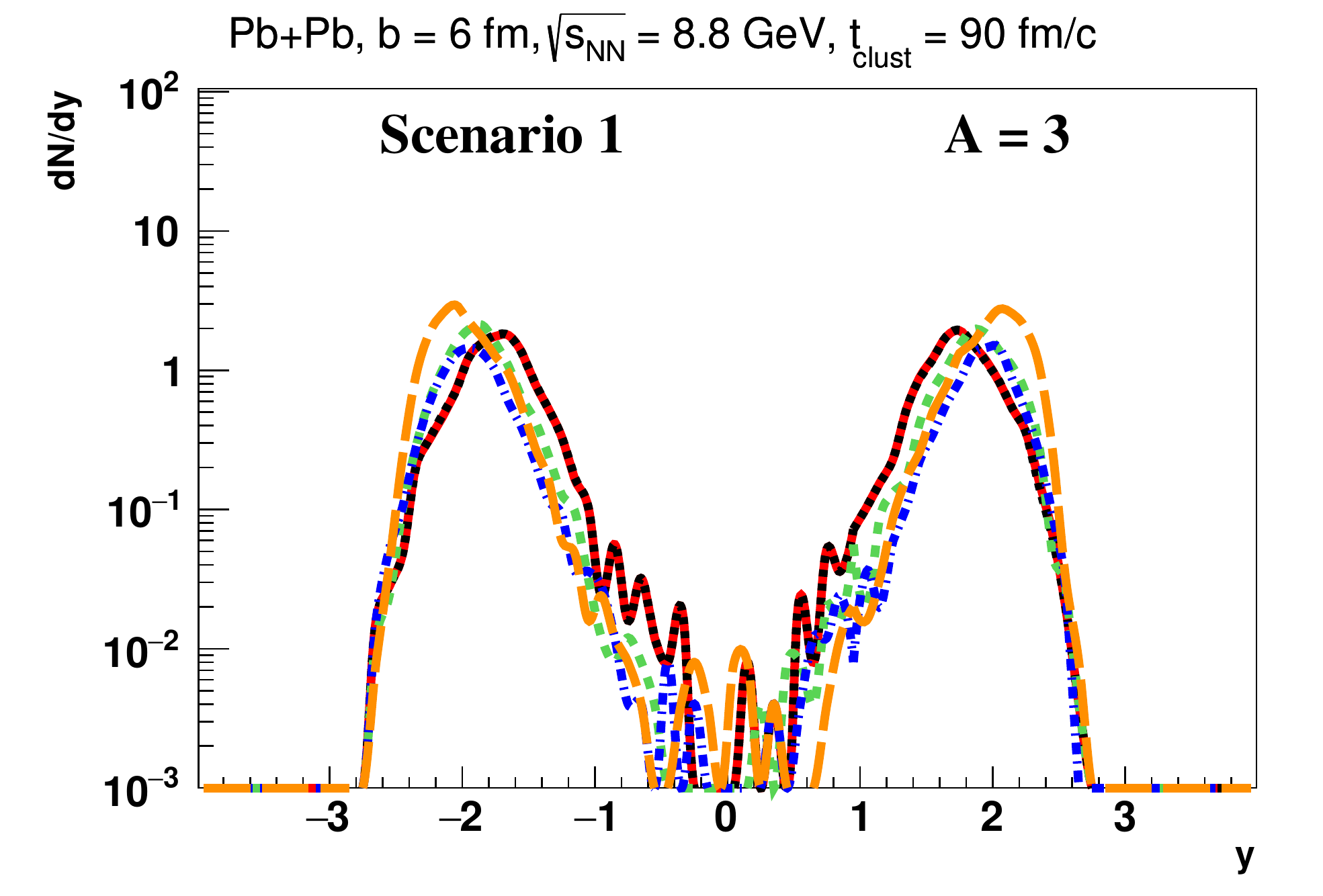} &
          \includegraphics{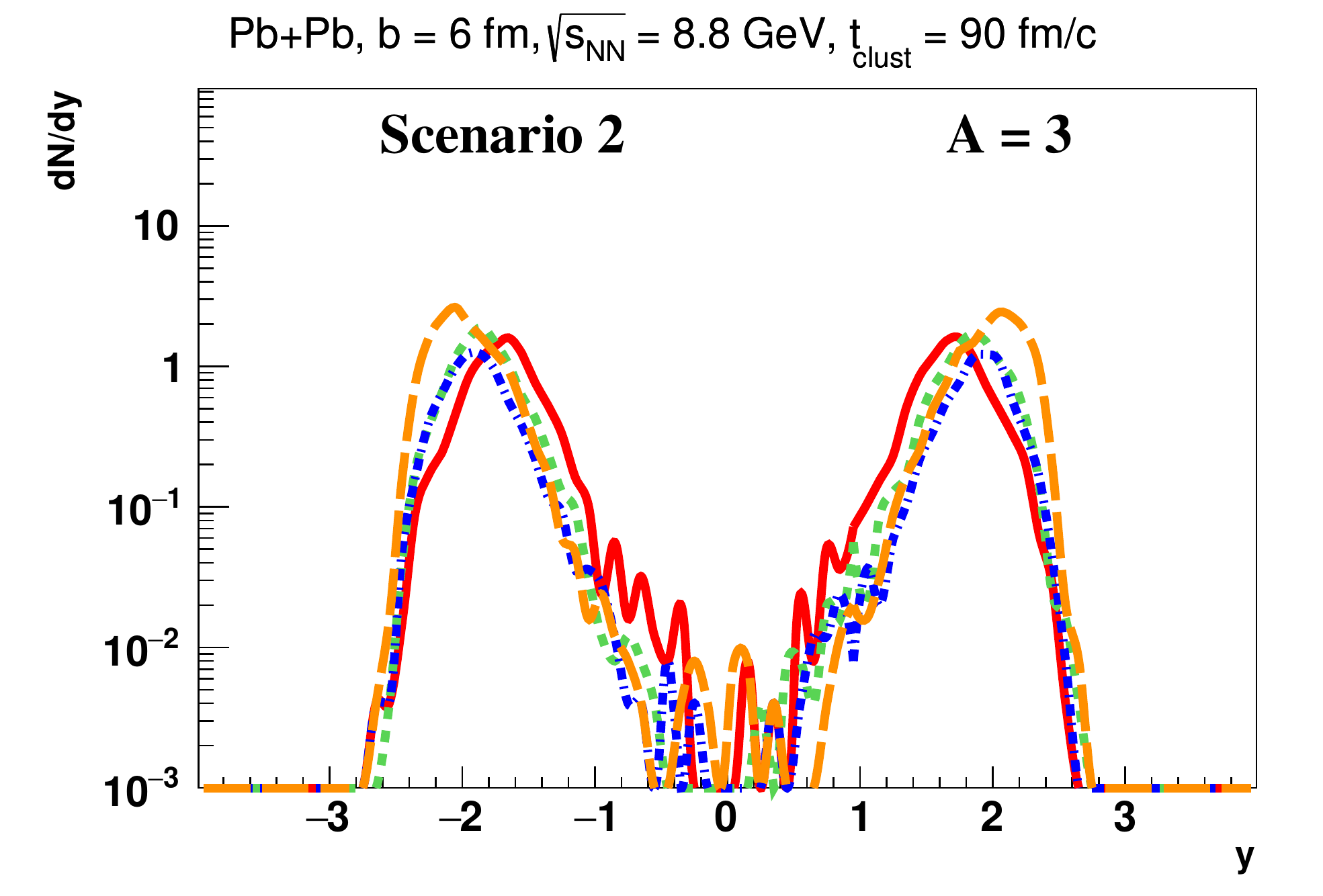} &
          \includegraphics{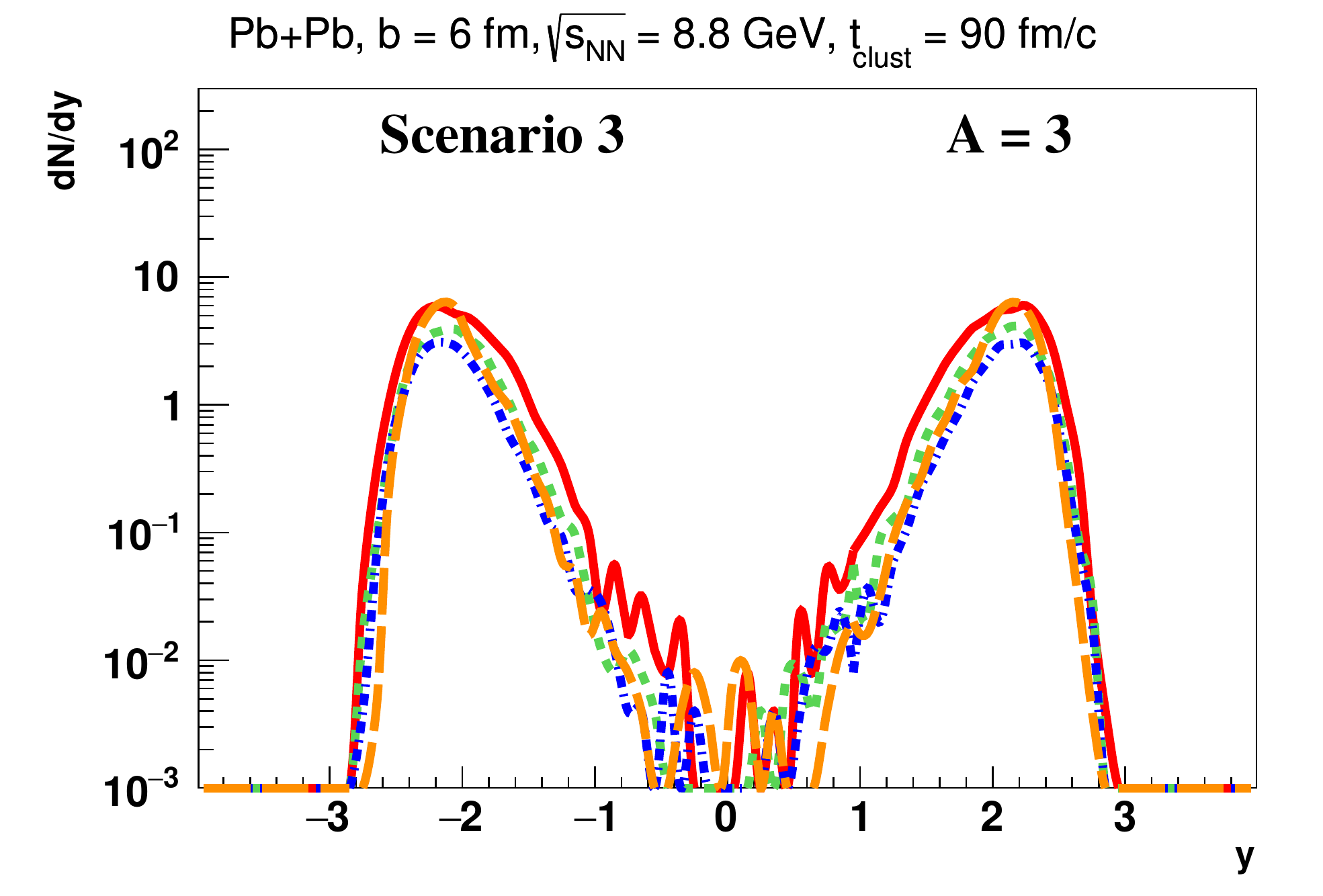} \\
          \includegraphics{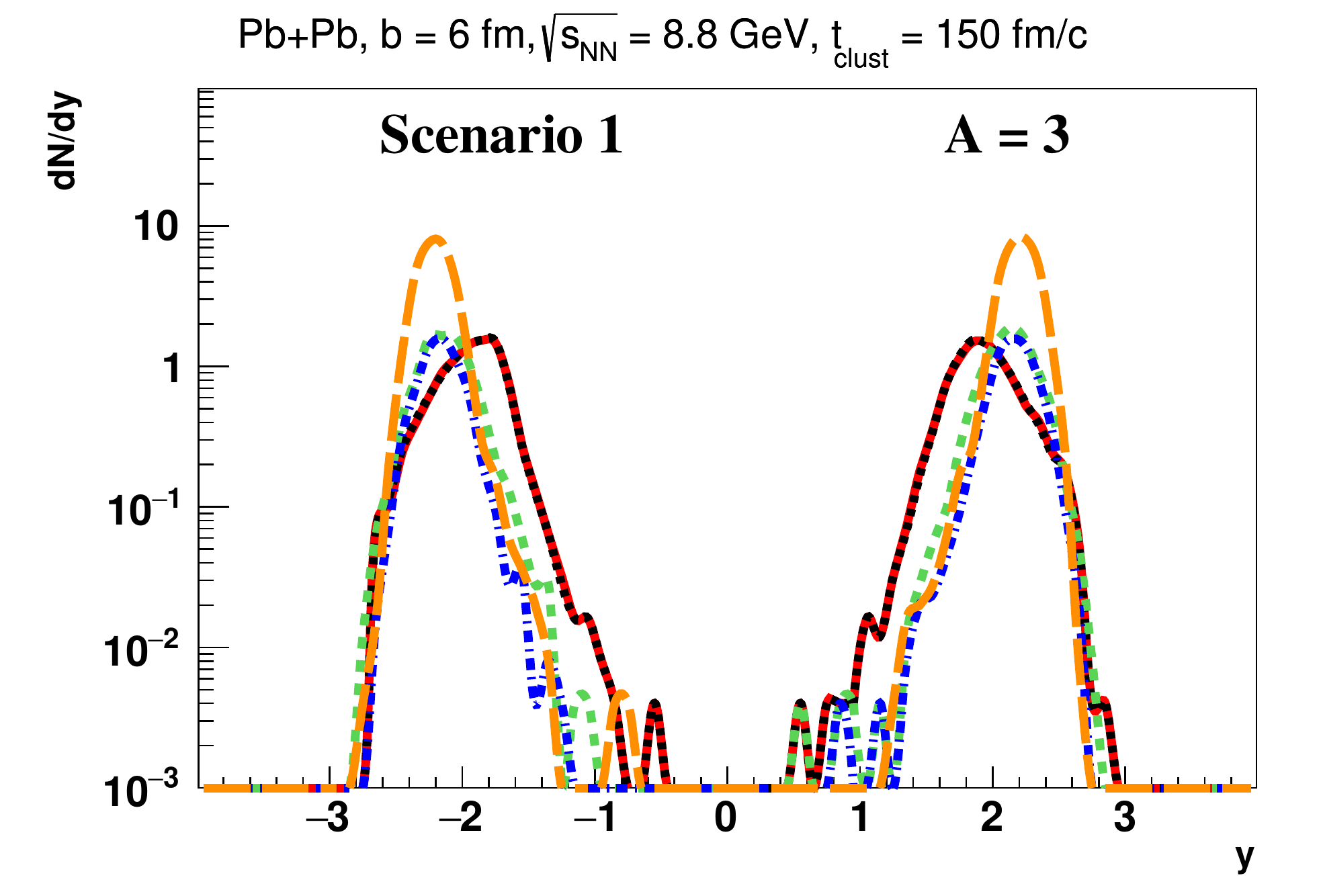} &
          \includegraphics{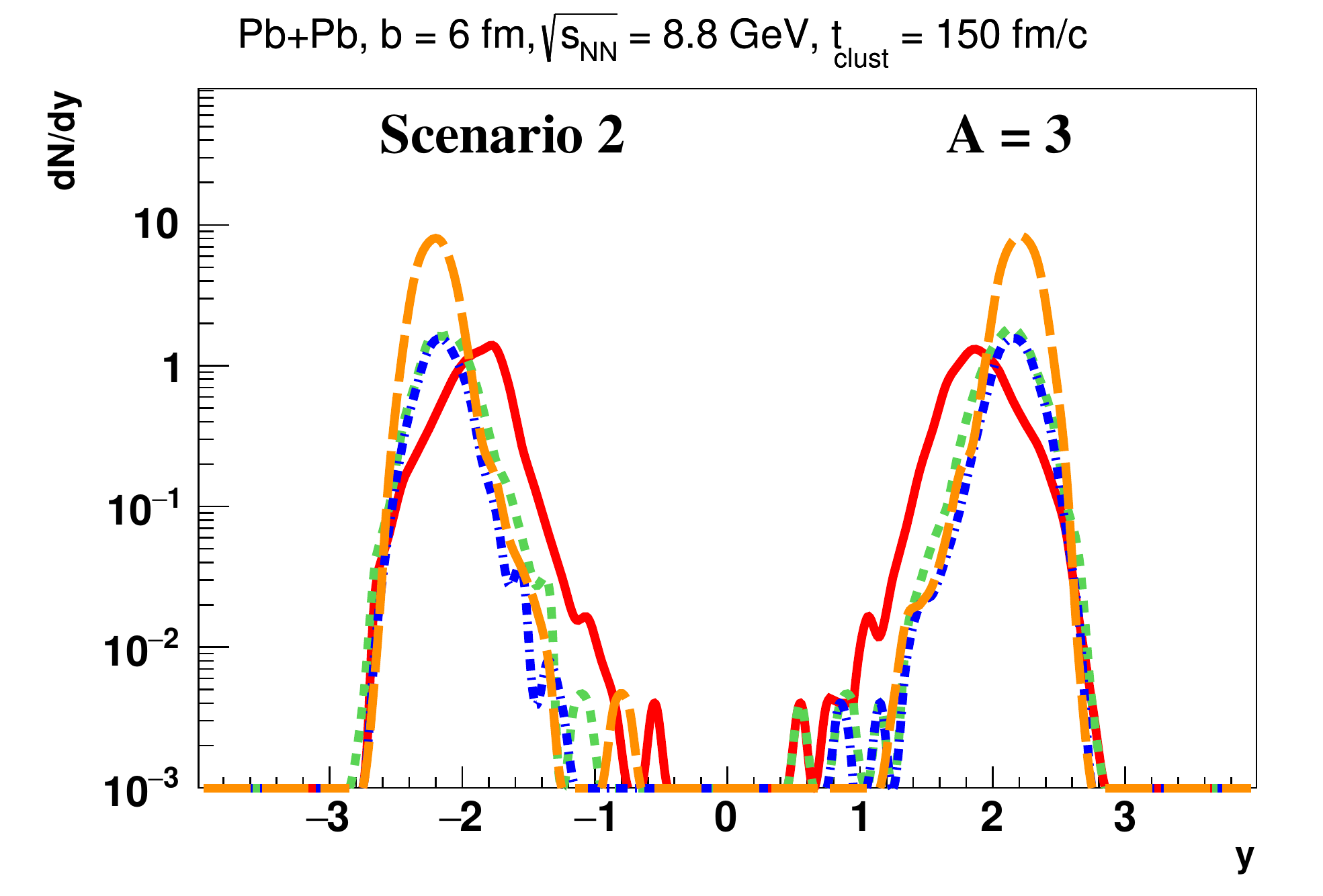} &
          \includegraphics{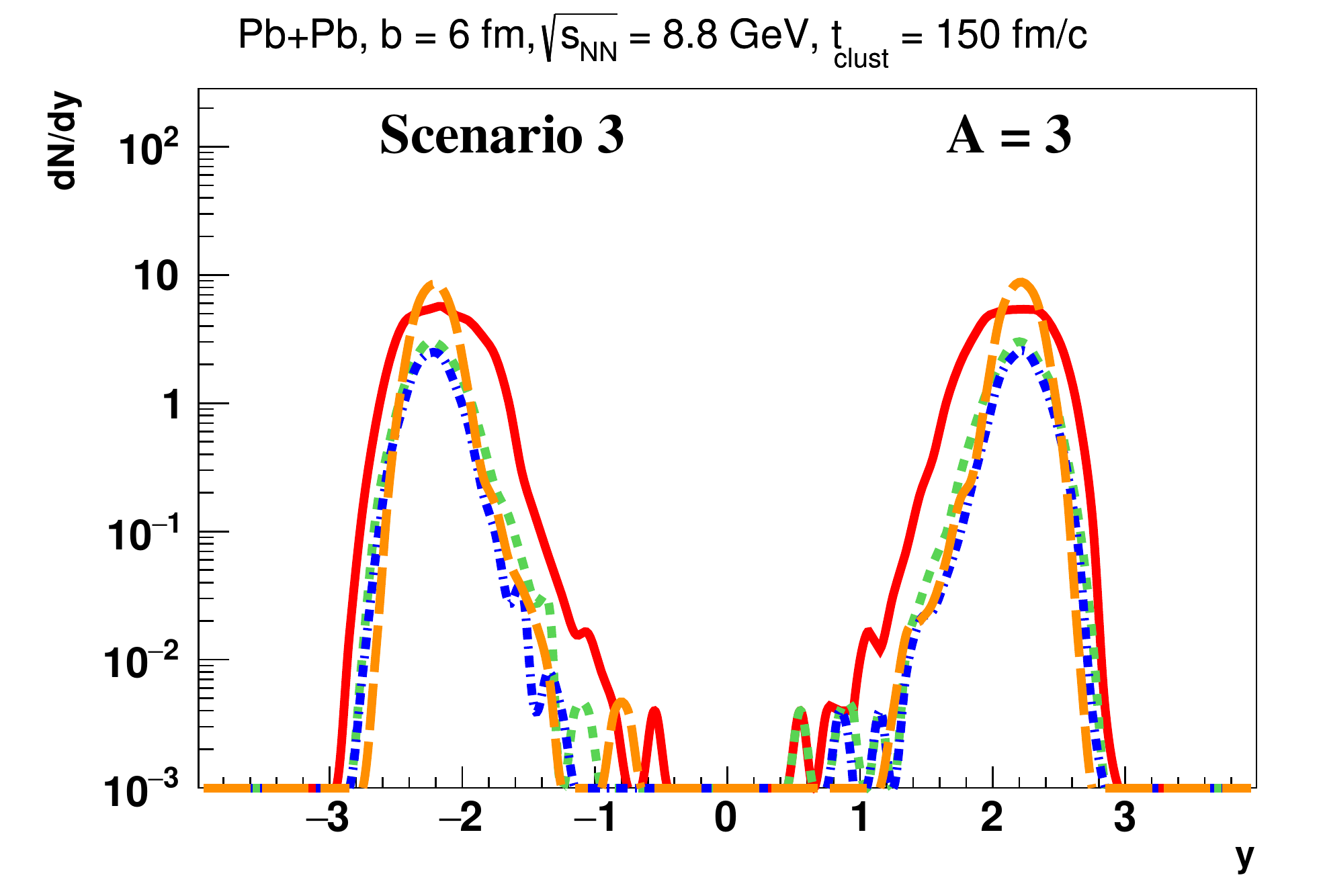} \\
        \end{tabular}
    }
\caption{\label{fig:8.8dndyA3} Rapidity distributions of clusters with the mass number $A = 3$ at $t_{clust} = 40, 90, 150$ fm/c in semi-peripheral ($b=6$ fm) $Pb+Pb$ collisions at $\sqrt{s}=8.8$ GeV. The left column: "Scenario 1", the center column: "Scenario 2", the right column: "Scenario 3". The color coding is the same as in Fig.~\ref{fig:2.52dndyA2}.}
\end{figure*}

\begin{figure*}
    \resizebox{\textwidth}{!}{
        \begin{tabular}{ccc}
          \includegraphics{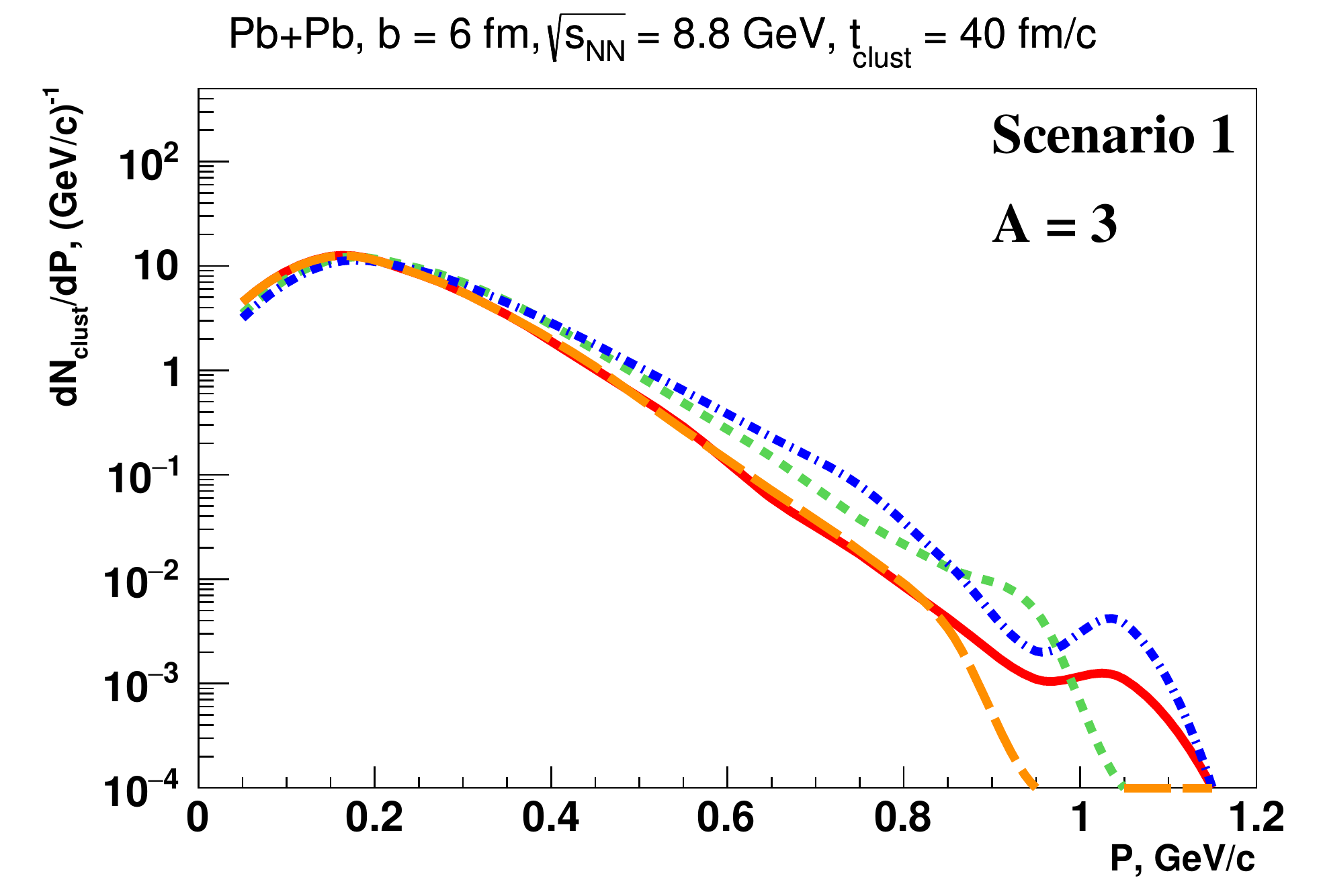} &
          \includegraphics{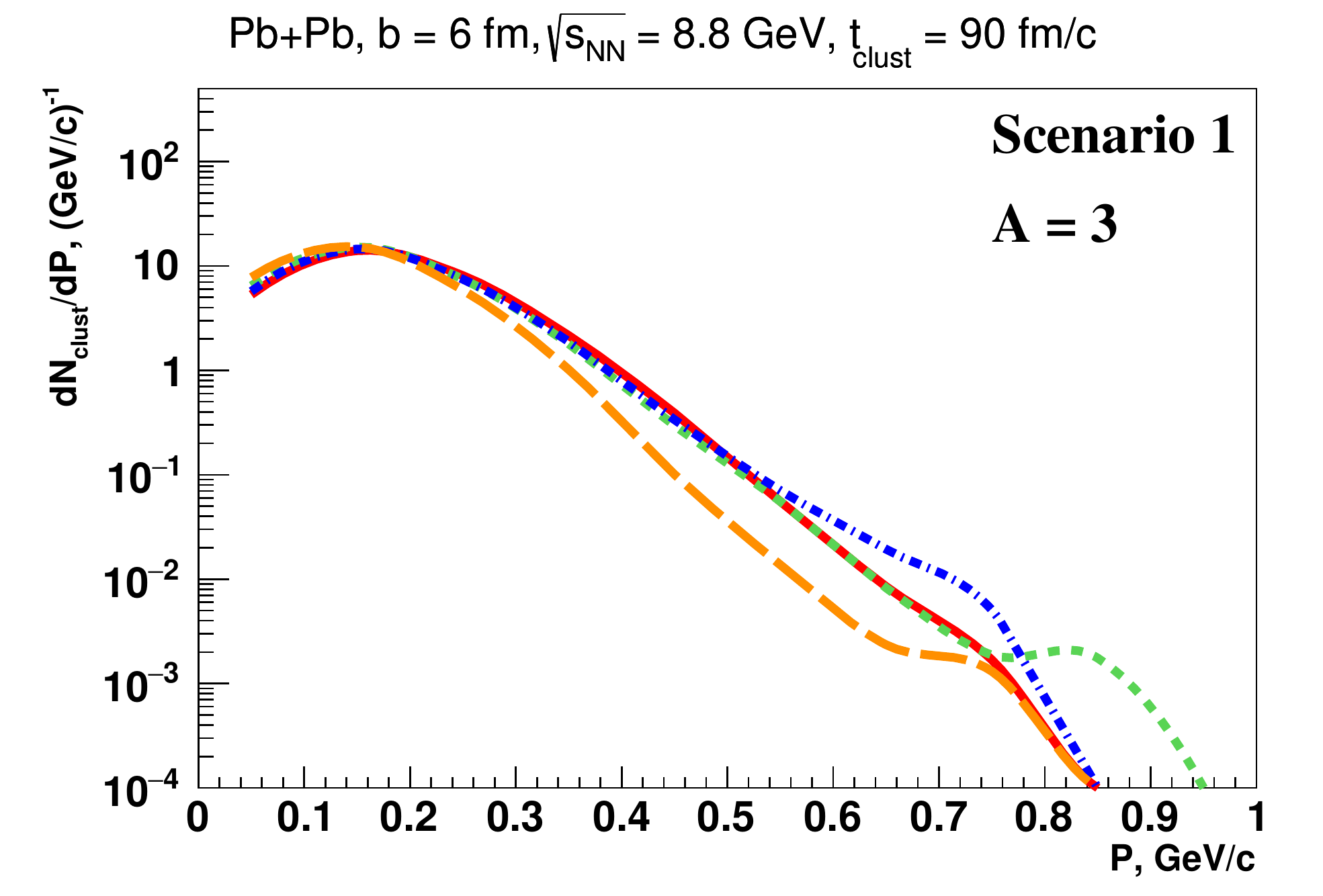} &
          \includegraphics{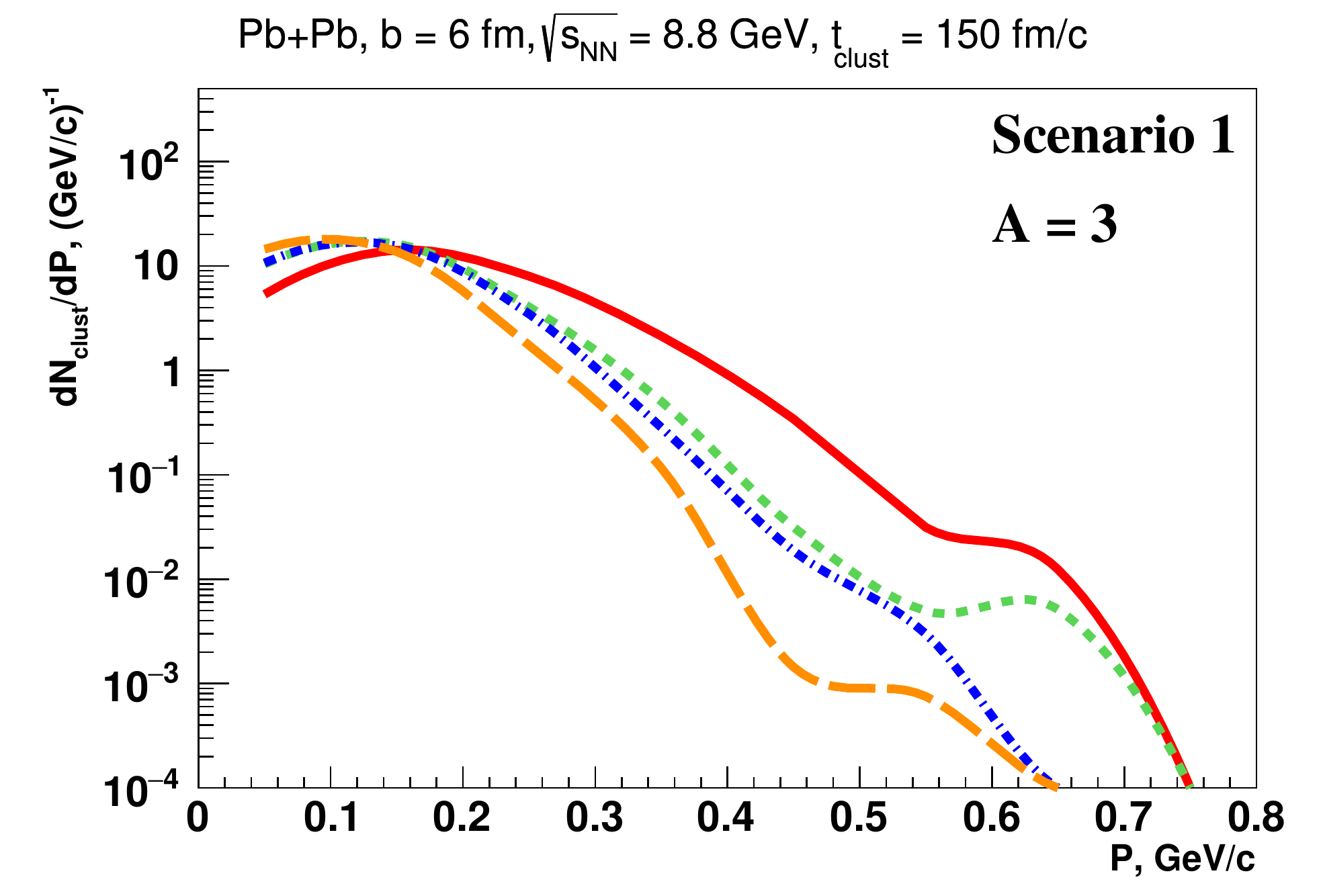} \\
        \end{tabular}
    }
\caption{\label{fig:8.8dndpA3} Momentum spectra of baryons ($p$, $n$, $\Lambda$ and $\Sigma^{0}$) in $A = 3$ clusters in semi-peripheral ($b=6$ fm) $Pb+Pb$ collisions at $\sqrt{s}=8.8$ GeV (integrated over all rapidity range). Momentum is calculated in the cluster center of mass frame. The left column: $t_{clust} = 40$ fm/c, the center column: $t_{clust} = 90$ fm/c, the right column: $t_{clust} = 150$ fm/c. The color coding is the same as in Fig.~\ref{fig:2.52dndyA2}.}
\end{figure*}

\begin{figure*}
    \resizebox{\textwidth}{!}{
        \includegraphics{plots/scenario1/header.pdf} 
    } \\
    \resizebox{\textwidth}{!}{
        \begin{tabular}{ccc}
          \includegraphics{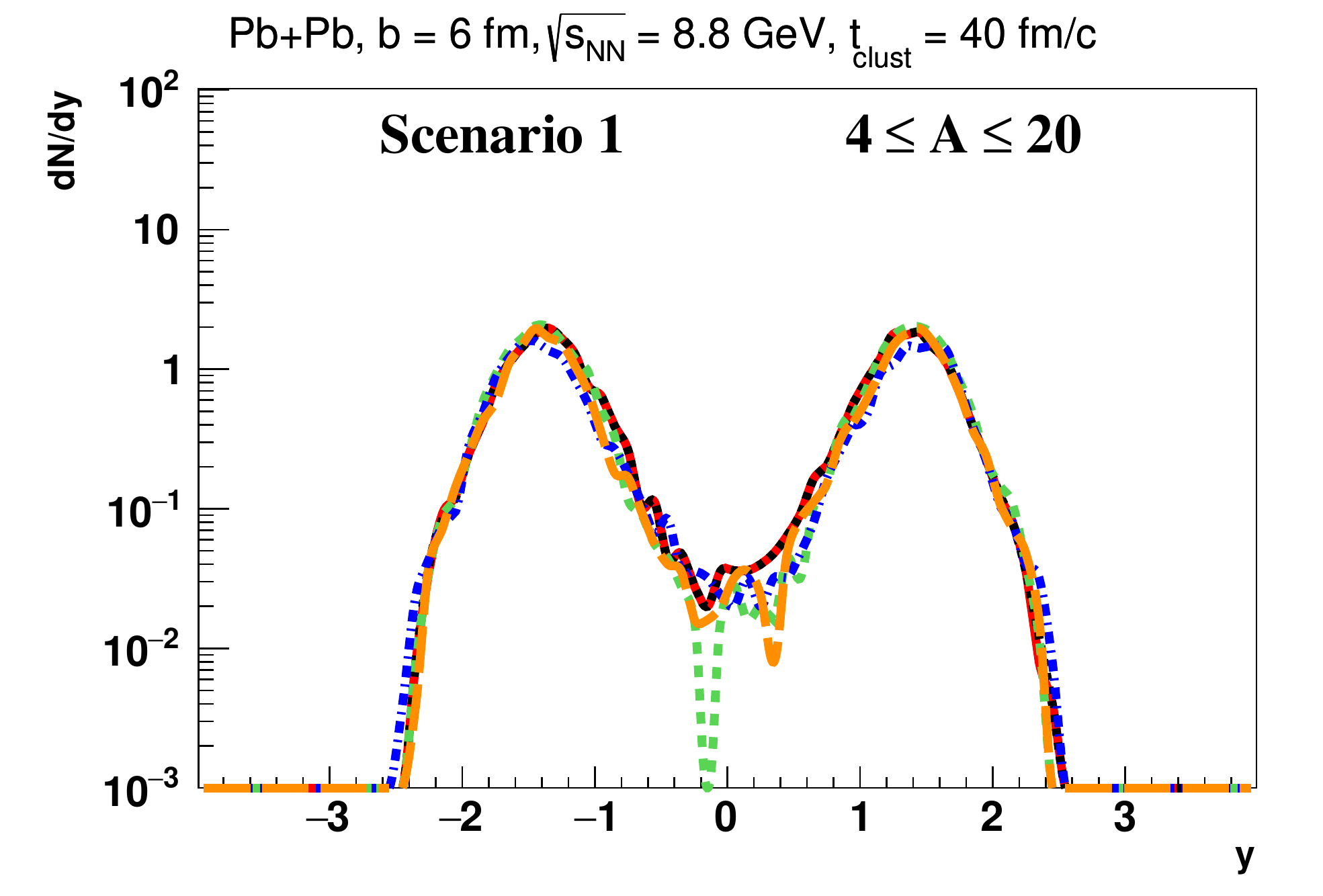} &
          \includegraphics{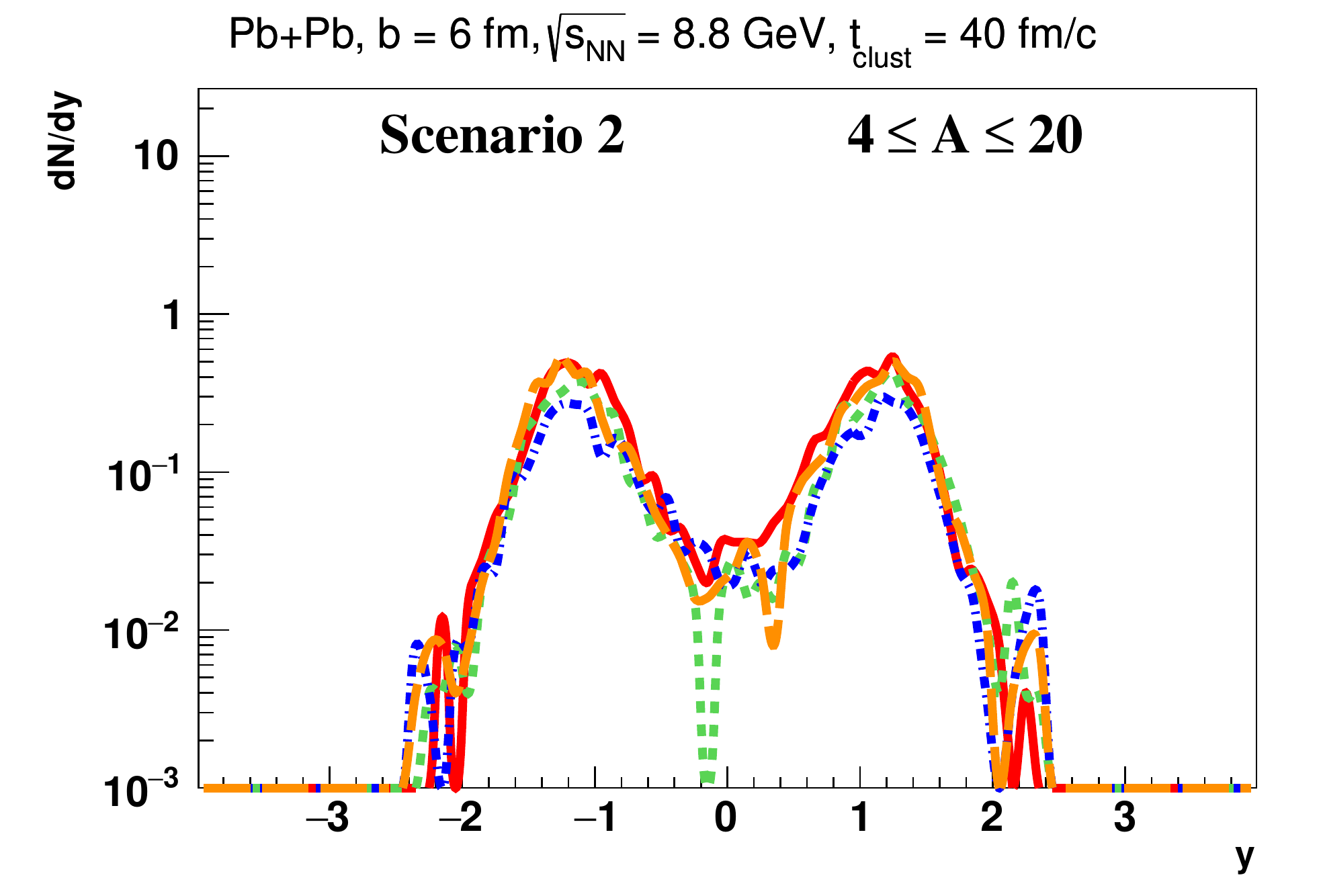} &
          \includegraphics{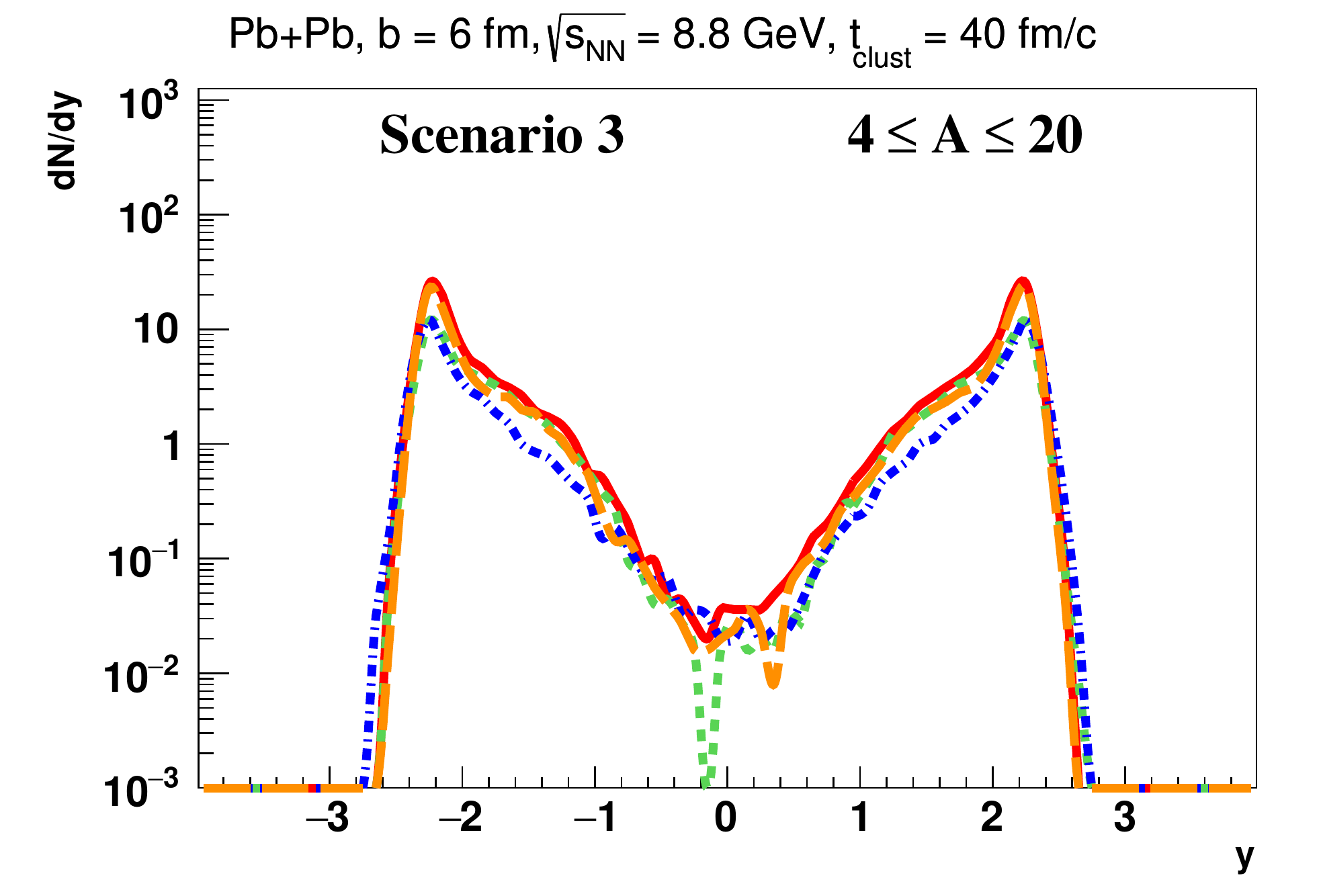} \\
          \includegraphics{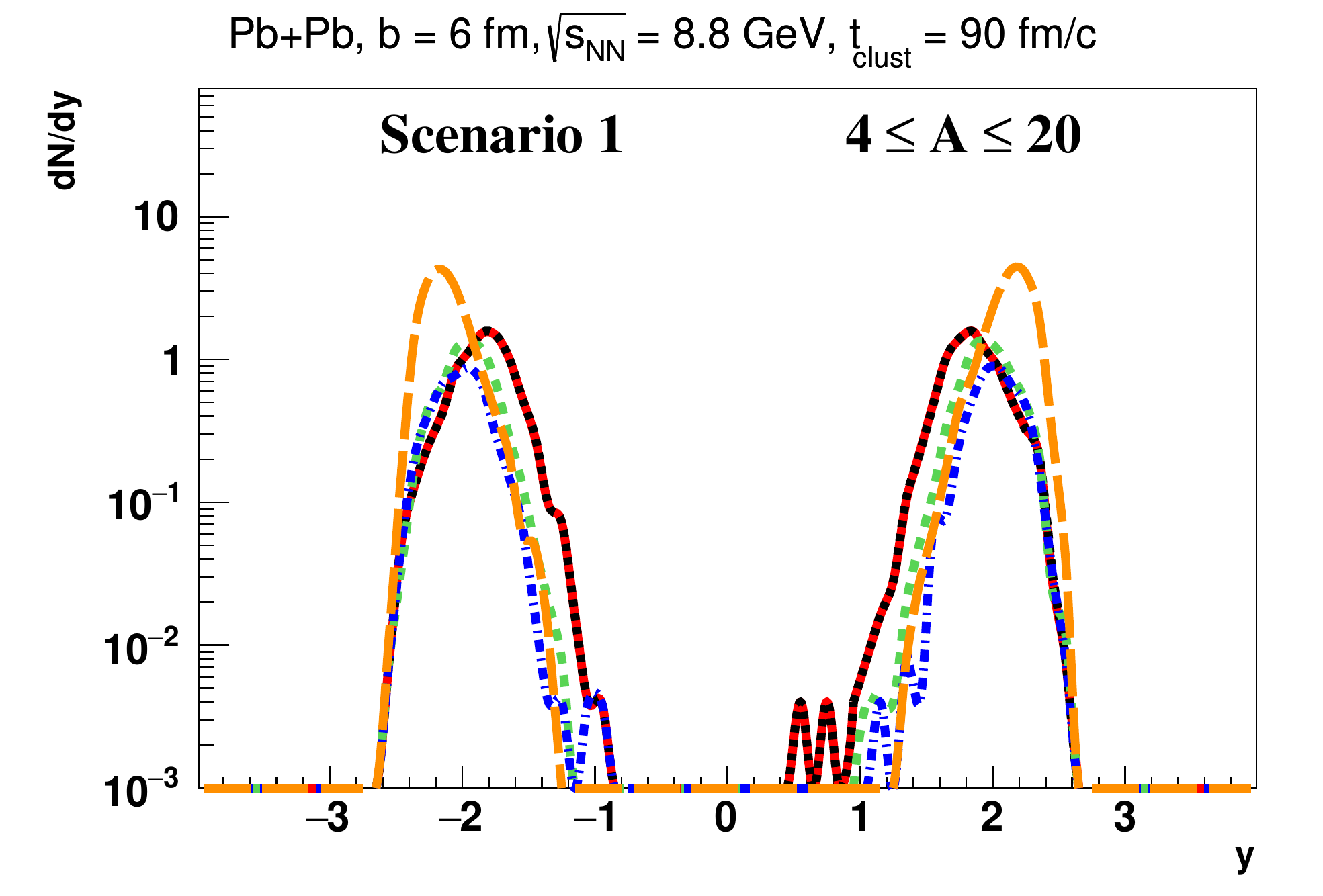} &
          \includegraphics{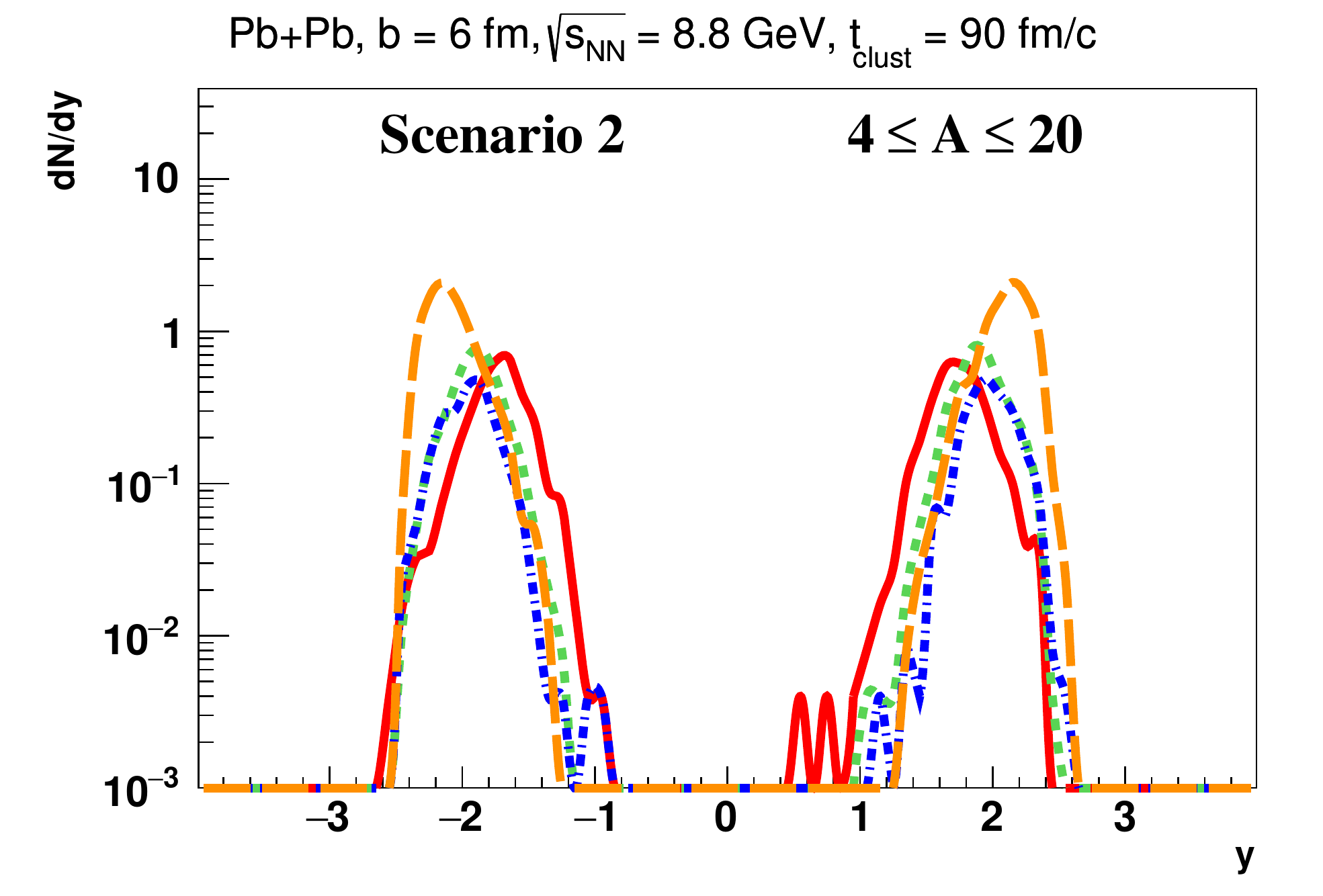} &
          \includegraphics{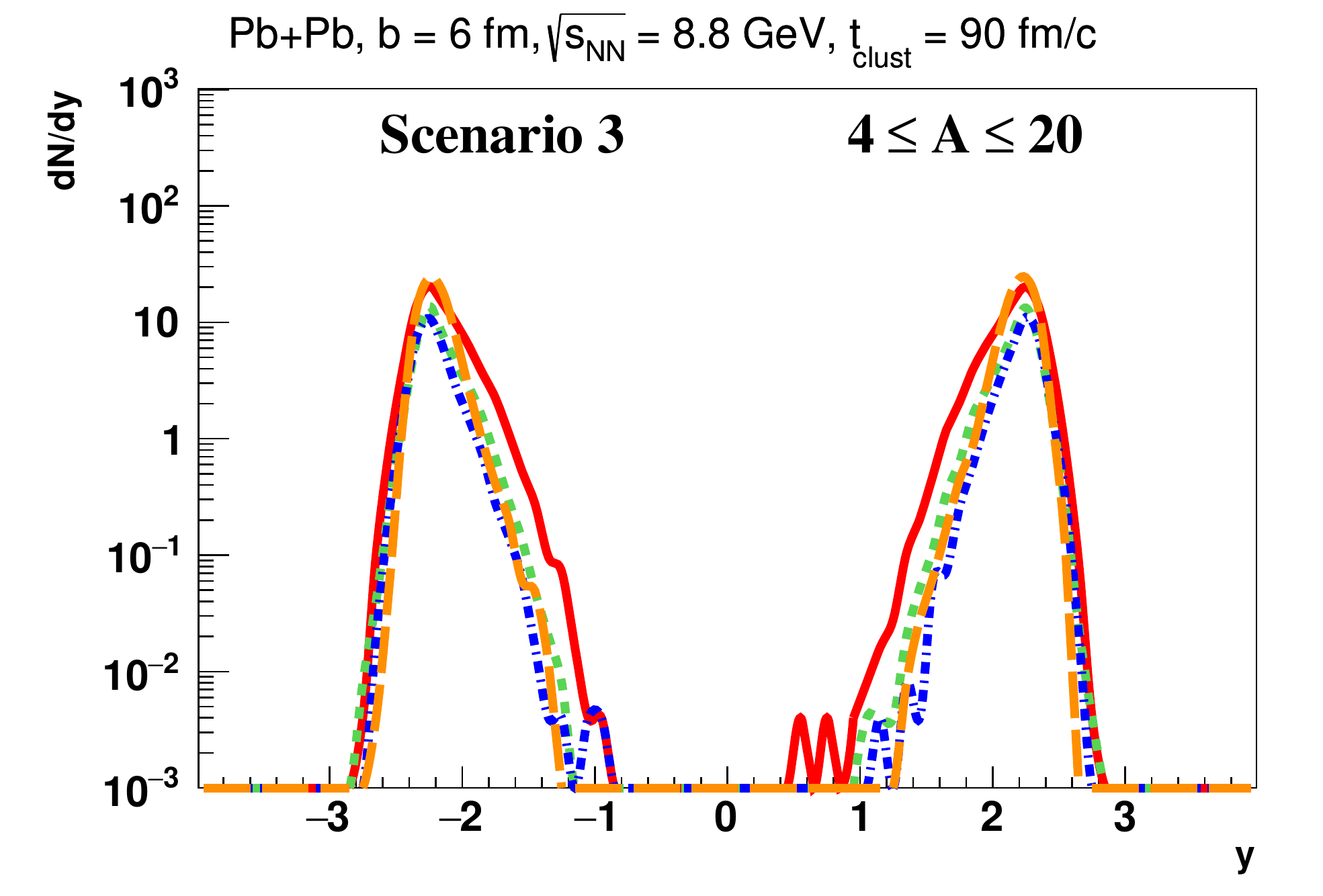} \\
          \includegraphics{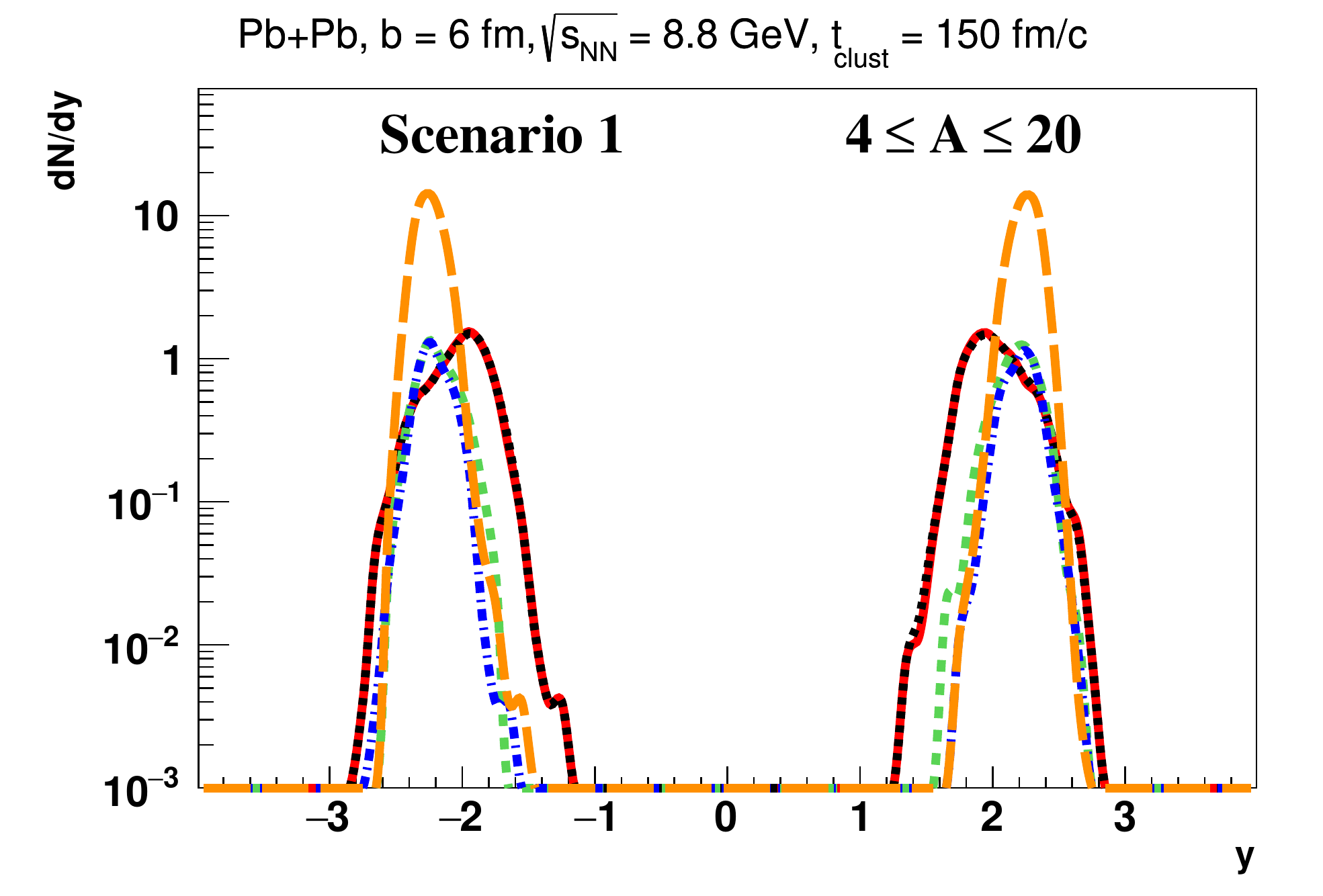} &
          \includegraphics{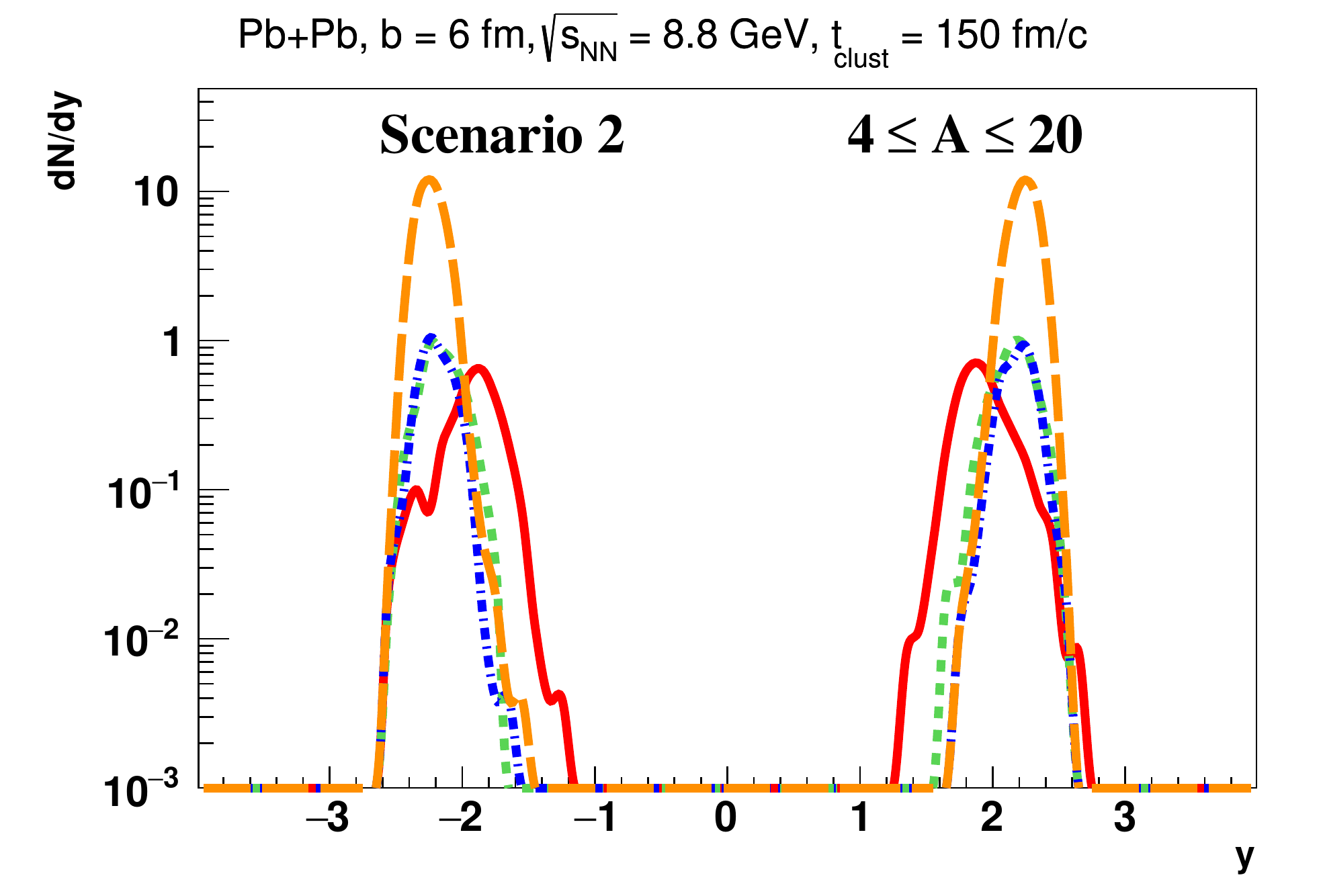} &
          \includegraphics{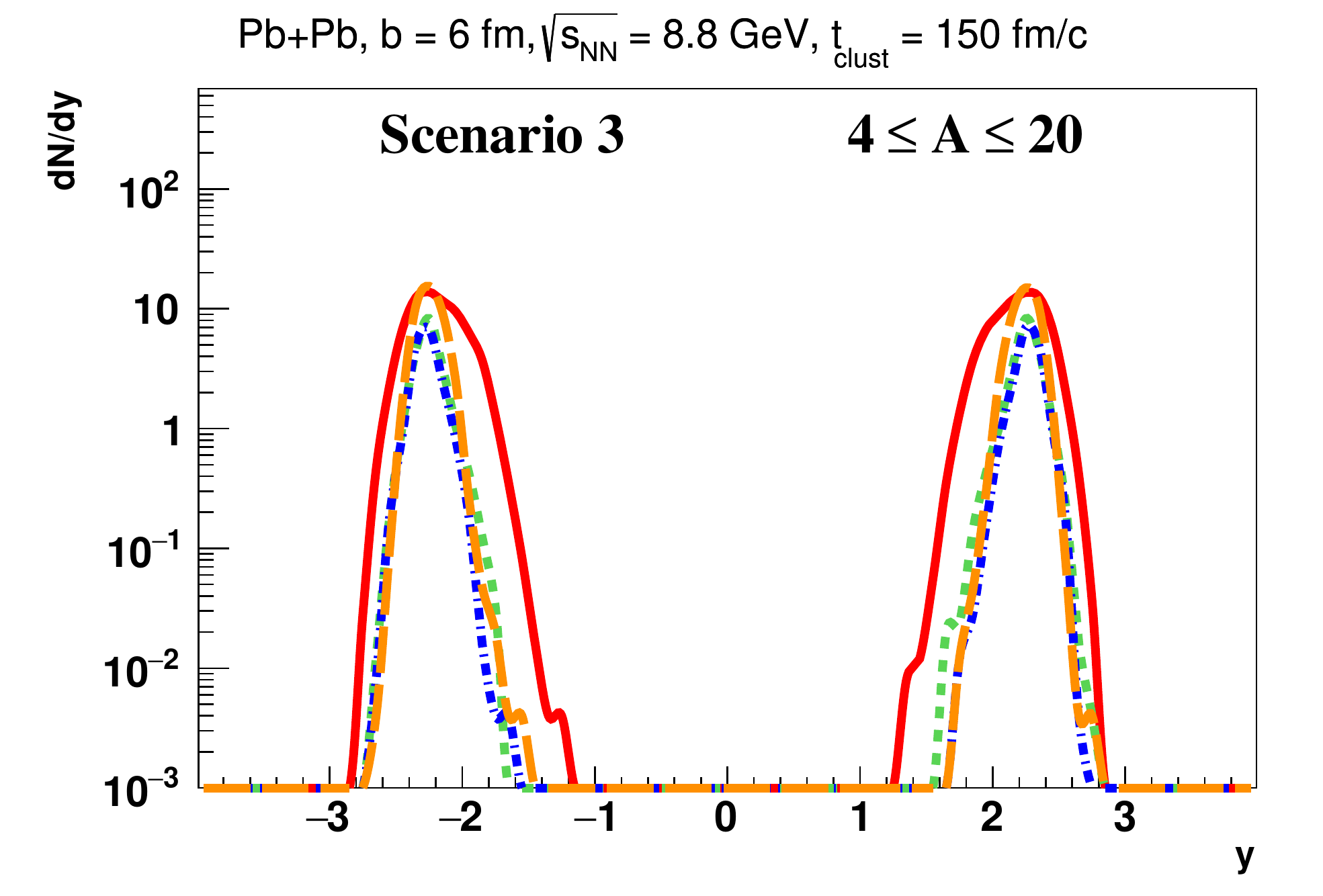} \\
        \end{tabular}
    }
\caption{\label{fig:8.8dndyA4_20} The rapidity distributions of clusters with the mass number $4 \leq A \leq 20$ at $t_{clust} = 40, 90, 150$ fm/c in semi-peripheral ($b=6$ fm) $Pb+Pb$ collisions at $\sqrt{s}=8.8$ GeV. The left column: "Scenario 1", the center column: "Scenario 2", the right column: "Scenario 3". The color coding is the same as in Fig.~\ref{fig:2.52dndyA2}.}
\end{figure*}

\begin{figure*}
    \resizebox{\textwidth}{!}{
        \begin{tabular}{ccc}
          \includegraphics{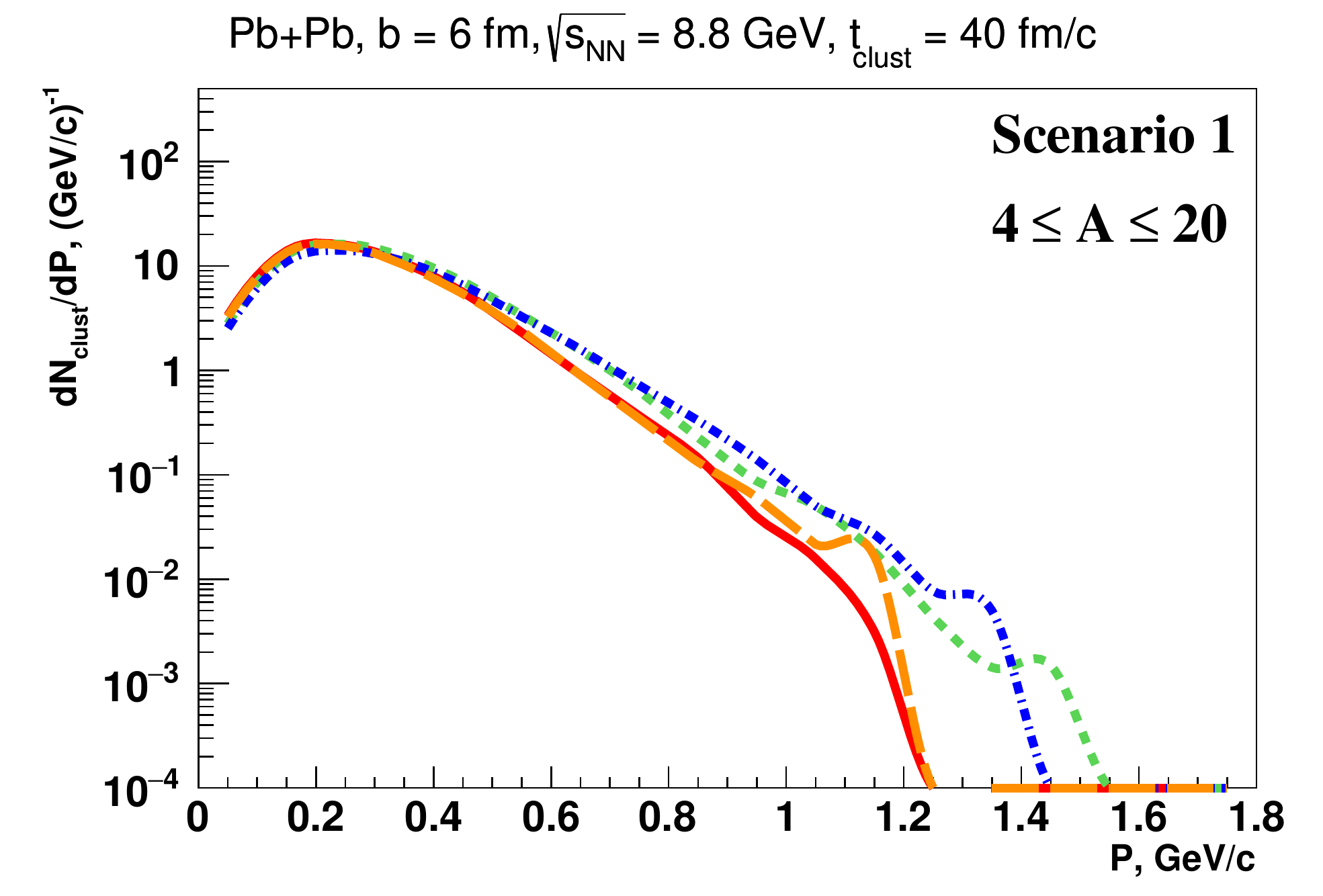} &
          \includegraphics{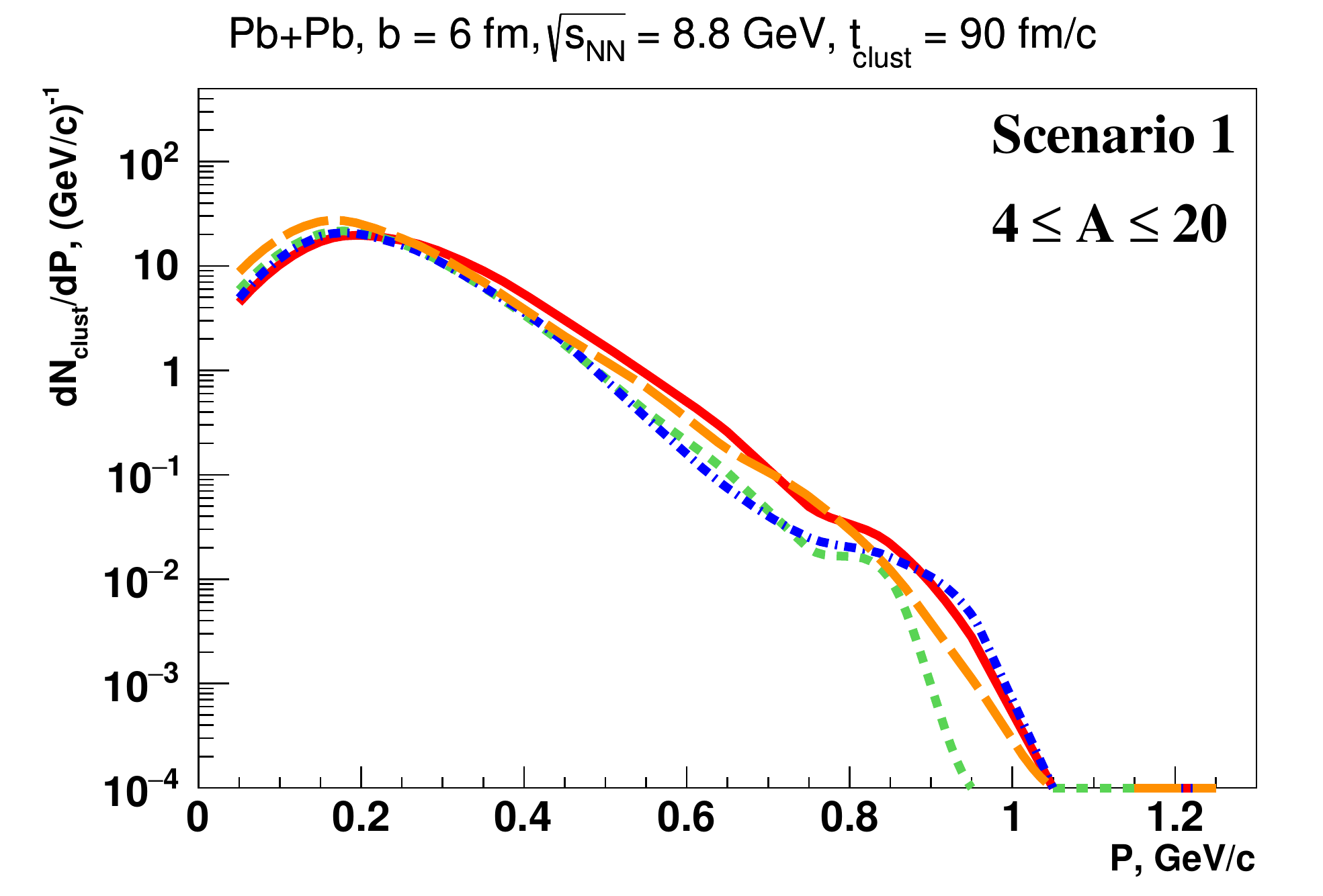} &
          \includegraphics{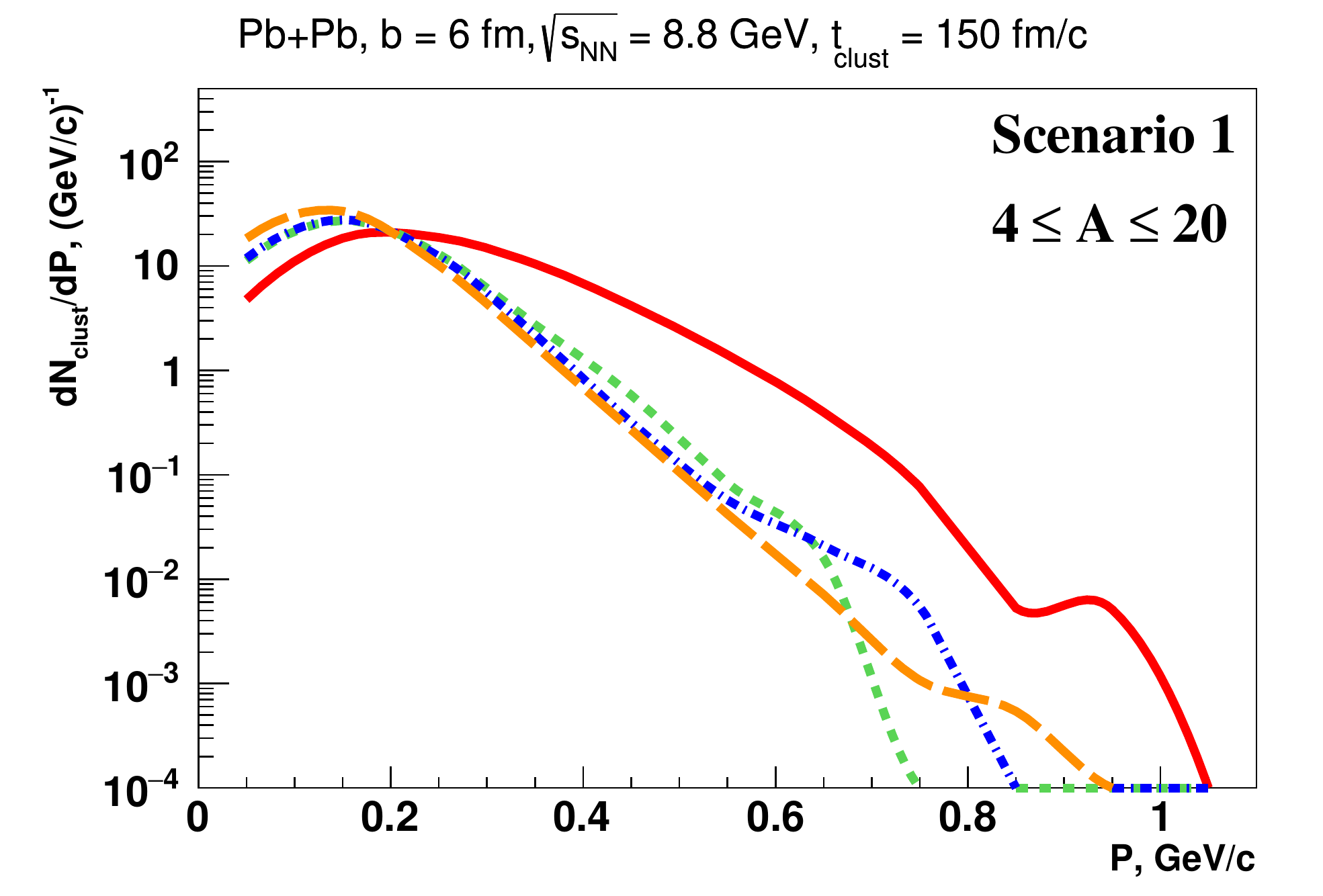} \\
        \end{tabular}
    }
\caption{\label{fig:8.8dndpA4_20} Momentum spectra of baryons ($p$, $n$, $\Lambda$ and $\Sigma^{0}$) in  $4 \leq A \leq 20$ clusters in semi-peripheral ($b=6$ fm) $Pb+Pb$ collisions at $\sqrt{s}=8.8$ GeV (integrated over all rapidity range). Momentum is calculated in the cluster center of mass frame. The left column: $t_{clust} = 40$ fm/c, the center column: $t_{clust} = 90$ fm/c, the right column: $t_{clust} = 150$ fm/c. The color coding is the same as in Fig.~\ref{fig:2.52dndyA2}.}
\end{figure*}

The picture becomes different at the higher energy: there PHQMD, SMASH and UrQMD show qualitatively similar results 
for the rapidity and momentum distributions in semi-peripheral ($b = 6$ fm) Pb+Pb collisions at $\sqrt{s_{NN}}=8.8$ GeV which are presented on Figures \ref{fig:8.8dndyA2} - ~\ref{fig:8.8dndpA4_20} for three scenarios. 
 Here the large difference in cluster production between the PHQMD and the other models at mid-rapidity
 becomes much less visible compared to the results for $\sqrt{s_{NN}}= 2.52$ GeV. That can be attributed to the fact
 that the dynamics at high energy is dominated by collisions rather than by potential interactions.
 Moreover, the QGP formation is included in the PHSD and the PHQMD explicitly, while the SMASH and the UrQMD are hadronic cascades only.

\begin{figure*}
    \resizebox{\textwidth}{!}{
        \includegraphics{plots/scenario1/header.pdf} 
    } \\
    \resizebox{\textwidth}{!}{
        \begin{tabular}{ccc}
          \includegraphics{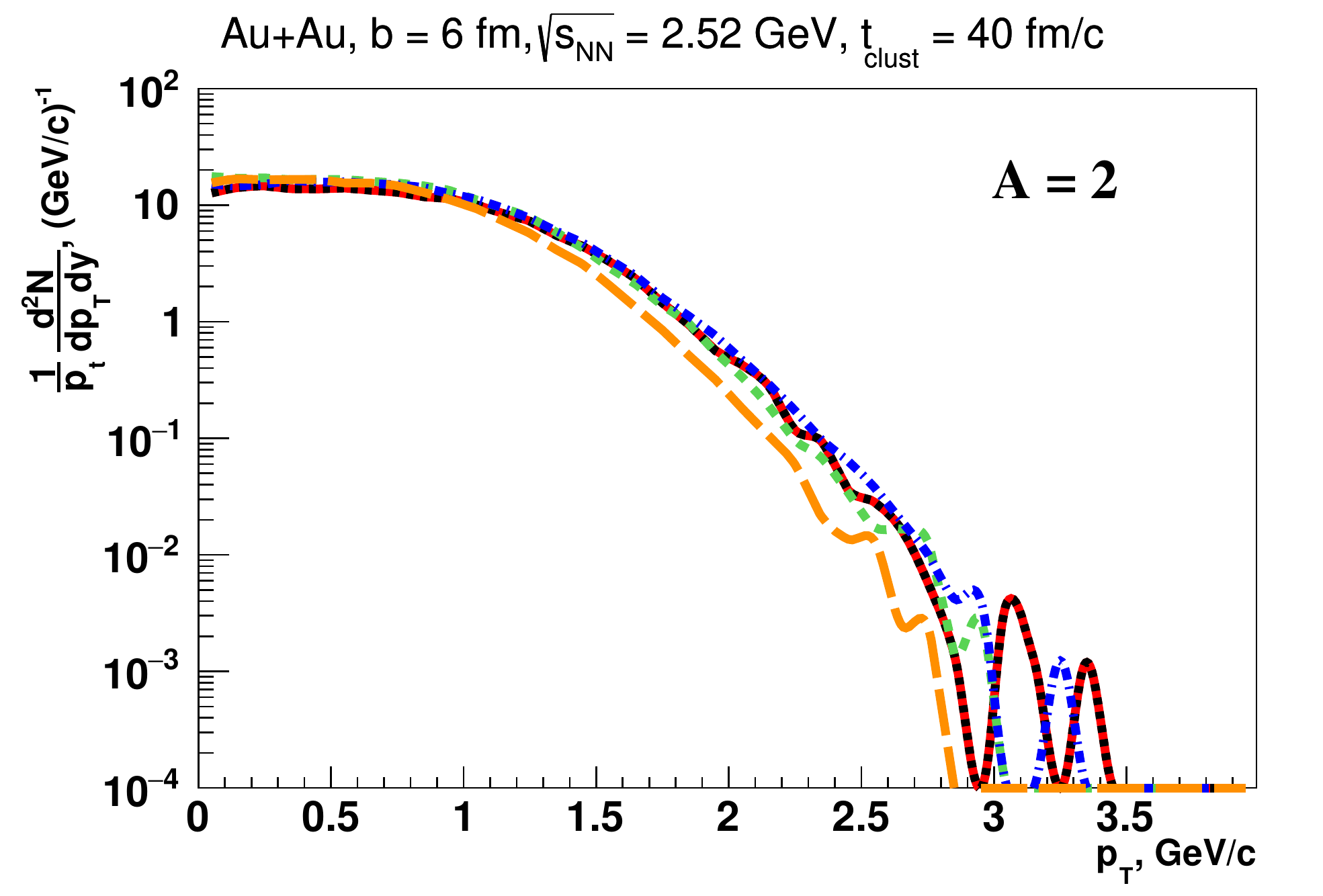} &
          \includegraphics{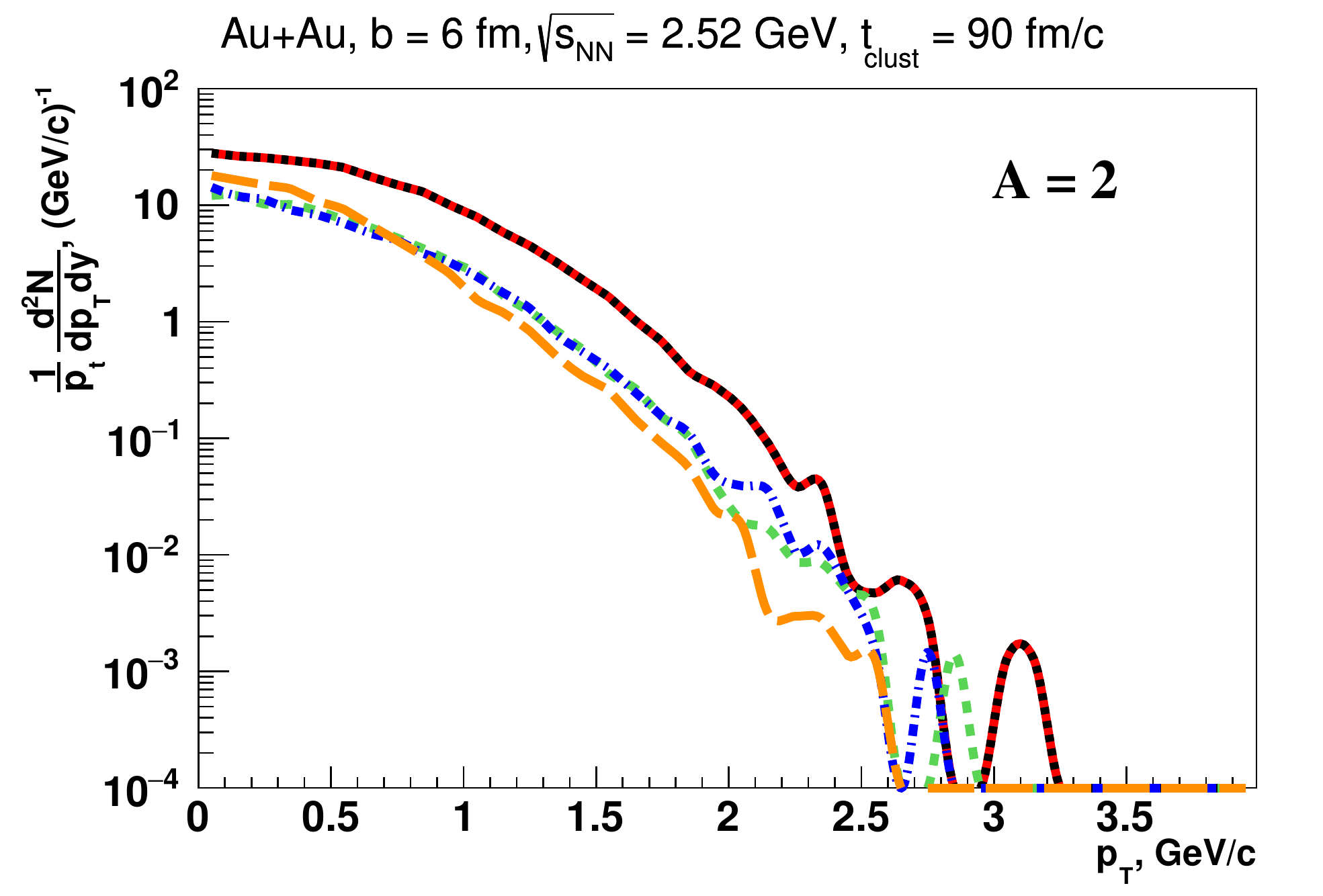} &
          \includegraphics{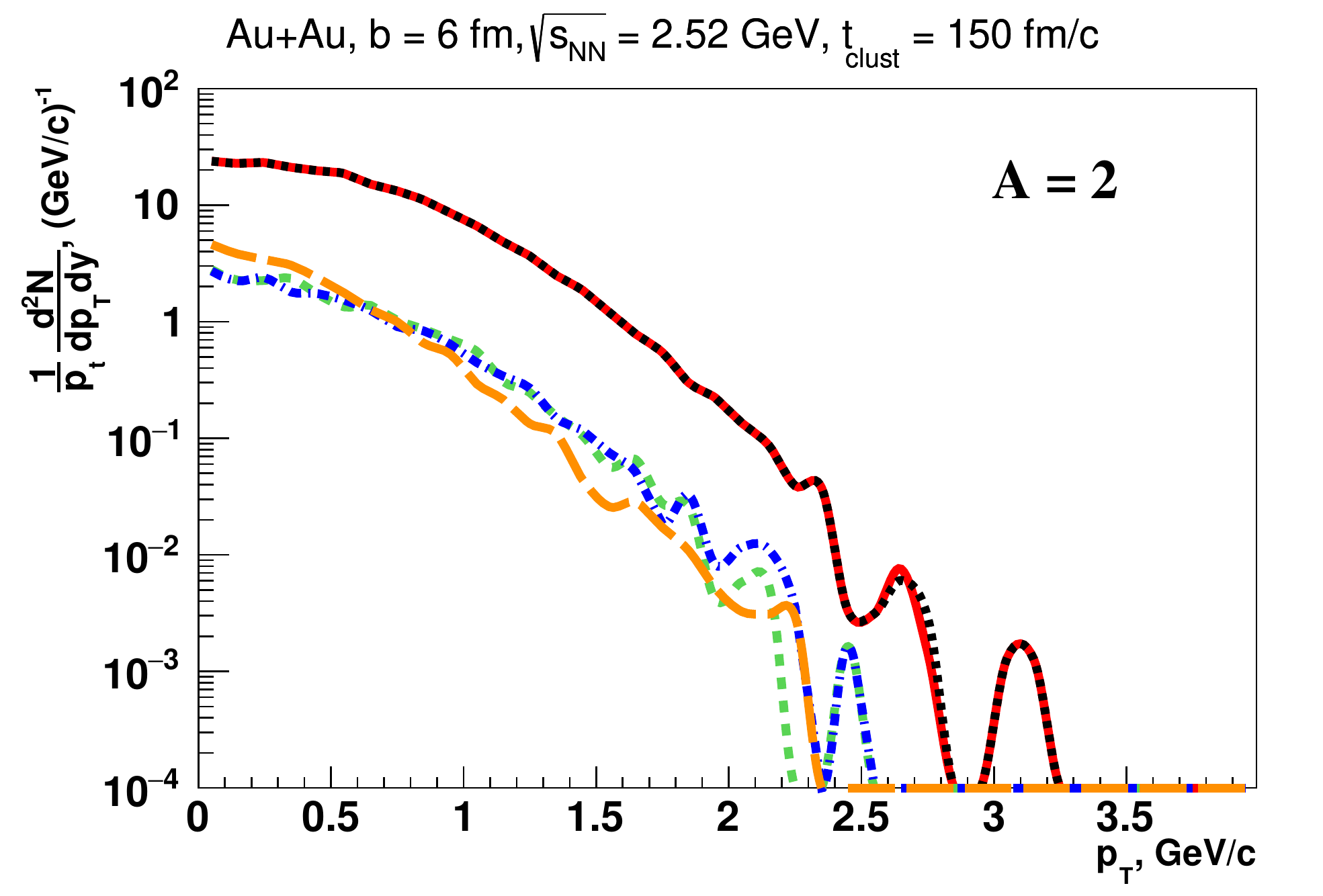} \\
          \includegraphics{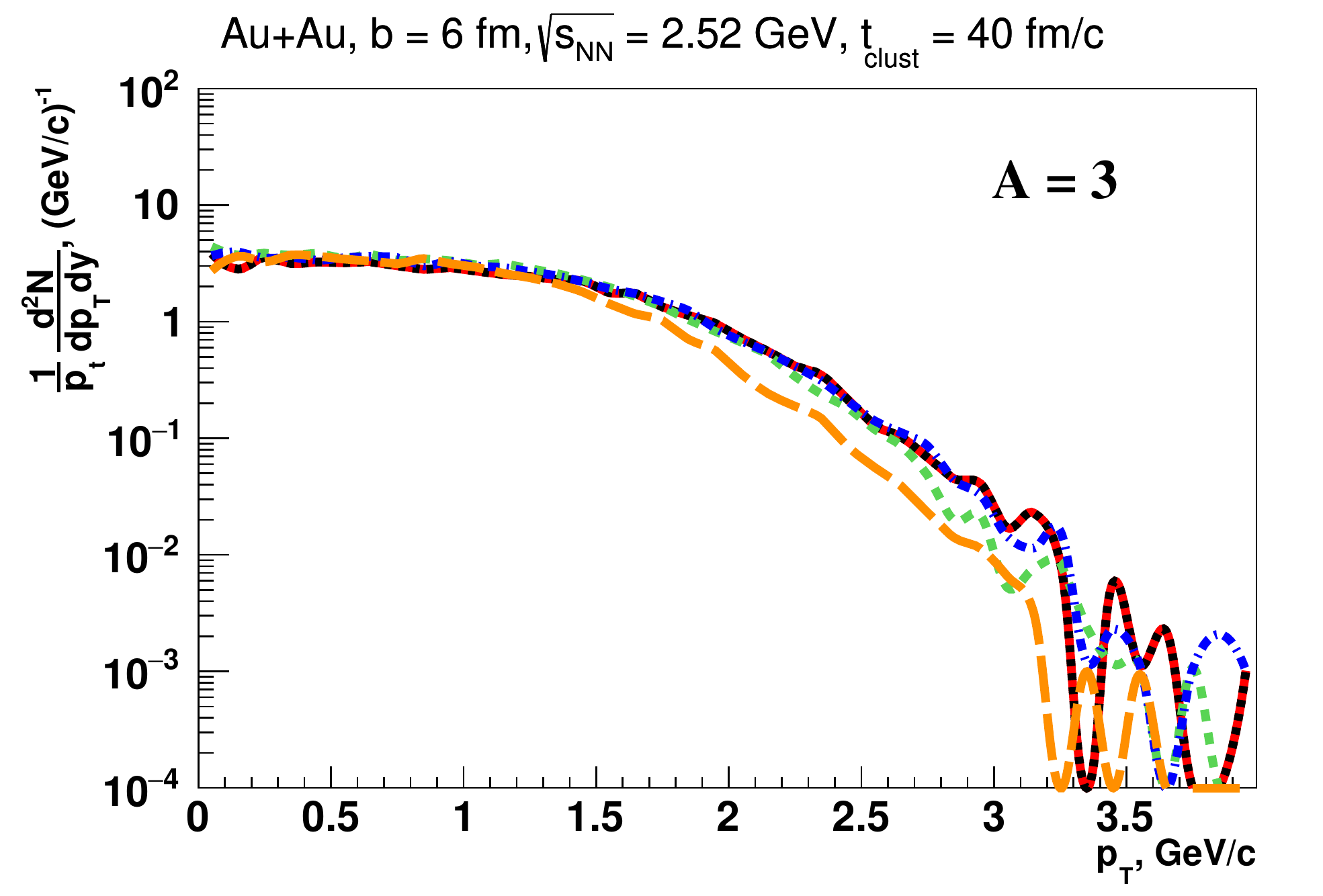} &
          \includegraphics{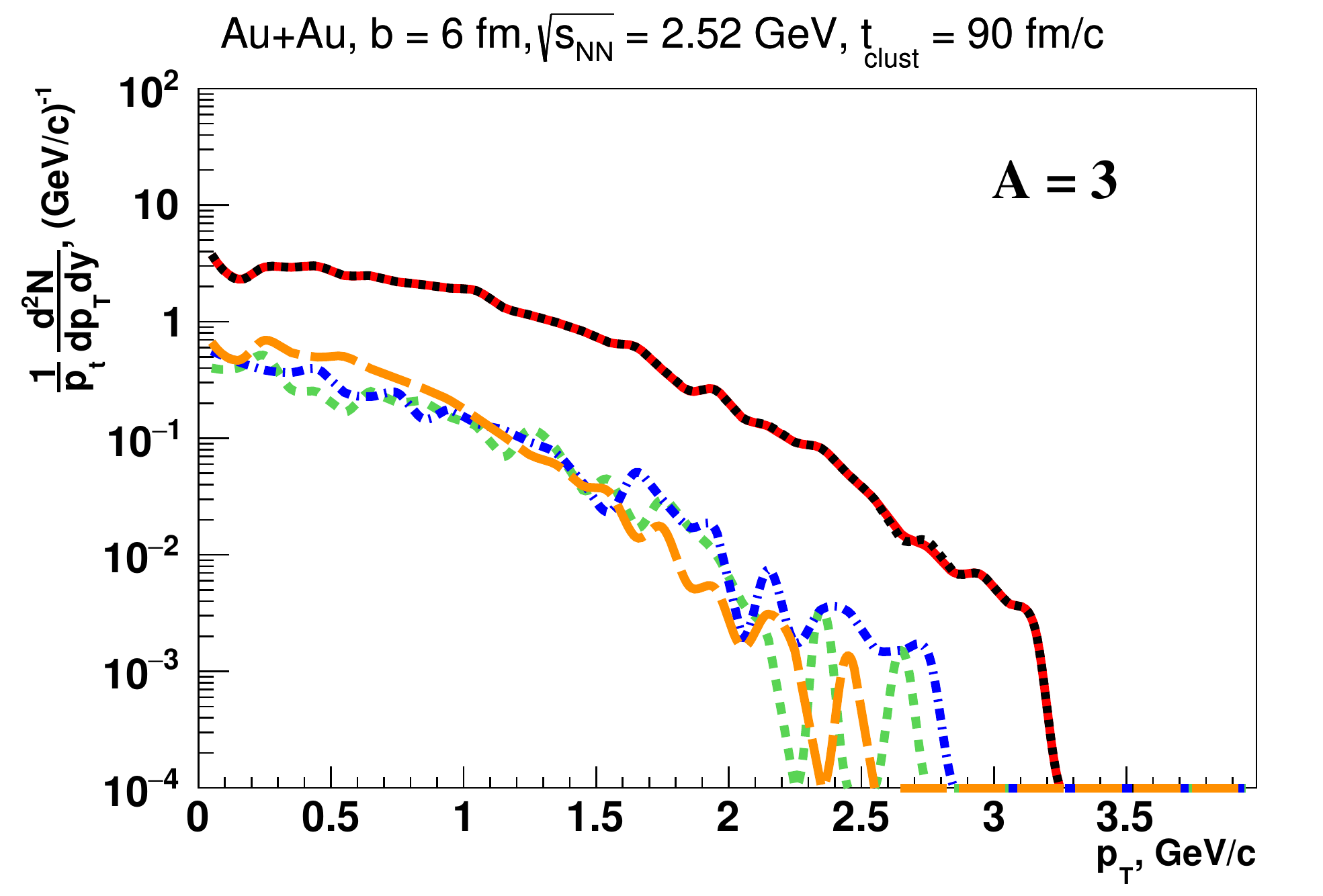} &
          \includegraphics{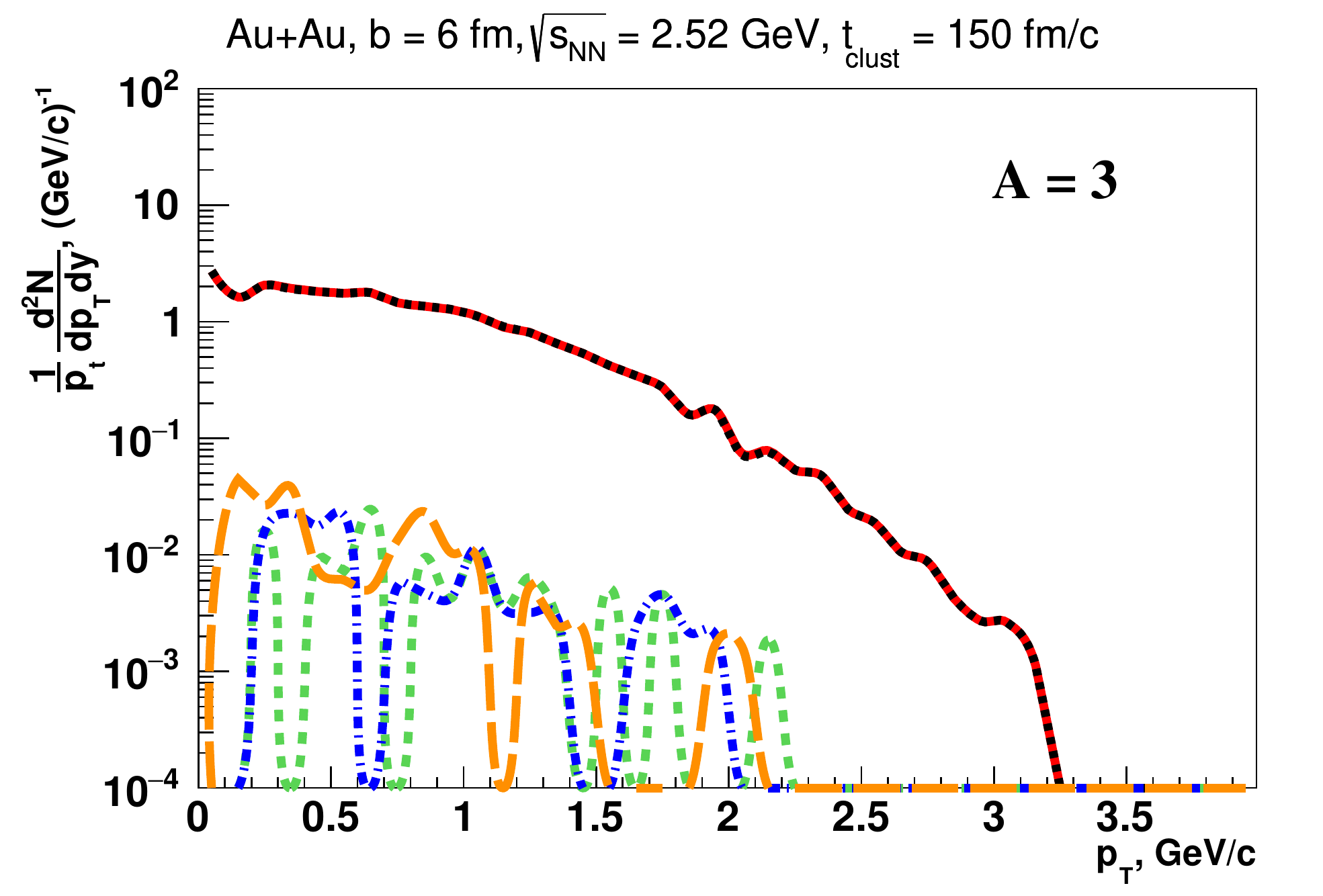} \\
        \end{tabular}
    }
\caption{\label{fig:clust_pt_2.52} The transverse momentum spectra of clusters with the mass number $A = 2$ (top row) and $A = 3$ (bottom row) at mid-rapidity $|y| < 0.5$ in semi-peripheral ($b = 6$ fm) $Au+Au$ collisions at $\sqrt{s_{NN}} = 2.52$ GeV. The left column -- $t_{clust} = 40$ fm/c, the middle column -- $t_{clust} = 90$ fm/c, the right column -- $t_{clust} = 150$ fm/c. The color coding is the same as in Fig.~\ref{fig:2.52dndyA2}.}
\end{figure*}

\begin{figure*}
    \resizebox{\textwidth}{!}{
        \includegraphics{plots/scenario1/header.pdf} 
    } \\
    \resizebox{\textwidth}{!}{
        \begin{tabular}{ccc}
          \includegraphics{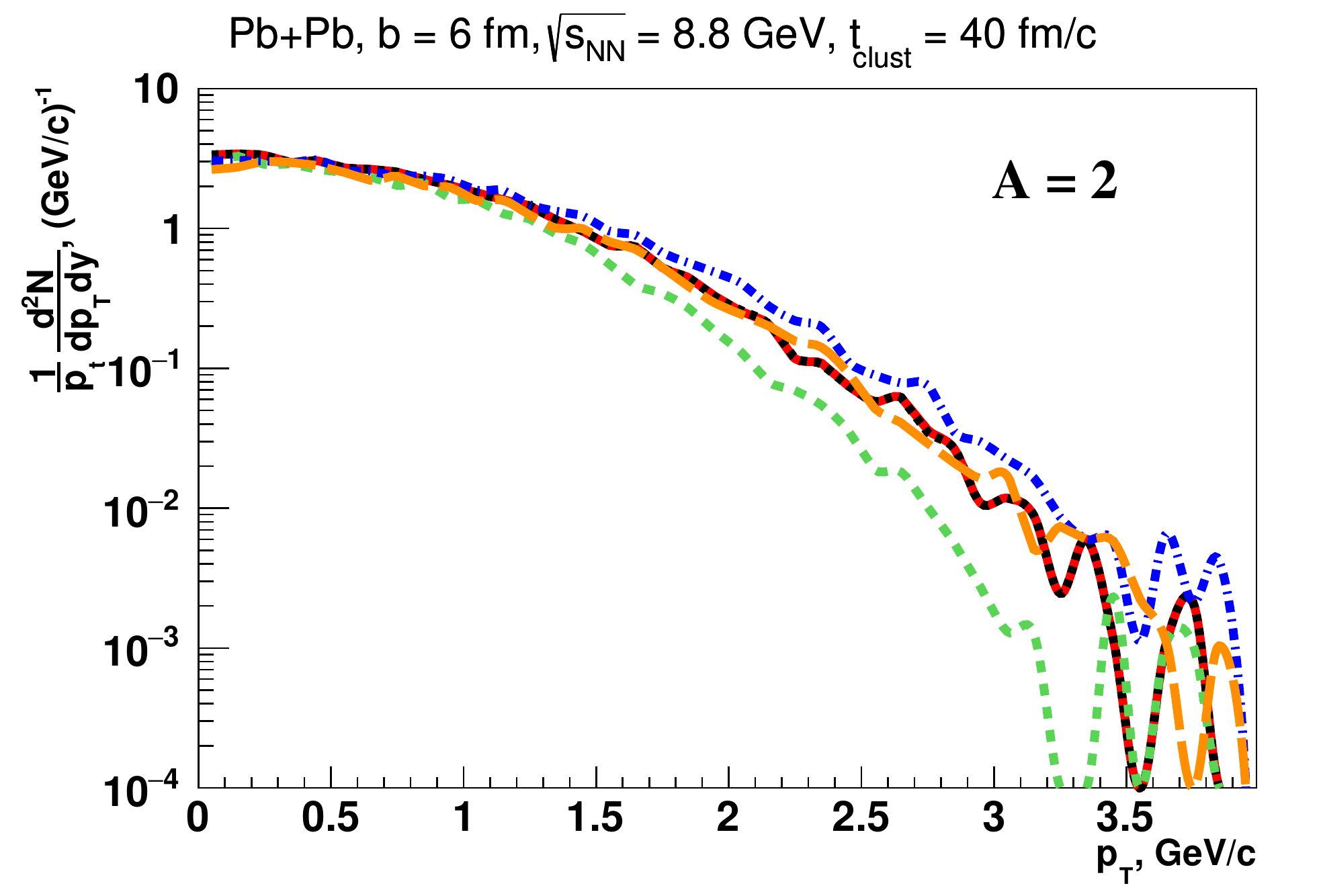} &
          \includegraphics{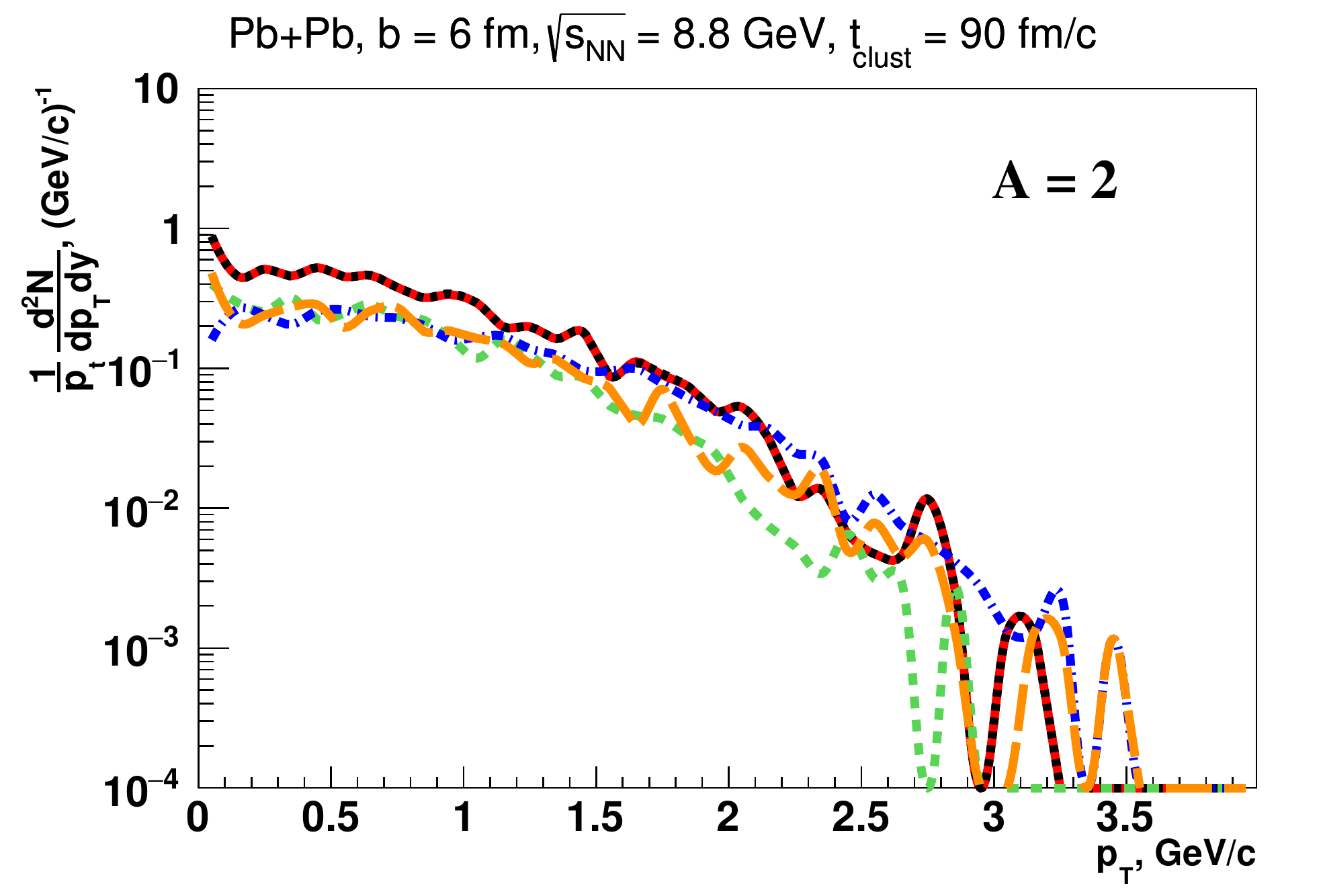} &
          \includegraphics{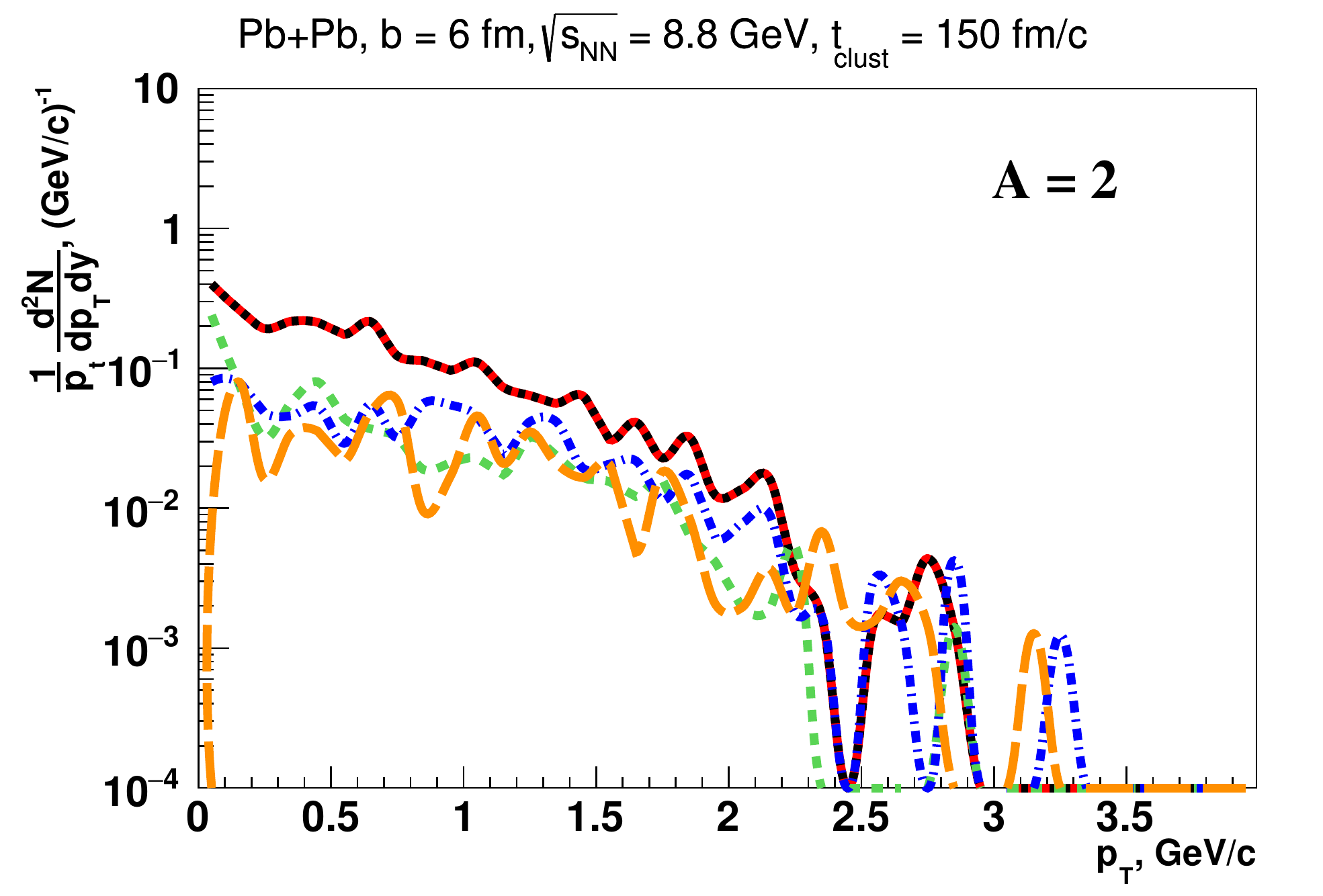} \\
        \end{tabular}
    }
\caption{\label{fig:clust_pt_8.8} The transverse momentum spectra of clusters with the mass number $A = 2$
at mid-rapidity $|y| < 0.5$ in semi-peripheral ($b = 6$ fm) $Pb+Pb$ collisions at $\sqrt{s_{NN}} = 8.8$ GeV. The left column -- $t_{clust} = 40$ fm/c, the middle column -- $t_{clust} = 90$ fm/c, the right column -- $t_{clust} = 150$ fm/c. The color coding is the same as in Fig.~\ref{fig:2.52dndyA2}.}

\end{figure*}

Finally, the transverse momentum $p_{T}$ spectra for clusters with the mass number $A = 2$ (top row) 
and $A = 3$ (bottom row) are presented in Fig.~\ref{fig:clust_pt_2.52} and Fig.~\ref{fig:clust_pt_8.8} 
for $\sqrt{s_{NN}} = 2.52$ GeV and $\sqrt{s_{NN}}=8.8$ GeV, respectively, at $t_{clust} = 40$ fm/c (left column), $t_{clust} = 90$ fm/c (central column) and $t_{clust} = 150$ fm/c (right column). 
The transverse momentum spectra  are calculated at mid-rapidity ($|y| < 0.5$).
The large splitting of the results for the $p_T$ spectra at later times is attributed to the differences in the rapidity distribution between the PHQMD and other models at low energies which are large at mid-rapidity.
We also mention that the slopes of the $p_{T}$ spectra from different models are in a good agreement. 
Unfortunately, $A = 3$ spectra were affected by insufficient statistics, especially at the higher energy, $\sqrt{s_{NN}}=8.8$ GeV.

\section{\label{sec:5}Conclusions}
In this study, the results on cluster dynamics within the novel model-independent cluster recognition library for the clusters finding "phase-space Minimum Spanning Tree" (psMST) have been presented. The psMST is a tool which can be applied to different transport approaches: the  input for psMST are the baryon coordinates and their 4-momenta at a selected time.  The psMST is based on the MST method for the cluster recognition by correlations in the coordinate space, however, it is extended for the possible inclusion of the momentum space information, which allows to study the momentum correlations of baryons in the clusters as well as to modify the criteria for the cluster recognition.

The psMST algorithm has been applied to  QMD-based (PHQMD) and mean field based (PHSD) transport approaches as well as to two cascade models, SMASH and UrQMD, which are used here without potentials.
PHSD incorporated the mean-field potential for baryons and the PHQMD follows the n-body quantum molecular dynamics based on density dependent 2-body interactions.

We find that the rapidity and momentum distributions of baryons ($p$, $n$, $\Lambda$ and $\Sigma^{0}$)
at early times are very similar within all four models. At later times
the rapidity distributions of clusters with $A=2, 3, [4\div 20]$ are rather different at low energies. The PHQMD with psMST predicts more clusters in the mid-rapidity region than the other models. 
This can be explained by the fact that the n-body quantum molecular dynamics allows to keep the
potential induced spacial correlations of baryons, contrary to the mean-field dynamics of the PHSD and the cascade versions
of SMASH and UrQMD. This observation shows the sensitivity of the clusters production 
at low energies to the realization of the potential interactions. This sensitivity is 
much more pronounced for the clusters than for the single particle observables.
At higher energies all models gives qualitatively similar results since the dynamics there is driven mainly by 
collisions and multiparticle productions rather then by potential interactions.

We note that the psMST library is an open source tool  \cite{psMSTlinkGitLab}, which can be used in a stand-alone mode or can be integrated into experimental software frameworks. Being applied to different transport models, the psMST might be useful for the simulations of the cluster production for the experimental studies of detector performances etc., which of a particular importance for the future experiments of NICA and FAIR.

\begin{acknowledgments}
We would like to acknowledge the theoretical support and discussions with Joerg Aichelin and Elena Bratkovskaya. We would also like to thank Hannah Elfner (Petersen) and Marcus Bleicher for comments and remarks on SMASH and UrQMD models.
Also we acknowledge the discussions with  Gabriele Coci, Susanne Glaessel, Vadim Kolesnikov and Vadim Voronyuk.
This work was supported by the  Russian Science Foundation, grant no. 19-42-04101 and by the RFBR according to the research project No. 18-02-40086. Furthermore, we acknowledge support by the Deutsche Forschungsgemeinschaft (DFG, German Research Foundation), grant  BR  4000/7-1;
by the Helmholtz Research Academy Hesse for FAIR (HFHF) 
and by Strong-2020, financed by the European community).
The author thanks the Giersch Science Center of the J. W. Goethe University for its hospitality.
\end{acknowledgments}

\bibliography{apssamp}

\end{document}